\magnification=\magstep1
\font \authfont               = cmr10 scaled\magstep4
\font \fivesans               = cmss10 at 5pt
\font \headfont               = cmbx12 scaled\magstep4
\font \markfont               = cmr10 scaled\magstep1
\font \ninebf                 = cmbx9
\font \ninei                  = cmmi9
\font \nineit                 = cmti9
\font \ninerm                 = cmr9
\font \ninesans               = cmss10 at 9pt
\font \ninesl                 = cmsl9
\font \ninesy                 = cmsy9
\font \ninett                 = cmtt9
\font \sevensans              = cmss10 at 7pt
\font \sixbf                  = cmbx6
\font \sixi                   = cmmi6
\font \sixrm                  = cmr6
\font \sixsans                = cmss10 at 6pt
\font \sixsy                  = cmsy6
\font \smallescriptfont       = cmr5 at 7pt
\font \smallescriptscriptfont = cmr5
\font \smalletextfont         = cmr5 at 10pt
\font \subhfont               = cmr10 scaled\magstep4
\font \tafonts                = cmbx7  scaled\magstep2
\font \tafontss               = cmbx5  scaled\magstep2
\font \tafontt                = cmbx10 scaled\magstep2
\font \tams                   = cmmib10
\font \tamss                  = cmmib10 scaled 700
\font \tamt                   = cmmib10 scaled\magstep2
\font \tass                   = cmsy7  scaled\magstep2
\font \tasss                  = cmsy5  scaled\magstep2
\font \tast                   = cmsy10 scaled\magstep2
\font \tasys                  = cmex10 scaled\magstep1
\font \tasyt                  = cmex10 scaled\magstep2
\font \tbfonts                = cmbx7  scaled\magstep1
\font \tbfontss               = cmbx5  scaled\magstep1
\font \tbfontt                = cmbx10 scaled\magstep1
\font \tbms                   = cmmib10 scaled 833
\font \tbmss                  = cmmib10 scaled 600
\font \tbmt                   = cmmib10 scaled\magstep1
\font \tbss                   = cmsy7  scaled\magstep1
\font \tbsss                  = cmsy5  scaled\magstep1
\font \tbst                   = cmsy10 scaled\magstep1
\font \tenbfne                = cmb10
\font \tensans                = cmss10
\font \tpfonts                = cmbx7  scaled\magstep3
\font \tpfontss               = cmbx5  scaled\magstep3
\font \tpfontt                = cmbx10 scaled\magstep3
\font \tpmt                   = cmmib10 scaled\magstep3
\font \tpss                   = cmsy7  scaled\magstep3
\font \tpsss                  = cmsy5  scaled\magstep3
\font \tpst                   = cmsy10 scaled\magstep3
\font \tpsyt                  = cmex10 scaled\magstep3
\vsize=24.2truecm
\hsize=15.3truecm
\hfuzz=2pt
\tolerance=500
\abovedisplayskip=3 mm plus6pt minus 4pt
\belowdisplayskip=3 mm plus6pt minus 4pt
\abovedisplayshortskip=0mm plus6pt minus 2pt
\belowdisplayshortskip=2 mm plus4pt minus 4pt
\predisplaypenalty=0
\clubpenalty=10000
\widowpenalty=10000
\frenchspacing
\newdimen\oldparindent\oldparindent=1.5em
\parindent=1.5em
\skewchar\ninei='177 \skewchar\sixi='177
\skewchar\ninesy='60 \skewchar\sixsy='60
\hyphenchar\ninett=-1
\def\newline{\hfil\break}%
\catcode`@=11
\def\folio{\ifnum\pageno<\z@
\uppercase\expandafter{\romannumeral-\pageno}%
\else\number\pageno \fi}
\catcode`@=12 
  \mathchardef\Gamma="0100
  \mathchardef\Delta="0101
  \mathchardef\Theta="0102
  \mathchardef\Lambda="0103
  \mathchardef\Xi="0104
  \mathchardef\Pi="0105
  \mathchardef\Sigma="0106
  \mathchardef\Upsilon="0107
  \mathchardef\Phi="0108
  \mathchardef\Psi="0109
  \mathchardef\Omega="010A
  \mathchardef\bfGamma="0\the\bffam 00
  \mathchardef\bfDelta="0\the\bffam 01
  \mathchardef\bfTheta="0\the\bffam 02
  \mathchardef\bfLambda="0\the\bffam 03
  \mathchardef\bfXi="0\the\bffam 04
  \mathchardef\bfPi="0\the\bffam 05
  \mathchardef\bfSigma="0\the\bffam 06
  \mathchardef\bfUpsilon="0\the\bffam 07
  \mathchardef\bfPhi="0\the\bffam 08
  \mathchardef\bfPsi="0\the\bffam 09
  \mathchardef\bfOmega="0\the\bffam 0A

\def\sq{\hbox{\rlap{$\sqcap$}$\sqcup$}}

\def\utw{\smash{\rlap{\lower5pt\hbox{$\sim$}}}}
\def\udtw{\smash{\rlap{\lower6pt\hbox{$\approx$}}}}

\def\diameter{{\ifmmode\mathchoice
{\ooalign{\hfil\hbox{$\displaystyle/$}\hfil\crcr
{\hbox{$\displaystyle\mathchar"20D$}}}}
{\ooalign{\hfil\hbox{$\textstyle/$}\hfil\crcr
{\hbox{$\textstyle\mathchar"20D$}}}}
{\ooalign{\hfil\hbox{$\scriptstyle/$}\hfil\crcr
{\hbox{$\scriptstyle\mathchar"20D$}}}}
{\ooalign{\hfil\hbox{$\scriptscriptstyle/$}\hfil\crcr
{\hbox{$\scriptscriptstyle\mathchar"20D$}}}}
\else{\ooalign{\hfil/\hfil\crcr\mathhexbox20D}}%
\fi}}


\def\bbbc{{\mathchoice {\setbox0=\hbox{$\displaystyle\rm C$}\hbox{\hbox
to0pt{\kern0.4\wd0\vrule height0.9\ht0\hss}\box0}}
{\setbox0=\hbox{$\textstyle\rm C$}\hbox{\hbox
to0pt{\kern0.4\wd0\vrule height0.9\ht0\hss}\box0}}
{\setbox0=\hbox{$\scriptstyle\rm C$}\hbox{\hbox
to0pt{\kern0.4\wd0\vrule height0.9\ht0\hss}\box0}}
{\setbox0=\hbox{$\scriptscriptstyle\rm C$}\hbox{\hbox
to0pt{\kern0.4\wd0\vrule height0.9\ht0\hss}\box0}}}}
\def\bbbe{{\mathchoice {\setbox0=\hbox{\smalletextfont e}\hbox{\raise
0.1\ht0\hbox to0pt{\kern0.4\wd0\vrule width0.3pt height0.7\ht0\hss}\box0}}
{\setbox0=\hbox{\smalletextfont e}\hbox{\raise
0.1\ht0\hbox to0pt{\kern0.4\wd0\vrule width0.3pt height0.7\ht0\hss}\box0}}
{\setbox0=\hbox{\smallescriptfont e}\hbox{\raise
0.1\ht0\hbox to0pt{\kern0.5\wd0\vrule width0.2pt height0.7\ht0\hss}\box0}}
{\setbox0=\hbox{\smallescriptscriptfont e}\hbox{\raise
0.1\ht0\hbox to0pt{\kern0.4\wd0\vrule width0.2pt height0.7\ht0\hss}\box0}}}}
\def\bbbq{{\mathchoice {\setbox0=\hbox{$\displaystyle\rm Q$}\hbox{\raise
0.15\ht0\hbox to0pt{\kern0.4\wd0\vrule height0.8\ht0\hss}\box0}}
{\setbox0=\hbox{$\textstyle\rm Q$}\hbox{\raise
0.15\ht0\hbox to0pt{\kern0.4\wd0\vrule height0.8\ht0\hss}\box0}}
{\setbox0=\hbox{$\scriptstyle\rm Q$}\hbox{\raise
0.15\ht0\hbox to0pt{\kern0.4\wd0\vrule height0.7\ht0\hss}\box0}}
{\setbox0=\hbox{$\scriptscriptstyle\rm Q$}\hbox{\raise
0.15\ht0\hbox to0pt{\kern0.4\wd0\vrule height0.7\ht0\hss}\box0}}}}
\def\bbbt{{\mathchoice {\setbox0=\hbox{$\displaystyle\rm
T$}\hbox{\hbox to0pt{\kern0.3\wd0\vrule height0.9\ht0\hss}\box0}}
{\setbox0=\hbox{$\textstyle\rm T$}\hbox{\hbox
to0pt{\kern0.3\wd0\vrule height0.9\ht0\hss}\box0}}
{\setbox0=\hbox{$\scriptstyle\rm T$}\hbox{\hbox
to0pt{\kern0.3\wd0\vrule height0.9\ht0\hss}\box0}}
{\setbox0=\hbox{$\scriptscriptstyle\rm T$}\hbox{\hbox
to0pt{\kern0.3\wd0\vrule height0.9\ht0\hss}\box0}}}}
\def\bbbs{{\mathchoice
{\setbox0=\hbox{$\displaystyle     \rm S$}\hbox{\raise0.5\ht0\hbox
to0pt{\kern0.35\wd0\vrule height0.45\ht0\hss}\hbox
to0pt{\kern0.55\wd0\vrule height0.5\ht0\hss}\box0}}
{\setbox0=\hbox{$\textstyle        \rm S$}\hbox{\raise0.5\ht0\hbox
to0pt{\kern0.35\wd0\vrule height0.45\ht0\hss}\hbox
to0pt{\kern0.55\wd0\vrule height0.5\ht0\hss}\box0}}
{\setbox0=\hbox{$\scriptstyle      \rm S$}\hbox{\raise0.5\ht0\hbox
to0pt{\kern0.35\wd0\vrule height0.45\ht0\hss}\raise0.05\ht0\hbox
to0pt{\kern0.5\wd0\vrule height0.45\ht0\hss}\box0}}
{\setbox0=\hbox{$\scriptscriptstyle\rm S$}\hbox{\raise0.5\ht0\hbox
to0pt{\kern0.4\wd0\vrule height0.45\ht0\hss}\raise0.05\ht0\hbox
to0pt{\kern0.55\wd0\vrule height0.45\ht0\hss}\box0}}}}
\def\bbbz{{\mathchoice {\hbox{$\sans\textstyle Z\kern-0.4em Z$}}
{\hbox{$\sans\textstyle Z\kern-0.4em Z$}}
{\hbox{$\sans\scriptstyle Z\kern-0.3em Z$}}
{\hbox{$\sans\scriptscriptstyle Z\kern-0.2em Z$}}}}
\def\qed{\ifmmode\sq\else{\unskip\nobreak\hfil
\penalty50\hskip1em\null\nobreak\hfil\sq
\parfillskip=0pt\finalhyphendemerits=0\endgraf}\fi}
\newfam\sansfam
\textfont\sansfam=\tensans\scriptfont\sansfam=\sevensans
\scriptscriptfont\sansfam=\fivesans
\def\sans{\fam\sansfam\tensans}
\def\stackfigbox{\if
Y\FIG\global\setbox\figbox=\vbox{\unvbox\figbox\box1}%
\else\global\setbox\figbox=\vbox{\box1}\global\let\FIG=Y\fi}
\def\placefigure{\dimen0=\ht1\advance\dimen0by\dp1
\advance\dimen0by5\baselineskip
\advance\dimen0by0.4true cm
\ifdim\dimen0>\vsize\pageinsert\box1\vfill\endinsert
\else
\if Y\FIG\stackfigbox\else
\dimen0=\pagetotal\ifdim\dimen0<\pagegoal
\advance\dimen0by\ht1\advance\dimen0by\dp1\advance\dimen0by1.4true cm
\ifdim\dimen0>\pagegoal\stackfigbox
\else\box1\vskip4true mm\fi
\else\box1\vskip4true mm\fi\fi\fi}
%
\def\begfig#1cm#2\endfig{\par
\setbox1=\vbox{\dimen0=#1true cm\advance\dimen0
by1true cm\kern\dimen0#2}\placefigure}
\def\begdoublefig#1cm #2 #3 \enddoublefig{\begfig#1cm%
\vskip-.8333\baselineskip\line{\vtop{\hsize=0.46\hsize#2}\hfill
\vtop{\hsize=0.46\hsize#3}}\endfig}
\def\begfigsidebottom#1cm#2cm#3\endfigsidebottom{\dimen0=#2true cm
\ifdim\dimen0<0.4\hsize\message{Room for legend to narrow;
begfigsidebottom changed to begfig}\begfig#1cm#3\endfig\else
\par\def\figure##1##2{\vbox{\noindent\petit{\bf
Fig.\ts##1\unskip.\ }\ignorespaces ##2\par}}%
\dimen0=\hsize\advance\dimen0 by-.8true cm\advance\dimen0 by-#2true cm
\setbox1=\vbox{\hbox{\hbox to\dimen0{\vrule height#1true cm\hrulefill}%
\kern.8true cm\vbox{\hsize=#2true cm#3}}}\placefigure\fi}
\def\begfigsidetop#1cm#2cm#3\endfigsidetop{\dimen0=#2true cm
\ifdim\dimen0<0.4\hsize\message{Room for legend to narrow; begfigsidetop
changed to begfig}\begfig#1cm#3\endfig\else
\par\def\figure##1##2{\vbox{\noindent\petit{\bf
Fig.\ts##1\unskip.\ }\ignorespaces ##2\par}}%
\dimen0=\hsize\advance\dimen0 by-.8true cm\advance\dimen0 by-#2true cm
\setbox1=\vbox{\hbox{\hbox to\dimen0{\vrule height#1true cm\hrulefill}%
\kern.8true cm\vbox to#1true cm{\hsize=#2 true cm#3\vfill
}}}\placefigure\fi}
\def\figure#1#2{\vskip1true cm\setbox0=\vbox{\noindent\petit{\bf
Fig.\ts#1\unskip.\ }\ignorespaces #2\smallskip
\count255=0\global\advance\count255by\prevgraf}%
\ifnum\count255>1\box0\else
\centerline{\petit{\bf Fig.\ts#1\unskip.\
}\ignorespaces#2}\smallskip\fi}

\def\begtab#1cm#2\endtab{\par
   \ifvoid\topins\midinsert\medskip\vbox{#2\kern#1true cm}\endinsert
   \else\topinsert\vbox{#2\kern#1true cm}\endinsert\fi}
\def\begpet{\vskip6pt\bgroup\petit}
\def\endpet{\vskip6pt\egroup}
\newcount\frpages
\newcount\frpagegoal
\def\freepage#1{\global\frpagegoal=#1\relax\global\frpages=0\relax
\loop\global\advance\frpages by 1\relax
\message{Doing freepage \the\frpages\space of
\the\frpagegoal}\null\vfill\eject
\ifnum\frpagegoal>\frpages\repeat}
\newdimen\refindent
\def\begrefchapter#1{\titlea{}{\ignorespaces#1}%
\bgroup\petit
\setbox0=\hbox{1000.\enspace}\refindent=\wd0}
\def\ref{\goodbreak
\hangindent\oldparindent\hangafter=1
\noindent\ignorespaces}
\def\refno#1{\goodbreak
\hangindent\refindent\hangafter=1
\noindent\hbox to\refindent{#1\hss}\ignorespaces}
\def\endref{\goodbreak\endpet}
\def\vec#1{{\textfont1=\tams\scriptfont1=\tamss
\textfont0=\tenbf\scriptfont0=\sevenbf
\mathchoice{\hbox{$\displaystyle#1$}}{\hbox{$\textstyle#1$}}
{\hbox{$\scriptstyle#1$}}{\hbox{$\scriptscriptstyle#1$}}}}
\def\petit{\def\rm{\fam0\ninerm}%
\textfont0=\ninerm \scriptfont0=\sixrm \scriptscriptfont0=\fiverm
 \textfont1=\ninei \scriptfont1=\sixi \scriptscriptfont1=\fivei
 \textfont2=\ninesy \scriptfont2=\sixsy \scriptscriptfont2=\fivesy
 \def\it{\fam\itfam\nineit}%
 \textfont\itfam=\nineit
 \def\sl{\fam\slfam\ninesl}%
 \textfont\slfam=\ninesl
 \def\bf{\fam\bffam\ninebf}%
 \textfont\bffam=\ninebf \scriptfont\bffam=\sixbf
 \scriptscriptfont\bffam=\fivebf
 \def\sans{\fam\sansfam\ninesans}%
 \textfont\sansfam=\ninesans \scriptfont\sansfam=\sixsans
 \scriptscriptfont\sansfam=\fivesans
 \def\tt{\fam\ttfam\ninett}%
 \textfont\ttfam=\ninett
 \normalbaselineskip=11pt
 \setbox\strutbox=\hbox{\vrule height7pt depth2pt width0pt}%
 \normalbaselines\rm
\def\vec##1{{\textfont1=\tbms\scriptfont1=\tbmss
\textfont0=\ninebf\scriptfont0=\sixbf
\mathchoice{\hbox{$\displaystyle##1$}}{\hbox{$\textstyle##1$}}
{\hbox{$\scriptstyle##1$}}{\hbox{$\scriptscriptstyle##1$}}}}}
\nopagenumbers
%
\let\header=Y
\let\FIG=N
\newbox\figbox
\output={\if N\header\headline={\hfil}\fi\plainoutput\global\let\header=Y
\if Y\FIG\topinsert\unvbox\figbox\endinsert\global\let\FIG=N\fi}
\let\lasttitle=N
\def\bookauthor#1{\vfill\eject
     \bgroup
     \baselineskip=22pt
     \lineskip=0pt
     \pretolerance=10000
     \authfont
     \rightskip 0pt plus 6em
     \centerpar{#1}\vskip2true cm\egroup}
\def\bookhead#1#2{\bgroup
     \baselineskip=36pt
     \lineskip=0pt
     \pretolerance=10000
     \headfont
     \rightskip 0pt plus 6em
     \centerpar{#1}\vskip1true cm
     \baselineskip=22pt
     \subhfont\centerpar{#2}\vfill
     \parindent=0pt
     \baselineskip=16pt
     \leftskip=2.2true cm
     \markfont Springer-Verlag\newline
     Berlin Heidelberg New York\newline
     London Paris Tokyo Singapore\bigskip\bigskip
     [{\it This is page III of your manuscript and will be reset by
     Springer.}]
     \egroup\let\header=N\eject}
\def\centerpar#1{{\parfillskip=0pt
\rightskip=0pt plus 1fil
\leftskip=0pt plus 1fil
\advance\leftskip by\oldparindent
\advance\rightskip by\oldparindent
\def\newline{\break}%
\noindent\ignorespaces#1\par}}
\def\part#1#2{\vfill\supereject\let\header=N
\centerline{\subhfont#1}%
\vskip75pt
     \bgroup
\textfont0=\tpfontt \scriptfont0=\tpfonts \scriptscriptfont0=\tpfontss
\textfont1=\tpmt \scriptfont1=\tbmt \scriptscriptfont1=\tams
\textfont2=\tpst \scriptfont2=\tpss \scriptscriptfont2=\tpsss
\textfont3=\tpsyt \scriptfont3=\tasys \scriptscriptfont3=\tenex
     \baselineskip=20pt
     \lineskip=0pt
     \pretolerance=10000
     \tpfontt
     \centerpar{#2}
     \vfill\eject\egroup\ignorespaces}
\newtoks\AUTHOR
\newtoks\HEAD
\catcode`\@=\active
\def\author#1{\bgroup
\baselineskip=22pt
\lineskip=0pt
\pretolerance=10000
\markfont
\centerpar{#1}\bigskip\egroup
{\def@##1{}%
\setbox0=\hbox{\petit\kern2.5true cc\ignorespaces#1\unskip}%
\ifdim\wd0>\hsize
\message{The names of the authors exceed the headline, please use a }%
\message{short form with AUTHORRUNNING}\gdef\leftheadline{%
\hbox to2.5true cc{\folio\hfil}AUTHORS suppressed due to excessive
length\hfil}%
\global\AUTHOR={AUTHORS were to long}\else
\xdef\leftheadline{\hbox to2.5true
cc{\noexpand\folio\hfil}\ignorespaces#1\hfill}%
\global\AUTHOR={\def@##1{}\ignorespaces#1\unskip}\fi
}\let\INS=E}
\def\address#1{\bgroup
\centerpar{#1}\bigskip\egroup
\catcode`\@=12
\vskip2cm\noindent\ignorespaces}
\let\INS=N%
\def@#1{\if N\INS\unskip\ $^{#1}$\else\if
E\INS\noindent$^{#1}$\let\INS=Y\ignorespaces
\else\par
\noindent$^{#1}$\ignorespaces\fi\fi}%
\catcode`\@=12
\headline={\petit\def\newline{ }\def\fonote#1{}\ifodd\pageno
\rightheadline\else\leftheadline\fi}
\def\rightheadline{\hfil Missing CONTRIBUTION
title\hbox to2.5true cc{\hfil\folio}}
\def\leftheadline{\hbox to2.5true cc{\folio\hfil}Missing name(s) of the
author(s)\hfil}
\nopagenumbers
\let\header=Y

\let\lasttitle=N
 \def\contribution#1{\vfill\supereject
 \ifodd\pageno\else\null\vfill\supereject\fi
 \let\header=N\bgroup
 \textfont0=\tafontt \scriptfont0=\tafonts \scriptscriptfont0=\tafontss
 \textfont1=\tamt \scriptfont1=\tams \scriptscriptfont1=\tams
 \textfont2=\tast \scriptfont2=\tass \scriptscriptfont2=\tasss
 \par\baselineskip=16pt
     \lineskip=16pt
     \tafontt
     \raggedright
     \pretolerance=10000
     \noindent
     \centerpar{\ignorespaces#1}%
     \vskip12pt\egroup
     \nobreak
     \parindent=0pt
     \everypar={\global\parindent=1.5em
     \global\let\lasttitle=N\global\everypar={}}%
     \global\let\lasttitle=A%
     \setbox0=\hbox{\petit\def\newline{ }\def\fonote##1{}\kern2.5true
     cc\ignorespaces#1}\ifdim\wd0>\hsize
     \message{Your CONTRIBUTIONtitle exceeds the headline,
please use a short form
with CONTRIBUTIONRUNNING}\gdef\rightheadline{\hfil CONTRIBUTION title
suppressed due to excessive length\hbox to2.5true cc{\hfil\folio}}%
\global\HEAD={HEAD was to long}\else
\gdef\rightheadline{\hfill\ignorespaces#1\unskip\hbox to2.5true
cc{\hfil\folio}}\global\HEAD={\ignorespaces#1\unskip}\fi
\catcode`\@=\active
     \ignorespaces}
 \def\contributionnext#1{\vfill\supereject
 \let\header=N\bgroup
 \textfont0=\tafontt \scriptfont0=\tafonts \scriptscriptfont0=\tafontss
 \textfont1=\tamt \scriptfont1=\tams \scriptscriptfont1=\tams
 \textfont2=\tast \scriptfont2=\tass \scriptscriptfont2=\tasss
 \par\baselineskip=16pt
     \lineskip=16pt
     \tafontt
     \raggedright
     \pretolerance=10000
     \noindent
     \centerpar{\ignorespaces#1}%
     \vskip12pt\egroup
     \nobreak
     \parindent=0pt
     \everypar={\global\parindent=1.5em
     \global\let\lasttitle=N\global\everypar={}}%
     \global\let\lasttitle=A%
     \setbox0=\hbox{\petit\def\newline{ }\def\fonote##1{}\kern2.5true
     cc\ignorespaces#1}\ifdim\wd0>\hsize
     \message{Your CONTRIBUTIONtitle exceeds the headline,
please use a short form
with CONTRIBUTIONRUNNING}\gdef\rightheadline{\hfil CONTRIBUTION title
suppressed due to excessive length\hbox to2.5true cc{\hfil\folio}}%
\global\HEAD={HEAD was to long}\else
\gdef\rightheadline{\hfill\ignorespaces#1\unskip\hbox to2.5true
cc{\hfil\folio}}\global\HEAD={\ignorespaces#1\unskip}\fi
\catcode`\@=\active
     \ignorespaces}
\def\motto#1#2{\bgroup\petit\leftskip=6.5true cm\noindent\ignorespaces#1
\if!#2!\else\medskip\noindent\it\ignorespaces#2\fi\bigskip\egroup
\let\lasttitle=M
\parindent=0pt
\everypar={\global\parindent=\oldparindent
\global\let\lasttitle=N\global\everypar={}}%
\global\let\lasttitle=M%
\ignorespaces}
\def\abstract#1{\bgroup\petit\noindent
{\bf Abstract: }\ignorespaces#1\vskip28pt\egroup
\let\lasttitle=N
\parindent=0pt
\everypar={\global\parindent=\oldparindent
\global\let\lasttitle=N\global\everypar={}}%
\ignorespaces}
\def\titlea#1#2{\if N\lasttitle\else\vskip-28pt
     \fi
     \vskip18pt plus 4pt minus4pt
     \bgroup
\textfont0=\tbfontt \scriptfont0=\tbfonts \scriptscriptfont0=\tbfontss
\textfont1=\tbmt \scriptfont1=\tbms \scriptscriptfont1=\tbmss
\textfont2=\tbst \scriptfont2=\tbss \scriptscriptfont2=\tbsss
\textfont3=\tasys \scriptfont3=\tenex \scriptscriptfont3=\tenex
     \baselineskip=16pt
     \lineskip=0pt
     \pretolerance=10000
     \noindent
     \tbfontt
     \rightskip 0pt plus 6em
     \setbox0=\vbox{\vskip23pt\def\fonote##1{}%
     \noindent
     \if!#1!\ignorespaces#2
     \else\setbox0=\hbox{\ignorespaces#1\unskip\ }\hangindent=\wd0
     \hangafter=1\box0\ignorespaces#2\fi
     \vskip18pt}%
     \dimen0=\pagetotal\advance\dimen0 by-\pageshrink
     \ifdim\dimen0<\pagegoal
     \dimen0=\ht0\advance\dimen0 by\dp0\advance\dimen0 by
     3\normalbaselineskip
     \advance\dimen0 by\pagetotal
     \ifdim\dimen0>\pagegoal\eject\fi\fi
     \noindent
     \if!#1!\ignorespaces#2
     \else\setbox0=\hbox{\ignorespaces#1\unskip\ }\hangindent=\wd0
     \hangafter=1\box0\ignorespaces#2\fi
     \vskip18pt plus4pt minus4pt\egroup
     \nobreak
     \parindent=0pt
     \everypar={\global\parindent=\oldparindent
     \global\let\lasttitle=N\global\everypar={}}%
     \global\let\lasttitle=A%
     \ignorespaces}
 \def\titleb#1#2{\if N\lasttitle\else\vskip-28pt
     \fi
     \vskip18pt plus 4pt minus4pt
     \bgroup
\textfont0=\tenbf \scriptfont0=\sevenbf \scriptscriptfont0=\fivebf
\textfont1=\tams \scriptfont1=\tamss \scriptscriptfont1=\tbmss
     \lineskip=0pt
     \pretolerance=10000
     \noindent
     \bf
     \rightskip 0pt plus 6em
     \setbox0=\vbox{\vskip23pt\def\fonote##1{}%
     \noindent
     \if!#1!\ignorespaces#2
     \else\setbox0=\hbox{\ignorespaces#1\unskip\enspace}\hangindent=\wd0
     \hangafter=1\box0\ignorespaces#2\fi
     \vskip10pt}%
     \dimen0=\pagetotal\advance\dimen0 by-\pageshrink
     \ifdim\dimen0<\pagegoal
     \dimen0=\ht0\advance\dimen0 by\dp0\advance\dimen0 by
     3\normalbaselineskip
     \advance\dimen0 by\pagetotal
     \ifdim\dimen0>\pagegoal\eject\fi\fi
     \noindent
     \if!#1!\ignorespaces#2
     \else\setbox0=\hbox{\ignorespaces#1\unskip\enspace}\hangindent=\wd0
     \hangafter=1\box0\ignorespaces#2\fi
     \vskip8pt plus4pt minus4pt\egroup
     \nobreak
     \parindent=0pt
     \everypar={\global\parindent=\oldparindent
     \global\let\lasttitle=N\global\everypar={}}%
     \global\let\lasttitle=B%
     \ignorespaces}
 \def\titlec#1#2{\if N\lasttitle\else\vskip-23pt
     \fi
     \vskip18pt plus 4pt minus4pt
     \bgroup
\textfont0=\tenbfne \scriptfont0=\sevenbf \scriptscriptfont0=\fivebf
\textfont1=\tams \scriptfont1=\tamss \scriptscriptfont1=\tbmss
     \tenbfne
     \lineskip=0pt
     \pretolerance=10000
     \noindent
     \rightskip 0pt plus 6em
     \setbox0=\vbox{\vskip23pt\def\fonote##1{}%
     \noindent
     \if!#1!\ignorespaces#2
     \else\setbox0=\hbox{\ignorespaces#1\unskip\enspace}\hangindent=\wd0
     \hangafter=1\box0\ignorespaces#2\fi
     \vskip6pt}%
     \dimen0=\pagetotal\advance\dimen0 by-\pageshrink
     \ifdim\dimen0<\pagegoal
     \dimen0=\ht0\advance\dimen0 by\dp0\advance\dimen0 by
     2\normalbaselineskip
     \advance\dimen0 by\pagetotal
     \ifdim\dimen0>\pagegoal\eject\fi\fi
     \noindent
     \if!#1!\ignorespaces#2
     \else\setbox0=\hbox{\ignorespaces#1\unskip\enspace}\hangindent=\wd0
     \hangafter=1\box0\ignorespaces#2\fi
     \vskip6pt plus4pt minus4pt\egroup
     \nobreak
     \parindent=0pt
     \everypar={\global\parindent=\oldparindent
     \global\let\lasttitle=N\global\everypar={}}%
     \global\let\lasttitle=C%
     \ignorespaces}
 \def\titled#1{\if N\lasttitle\else\vskip-\baselineskip
     \fi
     \vskip12pt plus 4pt minus 4pt
     \bgroup
\textfont1=\tams \scriptfont1=\tamss \scriptscriptfont1=\tbmss
     \bf
     \noindent
     \ignorespaces#1\ \ignorespaces\egroup
     \ignorespaces}
\let\ts=\thinspace
\def\footnoterule{\kern-3pt\hrule width 2true cm\kern2.6pt}
\newcount\footcount \footcount=0
\def\advftncnt{\advance\footcount by1\global\footcount=\footcount}
\def\fonote#1{\advftncnt$^{\the\footcount}$\begingroup\petit
\parfillskip=0pt plus 1fil
\def\textindent##1{\hangindent0.5\oldparindent\noindent\hbox
to0.5\oldparindent{##1\hss}\ignorespaces}%
\vfootnote{$^{\the\footcount}$}{#1\vskip-9.69pt}\endgroup}
\def\item#1{\par\noindent
\hangindent6.5 mm\hangafter=0
\llap{#1\enspace}\ignorespaces}

\def\titleao#1{\vfill\supereject
\ifodd\pageno\else\null\vfill\supereject\fi
\let\header=N
     \bgroup
\textfont0=\tafontt \scriptfont0=\tafonts \scriptscriptfont0=\tafontss
\textfont1=\tamt \scriptfont1=\tams \scriptscriptfont1=\tamss
\textfont2=\tast \scriptfont2=\tass \scriptscriptfont2=\tasss
\textfont3=\tasyt \scriptfont3=\tasys \scriptscriptfont3=\tenex
     \baselineskip=18pt
     \lineskip=0pt
     \pretolerance=10000
     \tafontt
     \centerpar{#1}%
     \vskip75pt\egroup
     \nobreak
     \parindent=0pt
     \everypar={\global\parindent=\oldparindent
     \global\let\lasttitle=N\global\everypar={}}%
     \global\let\lasttitle=A%
     \ignorespaces}






\def\leaderfill{\kern0.5em\leaders\hbox to 0.5em{\hss.\hss}\hfill\kern
0.5em}
\newdimen\chapindent
\newdimen\sectindent
\newdimen\subsecindent
\newdimen\thousand
\setbox0=\hbox{\bf 10. }\chapindent=\wd0
\setbox0=\hbox{10.10 }\sectindent=\wd0
\setbox0=\hbox{10.10.1 }\subsecindent=\wd0
\setbox0=\hbox{\enspace 100}\thousand=\wd0
\def\contpart#1#2{\medskip\noindent
\vbox{\kern10pt\leftline{\textfont1=\tams
\scriptfont1=\tamss\scriptscriptfont1=\tbmss\bf
\advance\chapindent by\sectindent
\hbox to\chapindent{\ignorespaces#1\hss}\ignorespaces#2}\kern8pt}%
\let\lasttitle=Y\par}
\def\contcontribution#1#2{\if N\lasttitle\bigskip\fi
\let\lasttitle=N\line{{\textfont1=\tams
\scriptfont1=\tamss\scriptscriptfont1=\tbmss\bf#1}%
\if!#2!\hfill\else\leaderfill\hbox to\thousand{\hss#2}\fi}\par}
\def\conttitlea#1#2#3{\line{\hbox to
\chapindent{\strut\bf#1\hss}{\textfont1=\tams
\scriptfont1=\tamss\scriptscriptfont1=\tbmss\bf#2}%
\if!#3!\hfill\else\leaderfill\hbox to\thousand{\hss#3}\fi}\par}
\def\conttitleb#1#2#3{\line{\kern\chapindent\hbox
to\sectindent{\strut#1\hss}{#2}%
\if!#3!\hfill\else\leaderfill\hbox to\thousand{\hss#3}\fi}\par}
\def\conttitlec#1#2#3{\line{\kern\chapindent\kern\sectindent
\hbox to\subsecindent{\strut#1\hss}{#2}%
\if!#3!\hfill\else\leaderfill\hbox to\thousand{\hss#3}\fi}\par}
\long\def\lemma#1#2{\removelastskip\vskip\baselineskip\noindent{\tenbfne
Lemma\if!#1!\else\ #1\fi\ \ }{\it\ignorespaces#2}\vskip\baselineskip}
\long\def\proposition#1#2{\removelastskip\vskip\baselineskip\noindent{\tenbfne
Proposition\if!#1!\else\ #1\fi\ \ }{\it\ignorespaces#2}\vskip\baselineskip}
\long\def\theorem#1#2{\removelastskip\vskip\baselineskip\noindent{\tenbfne
Theorem\if!#1!\else\ #1\fi\ \ }{\it\ignorespaces#2}\vskip\baselineskip}
\long\def\corollary#1#2{\removelastskip\vskip\baselineskip\noindent{\tenbfne
Corollary\if!#1!\else\ #1\fi\ \ }{\it\ignorespaces#2}\vskip\baselineskip}
\long\def\example#1#2{\removelastskip\vskip\baselineskip\noindent{\tenbfne
Example\if!#1!\else\ #1\fi\ \ }\ignorespaces#2\vskip\baselineskip}
\long\def\exercise#1#2{\removelastskip\vskip\baselineskip\noindent{\tenbfne
Exercise\if!#1!\else\ #1\fi\ \ }\ignorespaces#2\vskip\baselineskip}
\long\def\problem#1#2{\removelastskip\vskip\baselineskip\noindent{\tenbfne
Problem\if!#1!\else\ #1\fi\ \ }\ignorespaces#2\vskip\baselineskip}
\long\def\solution#1#2{\removelastskip\vskip\baselineskip\noindent{\tenbfne
Solution\if!#1!\else\ #1\fi\ \ }\ignorespaces#2\vskip\baselineskip}


\long\def\definition#1#2{\removelastskip\vskip\baselineskip\noindent{\tenbfne
Definition\if!#1!\else\
#1\fi\ \ }\ignorespaces#2\vskip\baselineskip}
\def\frame#1{\bigskip\vbox{\hrule\hbox{\vrule\kern5pt
\vbox{\kern5pt\advance\hsize by-10.8pt
\centerline{\vbox{#1}}\kern5pt}\kern5pt\vrule}\hrule}\bigskip}
\def\frameddisplay#1#2{$$\vcenter{\hrule\hbox{\vrule\kern5pt
\vbox{\kern5pt\hbox{$\displaystyle#1$}%
\kern5pt}\kern5pt\vrule}\hrule}\eqno#2$$}
\def\typeset{\petit\noindent This book was processed by the author using
the \TeX\ macro package from Springer-Verlag.\par}
\outer\def\byebye{\bigskip\bigskip\typeset
\footcount=1\ifx\speciali\undefined\else
\loop\smallskip\noindent special character No\number\footcount:
\csname special\romannumeral\footcount\endcsname
\advance\footcount by 1\global\footcount=\footcount
\ifnum\footcount<11\repeat\fi
\gdef\leftheadline{\hbox to2.5true cc{\folio\hfil}\ignorespaces
\the\AUTHOR\unskip: \the\HEAD\hfill}\vfill\supereject\end}
\voffset 0.5cm
%
%
\def\half{{\textstyle{\scriptstyle 1\over\scriptstyle 2}}}
\def\L{{\cal L}}
\def\erf{{\rm erf}}
\def\bra#1{\langle #1 |}
\def\ket#1{| #1 \rangle}
\def\ss{\scriptscriptstyle\rm}
\def\sqr#1#2{{\phantom{\vrule width.15em}
    \lower.05em\vbox{\hrule height.#2pt
    \hbox{\vrule width.#2pt height#1pt \kern#1pt
    \vrule width.#2pt}
    \hrule height.#2pt}
    \phantom{\vrule width.15em}}}
\def\dalem{\mathchoice\sqr64\sqr64\sqr43\sqr33}

\def\pmb#1{\setbox0=\hbox{#1}%
  \kern-.025em\copy0\kern-\wd0
  \kern.05em\copy0\kern-\wd0
  \kern-.025em\raise.0433em\box0 }
\def\del{\pmb{$\nabla$}}

\def\gs{\mathrel{\lower0.6ex\hbox{$\buildrel {\textstyle >}
 \over {\scriptstyle \sim}$}}}
\def\ls{\mathrel{\lower0.6ex\hbox{$\buildrel {\textstyle <}
 \over {\scriptstyle \sim}$}}}
\newcount\japequationnum
\def\japdisp#1$${\line{\hfill{$\displaystyle#1$}
    \global\advance\japequationnum by 1
    \hfill (\the\japequationnum )}$$}
\everydisplay{\japdisp}
\def\boxit#1{
\eqalign{\vbox{\hrule height1pt\hbox{\vrule width1pt\kern0.3cm
\vbox{\kern0.3cm\hbox{$\displaystyle #1$}\kern0.3cm}\kern0.3cm\vrule width1pt
}\hrule height1pt}}}
\newcount\japfignum
\japfignum=0
\def\fig#1#2{\global\advance \japfignum by 1
\begfig #2 cm
\figure{\the\japfignum}{#1}
\endfig
}
\newcount\tmpfig
\def\nextfig{\tmpfig=\japfignum \advance\tmpfig by 1 \the\tmpfig}
\def\lastfig{\tmpfig=\japfignum \the\tmpfig}
\def\llastfig{\tmpfig=\japfignum \advance\tmpfig by -1 \the\tmpfig}
\def\frac#1#2{{\textstyle{\scriptstyle #1\over\scriptstyle #2}}}

\def\kms{{\;\rm km\,s^{-1}}}

\def\hompc{{\;h\,\rm Mpc^{-1}}}

\contribution{Inflationary Cosmology  \newline and \newline
Structure Formation
}
\author{J.A. Peacock}
\address{Royal Observatory, Blackford Hill, Edinburgh EH9 3HJ, UK}
\abstract{These lectures cover the basics of inflationary models
for the early universe, concentrating particularly on the
generation of density fluctuations from scalar-field dynamics.
The subsequent gravitational dynamics of these fluctuations in
dark matter in a Friedmann model are described, leading to a
review of the current situation in confronting inflationary models
with the latest data on the clustering of galaxies and other
measures of large-scale structure.
}

\vglue -11.0cm
\centerline{Lectures delivered at the EADN School {\it The Structure of the Universe};}
\centerline{Leiden, July 1995.}
\vskip 11.0cm

\titlea{1}{General arguments for inflation}
\titleb{1.1}{The problems of classical cosmology}
The standard isotropic cosmology is a very
successful framework for interpreting observations,
but there are certain questions which, prior to the
early 1980s, had to be avoided.
These are encapsulated in a set of classical `problems',
as follows.

\titlec{}{The horizon problem}
Standard cosmology contains a particle horizon of comoving radius
$$
r_{\ss H}=\int_0^t {c\; dt \over R(t)},
$$
which converges because $R\propto t^{1/2}$ in the early
radiation-dominated phase.
At late times, the integral is dominated by the matter-dominated phase, for which
$$
D_{\ss H}=R_0 \, r_{\ss H} \simeq {6000\over \sqrt{\Omega z}}\; h^{-1}\,{\rm Mpc}.
$$
The horizon at last scattering was thus only $\sim 100$ Mpc in size,
subtending an angle of about 1 degree.
Why then are the large number of causally disconnected regions we see on
the microwave sky all at the same temperature?

\titlec{}{The flatness problem}
The $\Omega=1$ universe is unstable:
$$
[1-1/\Omega(z)] = f(z)\, [1-1/\Omega)],
$$
where $f(z)=(1+z)^{-1}$ in the matter-dominated
era, $\propto (1+z)^{-2}$ for radiation domination,
so that $f(z)\simeq (1+z_{\rm eq})/(1+z)^2$ at early times.
To get $\Omega\simeq 1$ now requires a {\bf fine tuning}
of $\Omega$ in the past, which becomes more and more
precisely constrained as we increase the redshift 
at which the initial conditions are supposed to have been imposed.
Ignoring annihilation effects, $1+z=T_{\rm init}/2.7\;\rm K$;
$1+z_{\rm eq}\simeq 10^4$, and so the required fine-tuning is
$$
|\Omega(t_{\rm init})-1|\ls 10^{-22}\, [E_{\rm init}/{\rm GeV}]^2.
$$
At the Planck epoch, which is the natural initial time,
this requires a deviation of only 1 part in $10^{60}$.
This is satisfied if $\Omega=1$ exactly, but a
mechanism is still required to set up such an initial state.
This equation is especially puzzling if $\Omega\ne 1$ today:
how could the universe `know' to start with a deviation from
$\Omega=1$ just so tuned that the curvature starts to
become important only now after so many $e$-foldings of the expansion?

\titlec{}{The antimatter problem}
At $kT\gs m_pc^2$, there exist in equilibrium
roughly equal numbers of photons, protons and
antiprotons. Today, $N_p/N_\gamma\sim 10^{-9}$,
but $N_{\bar p}\simeq 0$. Conservation of baryon number would
imply that $N_{p}/N_{\bar p}=1+O(10^{-9})$ at early times
Where did this initial asymmetry come from?

\titlec{}{The structure problem}
The Universe is not precisely homogeneous. We generally
presume that galaxies and clusters grew via
gravitational instability from some initial
perturbations. What is the origin of these?

This list can be extended to problems which come
closer to astrophysics than cosmology {\it per se.}
There is the question of the dark matter and
its composition, for example. However, the
above list encompasses problems which go back to the
initial conditions of the Big Bang. It seems clear
that conventional cosmological models need to be
set up in an extremely special configuration, and
this is certainly a deficiency of the theory. 
Critics can point out with some force that the big bang
model explains nothing about the origin of the universe
as we now perceive it, because all the most important
features are `predestined' by virtue of being built into
the assumed initial conditions near to $t=0$. 

\titlec{}{The expansion problem}
Even the most obvious fact of the cosmological expansion is unexplained.
Although general relativity forbids a static universe, this is
not enough to understand the expansion. The gravitational dynamics
of $R(t)$ are just those of a cannonball travelling vertically
in the Earth's gravity. Suppose we see a cannonball rising
at a given time $t=t_0$: it may be true to say that it has $r=r_0$ and $v=v_0$ 
at this time because at a time $\Delta t$ earlier it had
$r=r-v_0\Delta t$ and $v=v_0-g\Delta t$, but it
is hardly a satisfying explanation for the motion of a cannonball
which was fired by a cannon. Nevertheless, this is the only
level of explanation that classical cosmology offers:
the universe expands now because it did so in the past.
Although it is not usually included in the list, one might thus with justice add
an `expansion problem' as perhaps the most
fundamental of the catalogue of classical
cosmological problems. Certainly, early generations of
cosmologists were convinced that some specific mechanism was
required in order to explain how the universe was set in motion.

For many years, it was assumed that any solution to these difficulties
would have to await a theory of quantum gravity.
The classical singularity can be approached no
closer than the Planck time of $\sim 10^{-43}$ s, and so
the initial conditions for the classical evolution
following this time must have emerged from behind the
presently impenetrable barrier of the quantum gravity epoch.
There remains a significant possibility that this policy
of blaming everything on quantum gravity may be correct, and this
is what lies behind modern developments in {\bf quantum cosmology}.
This field is not really suitable for treatment at the present
level, and it deals with the subtlest possible questions concerning
the meaning of the wave function for the entire universe.
The eventual aim is to understand if there could be a way
in which the universe could be spontaneously created
as a quantum-mechanical fluctuation, and if so
whether it would have the initial properties which observations seem to require.
Since this programme has to face up to the challenge
of quantizing gravity, it is fair to say that there are as
yet no definitive answers, despite much thought-provoking
work. See chapter 11 of Kolb \& Turner (1990) for an introduction.

However, the great development of cosmology in the
1980s was the realization that
the explanation of the initial-condition puzzles might involve
physics at lower energies: `only' $10^{15}$ GeV.
Although this idea, now known as inflation, cannot be considered to be firmly established,
the ability to treat gravity classically puts the
discussion on a much less speculative foundation.
What has emerged is a general picture of the early universe
of compelling simplicity, which moreover may be subject to observational verification.
What follows is an outline of the main features of inflation;
for more details see {\it e.g.} chapter 8 of Kolb \& Turner (1990);
Brandenberger (1990); Liddle \& Lyth (1993).

\titleb{1.2}{An overview of inflation}
\titlec{}{Equation of state for inflation}
The list of problems with conventional cosmology provides
a strong hint that the equation of state of the universe may
have been modified at very early times. To solve the horizon
problem and allow causal contact over the whole of the region
observed at last scattering requires a universe that expands
`faster than light' near $t=0$: $R\propto t^\alpha$; $\alpha>1$.
If such a phase existed, the integral for the comoving horizon
would have diverged, and there would be no difficulty in understanding
the overall homogeneity of the universe -- this could then be
established by causal processes. 
Indeed, it is tempting to assert that the observed homogeneity 
{\it proves\/} that such causal contact must once have occurred.
This phase of accelerated
expansion is the most general feature of what has become
known as the {\bf inflationary universe}.

What condition does this place of the equation of state?
In the integral for $r_{\ss H}$, we can replace $dt$ by $dR/\dot R$, which the Friedmann
equation says is $\propto dR/\sqrt{\rho R^2}$ at early times.
Thus, the horizon diverges provided the equation of state is such that $\rho R^2$
vanishes or is finite as $R\rightarrow 0$.
For a perfect fluid with $p=(\Gamma-1)\epsilon$ as the relation between
pressure and energy density, we have the adiabatic
dependence $p\propto R^{-3\Gamma}$, and the same dependence for
$\rho$ if the rest-mass density is negligible. A period of
inflation therefore needs 
$$
\boxit{
\Gamma < 2/3 \Rightarrow \rho c^2 + 3p <0.
}
$$
An alternative way of seeing that this criterion is
sensible is that the `active mass density' $\rho +3p/c^2$ then vanishes.
Since this quantity forms the rhs of Poisson's equation
generalized to relativistic fluids, it is no surprise that the
vanishing of $\rho +3p/c^2$
allows a coasting solution with $R\propto t$.

Such a criterion can also solve the flatness problem. Consider
the Friedmann equation:
$$
{\dot R}^2= {8\pi G\rho R^2\over 3} - kc^2.
$$
As we have seen,
the density term on the rhs must exceed the curvature term
by a factor of at least $10^{60}$ at the Planck time, and
yet a more natural initial condition might be to have the
matter and curvature terms being of comparable
order of magnitude. However, an inflationary phase in which
$\rho R^2$ increases as the universe expands can clearly
make the curvature term relatively as small as required,
provided inflation persists for sufficiently long.

\titlec{}{De sitter space and inflation}
We have seen that inflation will require an equation of
state with negative pressure, and the only familiar
example of this is the $p=-\rho c^2$ relation which
applies for vacuum energy -- in other words we are
led to consider inflation as happening in a universe
dominated by a cosmological constant.
As usual, any initial expansion will redshift
away matter and radiation contributions to the density,
leading to increasing dominance by the vacuum term.
If the radiation and vacuum densities are initially
of comparable magnitude, we quickly reach a state
where the vacuum term dominates.
The Friedmann equation  in the vacuum-dominated case
has three solutions:
$$
\eqalign{
R&\propto \sinh Ht\quad k=-1\cr
 &\propto \cosh Ht\quad k=+1\cr
 &\propto \exp Ht \quad k=0,\cr
}
$$
where $H=\sqrt{\Lambda c^2/3}= \sqrt{8\pi G\rho_{\rm vac}/3}$.
Thus, all solutions evolve towards the exponential $k=0$ solution,
known as {\bf de Sitter space}.
Note that $H$ is not the Hubble parameter at an
arbitrary time (unless $k=0$), but it becomes
so exponentially fast as the hyperbolic trigonometric
functions tend to the exponential.

Because de Sitter space clearly has $H^2$ and $\rho$ in the
right ratio for $\Omega=1$ (obvious, since $k=0$),
the density parameter in all models tends to unity
as the Hubble parameter tends to $H$.
If we assume that the initial conditions are
not fine-tuned ({\it i.e.} $\Omega =O(1)$ initially), then
maintaining the expansion for a factor $f$ produces
$$
\Omega = 1+O(f^{-2}).
$$
This can solve the flatness problem, provided $f$ is large enough. 
To obtain $\Omega$ of order unity today requires
$|\Omega-1|\ls 10^{-52}$ at the GUT epoch, and so 
$$
\boxit{
\ln f\gs 60
}
$$
e-foldings of expansion are needed; it will be proved
below that this is also exactly the number needed to
solve the horizon problem.
It then seems almost inevitable that the process should go to
completion and yield $\Omega=1$ to measurable accuracy today.
There is only a rather small range of e-foldings 
($60\pm 2$, say) around the critical value for which
$\Omega$ today can be of order unity without being effectively
exactly unity, and it would constitute an unattractive
fine-tuning to require that the expansion hit this
narrow window exactly.

This gives the first of two strong {\bf predictions of inflation}:
that the universe must be spatially flat
$$
\boxit{
{\rm inflation} \Rightarrow k=0.
}
$$
Note that this need not mean the Einstein-de Sitter model;
the alternative possibility is that a vacuum contribution is
significant in addition to matter, so that $\Omega_m+\Omega_v=1$.
Astrophysical difficulties in finding evidence for $\Omega_m=1$
are thus one of the major motivations, through inflation,
for taking the idea of a large cosmological constant seriously.

\titlec{}{Reheating from inflation}
The discussion so far indicates a possible route
to solving the problems with initial conditions
in conventional cosmology, but has a critical
missing ingredient. The idea of inflation is to
set the universe expanding 
towards an effective $k=0$ state by using the repulsive
gravitational force of vacuum energy or some other
unknown state of matter which satisfies $p<-\rho c^2/3$.
There remains the difficulty of returning to a
normal equation of state: the universe is required
to undergo a {\bf cosmological phase transition}.
Such a suggestion would have seemed highly {\it ad hoc.}
in the 1960s when the horizon and flatness problems
were first clearly articulated by Dicke. The invention of
inflation by Guth (1981) had to await developments in
quantum field theory which provided a plausible basis
for this phase transition. This mechanism will be described
below; it has been deliberately put off so far to
emphasise the general character of many of the
arguments for inflation.

What will emerge is that it is possible for inflation
to erase its tracks in a very neat way. 
If we are dealing with quantum fields at a temperature $T$,
then an energy density $\sim T^4$
in natural units is expected in the form of vacuum
energy. The vacuum-driven expansion produces a universe
which is essentially devoid of normal matter and radiation;
these are all redshifted away by the expansion, and so the
temperature of the universe becomes $\ll T$. A phase
transition to a state of zero vacuum energy, if instantaneous,
would transfer the energy $T^4$ to normal matter and
radiation as a latent heat. The universe would therefore
be {\bf reheated}: it returns to the temperature $T$ at
which inflation was initiated, but with the correct
special initial conditions for the expansion. 
The transition in practical models is not instantaneous,
however, and so the reheating temperature is lower
than the temperature prior to inflation.

\titlec{}{Quantum fluctuations}
Note that de Sitter space contains an {\bf event horizon}, in that
the comoving distance that particles can travel between
a time $t_0$ and $t=\infty$ is finite
$$
r_{\ss EH}=\int_{t_0}^\infty {c\; dt \over R(t)}
$$
(do not confuse this with the particle horizon, where
the upper limit for the integral would be $t_0$).
With $R\propto \exp Ht$, the proper radius of the
horizon is $R_0 r_{\ss EH}=c/H$. 
Figure \nextfig\ illustrates the situation.
The exponential expansion literally makes distant
regions of space move faster than light, so that points
separated by $>c/H$ can never communicate with each other.

\fig{The event horizon in de Sitter space.
Particles outside the sphere at $r=c/H$ can never receive
light signals from the origin, not can an observer at the origin
receive information from outside the sphere. The exponential
expansion quickly accelerates any freely-falling observers to
the point where their recession from the origin is effectively
superluminal.
}{7.0}

As with black holes, it therefore
follows that thermal {\bf Hawking radiation} will be
created. These quantum fluctuations in de Sitter
spacetime provide the seeds for what will eventually
become galaxies and clusters.
The second main prediction of inflation is that such fluctuations
should exist in all fields, in particular that there should
exist a background of gravitational waves left as a relic
of inflation.

This idea of obtaining all structure in the universe
(including ourselves) from quantum fluctuations is a
magical idea of tremendous appeal. 
If it could be shown to be correct, it would rank as one of the
greatest possible intellectual advances.
We now have to look
at some  of the practical details to see how this concept might be made
to function in practice, and how it may be tested.

\titlea{2}{Inflation field dynamics}

The general concept of inflation rests on being able to
achieve a negative-pressure equation of state. 
This can be realised in a natural way by quantum fields
in the early universe. In order to understand what is 
going on, it is necessary to summarize some of the most important
concepts and jargon from this field.

\titleb{2.1}{Quantum fields and potentials}
\titlec{}{Lagrangians and fields}
Consider the formulation of classical
mechanics in terms of an {\bf action principle}. We write
the variational equation
$$
\boxit{
\delta\int L\; dt = 0,
}
$$
where the {\bf Lagrangian} $L$ is the difference of the kinetic
and potential energies, $L=T-V$, and the integral $\int L\; dt$
is the {\bf action}. What this says is that, for particles
described by coordinates $q_i(t)$, the path travelled by
each particle between given starting and finishing positions
is such that the action is extremal (not necessarily a minimum,
even though one often speaks of the principle of least action)
with respect to small variations in the paths. If we take
the `positions' $q_i(t)$ and `velocities' $\dot q_i(t)$ to be
independent variables, then expansion of $L$ in terms of
small variations in the paths and integration by parts leads
to {\bf Euler's equation} for each particle
$$
{d\over dt}\left({\partial L\over\partial \dot q_i}\right) =
  {\partial L\over\partial q_i}.
$$

A field may be regarded as a dynamical system, but with an
infinite number of coordinates, $q_i$, which are the field
values at each point in space.
How do we handle this?
A hint is provided by electromagnetism, where we are familiar
with writing the total energy in terms of a density which,
as we are dealing with generalized mechanics, we may formally
call the Hamiltonian density.
This suggests that we write the Lagrangian in terms of
a {\bf Lagrangian density} $\L$: $L=\int{\L}\; dV$. This
quantity is of such central importance in quantum field theory,
that it is usually referred to (incorrectly) simply as 
`the Lagrangian'. In these terms, our action principle now
takes the pleasingly relativistic form
$$
\delta\int{\L}\; d^4 x^\mu =0,
$$
although note that to be correct in general relativity, the Lagrangian
needs to take the form of a invariant scalar times the
Jacobian $\sqrt{-g}$.

We can now apply the variational principle as before,
considering $\L$ to be a function of a general coordinate, $\phi$
which is the field, and the `4-velocity' $\partial_\mu \phi$.
This yields the {\bf Euler-Lagrange equation}
$$
\boxit{
{\partial\over\partial x^\mu}\left({\partial\L\over\partial(\partial_\mu\phi)}\right)
  = {\partial\L\over\partial\phi}.
}
$$
A Lagrangian then immediately gives a wave equation,
as the following examples illustrate.
The forms of
most Lagrangians are often quite similar:
if we want a wave equation linear in second derivatives
of the field,  then the Lagrangian must
contain terms quadratic in first derivatives of the field.

(1) Waves on a string.
A good classical example of the Euler-Lagrange
formalism is provided by waves on a string. If the density
per unit length is $\sigma$, the tension is $T$, and we call
the transverse displacement of the string $y$, then
the (one-dimensional) Lagrangian density is
$$
\L = \half\sigma\dot y^2 - \half T y'{}^2
$$
(at least for small displacements). The potential term
comes from the work done in stretching the string. Inserting
this in the Euler-Lagrange equation yields a familiar result:
$$
\sigma \ddot y - T y'' =0.
$$
This is just the wave equation, and it tells us that the speed
of sound on a plucked string is $\sqrt{T/\sigma}$.

(2) Complex scalar field.
Here we want a Lagrangian
which will yield the Klein-Gordon equation $(\dalem+\mu^2)\phi=0$. The
required form is
$$
\L = (\partial^\mu\phi)(\partial_\mu\phi)^* - \mu^2\phi\phi^*,
$$
where the only subtlety is that $phi$ and $\phi^*$ are treated
as independent variables (rather than the real and imaginary
parts of $\phi$). For a real field, the Lagrangian becomes
$$
\L = \half(\partial^\mu\phi\partial_\mu\phi - \mu^2\phi^2).
$$

\titlec{}{Noether's theorem}
{\pretolerance 10000
The existence of global symmetries of the Lagrangian is closely connected with
conservation laws in physics. In classical mechanics,
conservation of energy and momentum arise by considering
Euler's equation
$$
{d\over dt}\left({\partial L\over\partial \dot x_i}\right) -
{\partial L\over\partial x_i} =0.
$$
If $L$ is independent of position, then we obtain conservation of momentum
(or angular momentum, if $x$ is an angular coordinate):
$p_i\equiv \partial L/\partial\dot x_i = {\rm constant}$.
If $L$ has no explicit dependence on $t$, then
$$
{dL\over dt} = \sum \left({\partial L\over\partial q_i}\dot q_i
  + {\partial L\over\partial\dot q_i}\ddot q\right) =
  \sum (\dot p_i \dot q_i + p_i \ddot q_i),
$$
which leads us to define the {\bf Hamiltonian} as a further constant of the motion
$$
H\equiv \sum p_i\dot q_i - L = {\rm constant}.
$$
}

Something rather similar happens in the case of quantum (or classical) 
field theory: the existence of a global symmetry leads directly
to a conservation law. The difference between discrete dynamics
and field dynamics where the Lagrangian is a {\it density\/}
is that the result is expressed as a {\bf conserved current} rather
than a simple constant of the motion. 
In what follows, we symbolize the field by $\phi$, but this is
not to imply that there is any restriction to scalar fields.
If several fields are involved ({\it e.g.} $\phi$ and $\phi^*$
for a complex scalar field), they should be summed over.

Suppose the Lagrangian is
independent of explicit dependence on spacetime ({\it i.e.}
depends on $x^\mu$ only implicitly through the fields and their
4-derivatives). The algebra is similar to that above, and
we obtain 
$$
\boxit{
{d\over dx_\nu}\left[{\partial\L\over\partial (\partial^\nu\phi)}\,
  {\partial\phi\over\partial x^\mu} - \L \, g_{\mu\nu} \right] = 0.
}
$$
This has produced a conserved tensor: the term in square brackets
is to be identified with the energy-momentum tensor of the field,
$T_{\mu\nu}$. 

\titlec{}{Natural units}
To simplify the appearance of equations, it is
universal practice in quantum field theory to
adopt {\bf natural units} where we take
$$
\boxit{
\hbar=c=\mu_0=\epsilon_0=1.
}
$$
This convention makes the meaning of equations clearer
by reducing the algebraic clutter, and is also useful in
the construction of intuitive arguments for the
order of magnitude of quantities in field theory.

The adoption of natural units corresponds to fixing the units
of charge, mass, length and time relative to each other.
This leaves one free unit, usually taken to be energy.
Natural units are thus one step short of the Planck
system, in which $G=1$ also, so that all units are
fixed and all physical quantities are dimensionless.
In natural units, the following dimensional equalities hold:
$$
\eqalign{
[E] &= [m]\cr
[L] &= [m]^{-1}\cr
}
$$
Hence, the dimensions of energy density are
$$
[\L]=[m]^4,
$$
with units usually quoted in GeV$^4$. Thus, when we deal
with quadratic derivative terms $\partial^\mu\phi\partial_\mu\phi$
and quadratic mass terms
($m^2\phi^2$, $m^2A^\mu A_\mu/2$ {\it etc.}), the dimensions of the fields
must clearly be 
$$
[\phi]=[A^\mu]=[m].
$$

\titleb{2.2}{Equations of motion}
\titlec{}{Quantum fields at high temperatures}
The critical fact we shall need from quantum field
theory is that quantum fields can produce an energy
density which mimics a cosmological constant.
The discussion will be restricted to the case of 
a scalar field $\phi$ (complex in general, but often illustrated
using the case of a single real field). 
The restriction to scalar fields 
is not simply for reasons of simplicity, but because
the scalar sector of particle physics is relatively
unexplored. While vector fields such as electromagnetism
are well understood, it is expected in many theories of
unification that additional scalar fields such as the
Higgs field will exist. We now need to look at what these
can do for cosmology.

The Lagrangian density for a scalar field is as usual of
the form of a kinetic minus a potential term:
$$
{\cal L}=\half\partial_\mu\phi \, \partial^\mu\phi - V(\phi).
$$
In familiar examples of quantum fields, the potential would be
$$
V(\phi) = m^2\, \phi^2/2,
$$
where $m$ is the mass of the field in natural
units. However, it will be better to keep the potential
function general at this stage.
As usual, Noether's theorem gives the energy-momentum
tensor for the field as
$$
T^{\mu\nu}=\partial^\mu\phi\partial^\nu\phi-g^{\mu\nu}{\cal L}.
$$
From this, we can read off the energy density and pressure:
$$
\eqalign{
\rho&=\half\dot\phi^2 +V(\phi)+\half(\nabla\phi)^2 \cr
p   &=\half\dot\phi^2 -V(\phi)-
  {\textstyle{\scriptstyle 1\over\scriptstyle 6}}(\nabla\phi)^2. \cr
}
$$
If the field is constant both spatially and temporally,
the equation of state is then $p=-\rho$, as required if
the scalar field is to act as a cosmological constant;
note that derivatives of the field spoil this identification.

If $\phi$ is a (complex) Higgs field, then the
symmetry-breaking Mexican hat potential might be assumed:
$$
V(\phi) = -\mu^2|\phi|^2 +\lambda|\phi|^4.
$$
At the classical level, such potentials determine
where $|\phi|$ will be found in equilibrium:
at the potential minimum. 
In quantum terms, this goes over to the
{\bf vacuum expectation value} $\bra{0}\phi\ket{0}$.
However, these potentials do not include
the inevitable fluctuations which will arise in
thermal equilibrium.
We know how to treat these in classical systems:
at non-zero temperature a system
of fixed volume will minimize not its potential
energy, but the {\bf Helmholtz free energy}
$F=V-TS$, $S$ being the entropy. The calculation of
the entropy is technically complex, since it involves
allowance for the quantum interactions with a thermal bath
of background particles. However, the main result can be
justified, as follows. The effect of the thermal interaction
must be to add an interaction term to the Lagrangian
$\L_{\rm int}(\phi, \psi)$, where $\psi$ is a thermally-fluctuating field
which corresponds to the heat bath.
In general, we would expect $\L_{\rm int}$ to have a quadratic
dependence on $|\phi|$ around the origin: $\L_{\rm int}\propto
|\phi|^2$ (otherwise we would need to explain why the
second derivative either vanished or diverged); the coefficient of proportionality will
be an effective mass$^2$ that depends on the thermal fluctuations
in $\psi$. 
On dimensional grounds, this coefficient must be
proportional to $T^2$, although a more detailed analysis would
be required to get the constant of proportionality.

There is thus a temperature-dependent
{\bf effective potential} which we have to minimize:
$$
V_{\rm eff}(\phi,T) = V(\phi,0) + aT^2|\phi|^2.
$$
The effect of this on the symmetry-breaking potential
is dramatic, as illustrated in Figure \nextfig.
At very high temperatures, the potential will be
parabolic, with a minimum at $|\phi|=0$, whereas
at $T=0$, the ground state is at
$|\phi|=\sqrt{\mu^2/(2\lambda)}$ and the symmetry is broken.
In between, there must be three critical temperatures:
at $T_1$, a second minimum appears in $V_{\rm eff}$
at $|\phi|\ne 0$, and this will be the global minimum
for some $T_2<T_1$.
For $T<T_2$, the state at $|\phi|=0$ is known as the
{\bf false vacuum}, whereas the global minimum is known
as the {\bf true vacuum}.
Finally, at $T=T_3$, the curvature around the origin
changes sign and there is only one minimum in the
potential. The universe can no longer be trapped in
the false vacuum and can make a 
first-order phase transition to the true vacuum state.

\fig{The temperature-dependent effective potential
illustrated at several temperatures. For $T>T_1$, only
the false vacuum is available; for $T<T_2$ the true vacuum
is energetically favoured and 
the potential approaches the zero-temperature form.
For $T<T_3$ the true vacuum is the only minimum.}
{7.0}

The crucial point to note for cosmology is that there
is an energy-density difference between the two vacuum states:
$$
\Delta V={\mu^4\over 2\lambda}
$$
If we say that the zero of energy is such that $V=0$ in the
true vacuum, this implies that the false-vacuum
symmetric state displayed an effective cosmological
constant. On dimensional grounds, this must be
an energy density $\sim m^4$ in natural units, where
$m$ is the energy at which the phase transition occurs.
For GUTs, $m\simeq 10^{15}$ GeV; in laboratory units,
this implies
$$
\rho_{\rm vac} = {10^{60} {\rm GeV}^4\over \hbar^3 c^5}
  \simeq 10^{80}\; {\rm kg\, m}^{-3}.
$$
The inevitability of such a colossal vacuum energy in models
with GUT-scale symmetry breaking was the major motivation for the
concept of inflation as originally envisaged by Guth (1981).
At first sight, the overall package looks highly appealing,
since the phase transition from false to true vacuum
both terminates inflation and also reheats the universe to
the GUT temperature, allowing the possibility that GUT-based
reactions which violate baryon-number conservation can generate
the observed matter/antimatter asymmetry.

However, while a workable inflationary
cosmology will very likely deploy these basic
elements of vacuum-driven expansion, fluctuation generation,
and reheating, it has become clear that such a model
must be more complex than Guth's initial proposal.
To explain where the problems arise, we need to look
in more detail at the functioning of the inflation mechanism.

\titlec{}{Dynamics of the inflation field}
Treating the field classically ({\it i.e.} considering
the expectation value $\langle\phi\rangle$, we get 
from energy-momentum conservation ($T^{\mu\nu}_{;\nu}=0$) the
equation of motion
$$
\boxit{
\ddot\phi+3H\dot\phi-\nabla^2\phi+dV/d\phi=0.
}
$$
This can also be derived more easily by 
the direct route of writing down the action
$S=\int \L\;\sqrt{-g}\; d^4x$ and applying the Euler-Lagrange
equation which arises from a stationary action
($\sqrt{-g}=R^3(t)$ for an FRW model, and this is the origin of
the Hubble drag $3H\dot\phi$ term).

The solution of the equation of motion becomes tractable if we
both ignore spatial inhomogeneities in $\phi$ and
make the {\bf slow-rolling approximation} that
$|\ddot\phi|\ll|3H\dot\phi|$, $|dV/d\phi|$.
Both these steps are required in order that
inflation can happen; we have shown above that the
vacuum equation of state only holds if in
some sense $\phi$ changes slowly both spatially
and temporally. Suppose there are characteristic
temporal and spatial scales $T$ and $X$ for the scalar field;
the conditions for inflation are that the negative-pressure
equation of state from $V(\phi)$ must dominate the
normal-pressure effects of time and space derivatives:
$$
V\gg \phi^2/T^2,\quad V\gg \phi^2/X^2,
$$
hence $|dV/d\phi|\sim V/\phi$ must be $\gg  \phi/T^2\sim\ddot\phi$.
The $\ddot\phi$ term can therefore be neglected in the
equation of motion, which then takes the slow-rolling form
$$
\boxit{
3H\dot\phi=-dV/d\phi.
}
$$
The conditions for inflation can be cast into useful dimensionless
forms. The basic $V\gg {\dot\phi}^2$ condition can now be rewritten
using the slow-roll relation as
$$
\boxit{
\epsilon \equiv {m_{\ss P}^2\over 16\pi} \,(V'/V)^2 \ll 1.
}
$$
Also, we can differentiate this expression to obtain the
criterion $V''\ll V'/m_{\ss P}$. Using slow-roll once more
gives $3H \dot\phi/m_{\ss P}$ for the rhs, which is in turn
$\ll 3H \sqrt{V}/m_{\ss P}$ because ${\dot\phi}^2\ll V$, giving finally
$$
\boxit{
\eta \equiv {m_{\ss P}^2\over 8\pi} \,(V''/V) \ll 1
}
$$
(recall that for de Sitter space $H=\sqrt{8\pi G V(\phi)/3}
\sim \sqrt{V}/m_{\ss P}$ in natural units).
These two criteria make perfect intuitive sense:
the potential must be flat in the sense of having small
derivatives if the field is to roll slowly enough
for inflation to be possible.

Similar arguments can be made for the spatial parts.
However, they are less critical: what matters is
the value of $\nabla\phi=\nabla_{\rm comoving}\,\phi/R$.
Since $R$ increases exponentially, these perturbations
are damped away: assuming $V$ is large enough for
inflation to start in the first place, inhomogeneities
rapidly become negligible.
This `stretching' of field gradients as we increase
the cosmological horizon beyond the
value predicted in classical cosmology also solves a related
problem which was historically important in motivating the
invention of inflation -- the {\bf monopole problem}.
Monopoles are point-like topological defects which would be
expected to arise in any phase transition at around the GUT scale
($t\sim 10^{-35}$ s). If they form at approximately one per
horizon volume at this time, then it follows that the
present universe would contain $\Omega \gg 1$ in monopoles
(see {\it e.g.} section 7.6 of Kolb \& Turner 1990).
This unpleasant conclusion is avoided if the horizon can
be made much larger than the classical one at the end of
inflation; the GUT fields have then been aligned over a
vast scale, so that topological defect formation becomes
extremely rare.

\titlec{}{Ending inflation}
Although spatial derivatives of the scalar field can thus be neglected,
the same is not always true for time derivatives.
Although they may be negligible initially, the relative importance
of time derivatives increases as $\phi$
rolls down the potential and $V$ approaches zero (leaving
aside the subtle question of how we know that the minimum is indeed at
zero energy). Even if the potential does not steepen,
sooner or later we will have $\epsilon\simeq 1$
or $|\eta|\simeq 1$ and the inflationary phase will cease.
Instead of rolling slowly
`downhill', the field will oscillate
about the bottom of the potential, but with the
oscillations becoming damped by the $3H\dot\phi$
friction term. Eventually, we will be left with
a stationary field which either continues to inflate
without end (if $V(\phi=0)>0$) or which simply has
zero density. This would be rather a boring
universe to inhabit, but fortunately there is a
more realistic way in which inflation can end.
We have neglected so far the couplings of the scalar
field to matter fields. Such couplings will 
cause the rapid oscillatory phase to produce particles,
leading to {\bf reheating}. Thus, even if the
minimum of $V(\phi)$ is at $V=0$, the universe
is left containing roughly as much energy density as
it started with, but now in the form of normal
matter and radiation -- which starts the usual
FRW phase, albeit with the desired special `initial'
conditions.

As well as being of interest for completing the picture
of inflation, it is essential to realise that these
closing stages of inflation are the only ones of observational
relevance. Inflation might well continue for a huge number of $e$-foldings,
all but the last few satisfying $\epsilon, \eta\ll 1$. However, the scales
which left the de Sitter horizon at these early times are now vastly
greater than our observable horizon, $c/H_0$, which exceeds the
de Sitter horizon by only a finite factor. If inflation terminated
by reheating to the GUT temperature, then the expansion factor to
today is
$$
a_{\ss GUT}^{-1}\simeq E_{\ss GUT}/E_\gamma.
$$
The comoving horizon size at the end of inflation was therefore
$$
d_{\ss H}(t_{\ss GUT})\simeq a_{\ss GUT}^{-1} \, [c/H_{\ss GUT}]
\simeq [E_{\ss P}/E_\gamma] \, E_{\ss GUT}^{-1},
$$
where the last expression in natural units uses $H\simeq \sqrt{V}/E_{\ss P}
\simeq E_{\ss GUT}^2/E_{\ss P}$. For a GUT energy of $10^{15} \rm GeV$, this
is about 10 m. We need enough $e$-foldings to have stretched this to the present
horizon size
$$
\boxit{
N_{\rm obs}=\ln \left[ { 3000h^{-1}\,{\rm Mpc} \over
[E_{\ss P}/E_\gamma] E_{\ss GUT}^{-1}} \right] \simeq 60.
}
$$
By construction, this is enough to solve the horizon problem, and it
is also the number of $e$-foldings needed to solve the flatness problem.
This is no coincidence, since we saw earlier that the criterion in this
case was
$$
N\gs \half\; \ln \left[ {a_{\rm eq}\over a_{\ss GUT}^2} \right].
$$
Now, $a_{\rm eq}=\rho_\gamma/\rho$, and $\rho=3H^2\Omega/8\pi G$.
In natural units, this translates to $\rho \sim E_{\ss P}^2 [c/H_0]^{-2}$,
or $a_{\rm eq}^{-1}\sim E_{\ss P}^2 [c/H_0]^{-2} / E_\gamma^4$.
The expression for $N$ is then identical to the one for $N$
in the case of the horizon problem: the same number of $e$-folds 
will always solve both.

Realizing that the observational regime corresponds 
only to the terminal
phases of inflation is both depressing and stimulating.
Depressing, because $\phi$ may well not move very much during
the last phases: our observations relate only to a small piece
of the potential, and we cannot hope to recover its form
without substantial {\it a priori} knowledge. Stimulating
because observations even on very large scales must relate
to a period where the simple concepts of exponential inflation
and scale-invariant density fluctuations were coming close
to breaking down. This opens the possibility of testing
inflation theories in a way that would not be possible with
data relating to only the simpler early phases.
These tests take the form of tilt and gravitational waves
in the final perturbation spectrum, to be discussed further below.

\titleb{2.3}{Inflationary models}
\titlec{}{Early inflation models}
These general principles contrast sharply
with Guth's initial idea, where the potential
was trapped at $\phi=0$, and eventually underwent
a first-order phase transition. This model
suffers from the problem that it predicts
far too large residual inhomogeneities after
inflation is over. This is easily seen: because the
transition is first-order, it proceeds by {\bf bubble nucleation}
where the vacuum tunnels between false and true
vacuum. However, the extent of these bubbles will spread
as a causal process, whereas outside the bubbles the
exponential expansion of inflation is continuing.
This means that it is very difficult for the bubbles to
percolate and eliminate the false vacuum everywhere,
as is needed for an end to inflation. Instead, inflation
continues indefinitely, with the bubbles of true
vacuum having only a small filling factor at any time.
This {\bf graceful exit problem} motivated variants
in which the potential is flatter near the origin,
so that the phase transition is second order and can
proceed smoothly everywhere.

However, there is also a more general problem with Guth's
model and its variants. If the initial conditions are
at a temperature $T_{\ss GUT}$, we expect
thermal fluctuations in $\phi$; the potential
should generally differ from its minimum by
an amount $V\sim T_{\ss GUT}^4$, which is of the
order of the difference between true and false vacua. How 
then is the special case needed to trap the potential near
$\phi=0$ to arise? We have returned to the sort of
fine-tuned initial conditions from which inflation 
was designed to save us. 

Combined with the
difficulties in achieving small inhomogeneities after
inflation is over, Guth's original inflation model 
thus turned out to have insuperable difficulties.
However, for many cosmologists the main concepts
of inflation have been too attractive to give up.
The price one pays for this is to decouple inflation
from standard particle physics (taking the liberty of
including GUTs in this category): inflation can in
principle be driven by the vacuum energy of any scalar field.
The ideological inflationist will then take the position
that such a field (the {\bf inflaton}) must have existed,
and that it is our task to work out its properties
from empirical cosmological evidence, rather than
{\it a priori\/} particle-physics considerations.
Barring a credible alternative way of understanding
the peculiarities of the initial conditions of the
big bang, there is much to be said for this point of view.

\titlec{}{Chaotic inflation models}
Most attention is currently paid to 
the more general models where the field
finds itself some way from its potential minimum.
This idea is termed {\bf chaotic inflation} (see
{\it e.g.} Linde 1989). 
The name originates because this class of models is
also quite philosophically different from other
inflation models; it does not require that there is
a single Friedmann model containing an inflation-driving
scalar field. Rather, there could be some primordial chaos,
within which conditions might vary. Some parts may
attain the conditions needed for inflation, in which
case they will expand hugely, leaving a universe inside a
single bubble -- which we inhabit. In principle this
bubble has an edge, but if inflation persists for sufficiently long, the
distance to this nastiness is so much greater than the
current particle horizon that its existence has no testable consequences.

A wide range of inflation models of this kind is possible, illustrating the
freedom which arises once the parameters of the
theory are constrained only by the requirement that
inflation be produced. Things become even less
constrained once it is realized that inflation
need not correspond to de Sitter space, even though
this was taken for granted in early discussions.
As discussed earlier, it is only necessary that
the universe enter a phase of `superluminal' expansion
in which the equation of state satisfies
$p < -\rho c^2/3$.

For a pure static field, we will have the usual $p=-\rho c^2$
vacuum equation of state, and so a significant deviation
from de Sitter space requires a large contribution from
$\dot\phi$ terms (although the slow-roll conditions
will often still be satisfied). Intuitively, this corresponds
to a potential which must be steep in some sense that is
determined by the desired time dependence of the scale factor.
Three special cases are of particular interest:

\item{(i)} {\bf Polynomial inflation.} If the potential is 
taken to be $V\propto \phi^\alpha$, then the scale-factor
behaviour is very close to exponential. This becomes 
less true as $\alpha$ increases, but investigations are
usually limited to $\phi^2$ and $\phi^4$ potentials on
the grounds that higher powers are nonrenormalizable.

\item{(ii)} {\bf Power-law inflation.} On the other hand,
$a(t)\propto t^p$ would suffice, provided $p>1$. The
potential required to produce this behaviour is

$$
V(\phi)\propto \exp\left[\sqrt{16\pi\over p \, m_{\ss P}^2}
\; \phi \right].
$$

\item{(iii)} {\bf Intermediate inflation.}
Another simple time-dependence which suffices for inflation
is $a(t)\propto \exp[(t/t_0)^f]$. In the slow-roll approximation,
the required potential here is $V(\phi)\propto \phi^{-\beta}$,
where $\beta=4(f^{-1}-1)$.

There are in addition a plethora of more specific models
with various degrees of particle-physics motivation.
Since at the time of writing none of these seem likely
to become permanent fixtures, these will mostly not
be described in detail. The above examples are more than enough
to illustrate the wide range of choice available.

\titlec{}{Criteria for inflation}
Successful inflation in any of these models requires $>60$
$e$-foldings of the expansion. The implications of this are
easily calculated using the slow-roll equation, which gives
the number of $e$-foldings between $\phi_1$ and $\phi_2$
as
$$
N=\int H\; dt = -{8\pi\over m_{\ss P}^2} \, \int_{\phi_1}^{\phi_2} {V\over V'}\; d\phi
$$
For any potential which is relatively smooth, $V'\sim V/\phi$, and
so we get $N\sim (\phi_{\rm start}/m_{\ss P})^2$, assuming that
inflation terminates at a value of $\phi$ rather smaller
than the start. The criterion for successful inflation is thus
that the initial value of the field exceeds the Planck scale:
$$
\boxit{
\phi_{\rm start} \gg m_{\ss P}.
}
$$
By the same argument, it is easily seen that this is also
the criterion needed to make the slow-roll parameters $\epsilon$
and $\eta$ be $\ll 1$. To summarise, any model in which the
potential is sufficiently flat that slow-roll inflation
can commence, will probably achieve the critical 60 $e$-foldings.
Counterexamples can of course be constructed, but they have to be
somewhat special cases.

It is interesting to review this conclusion 
for some of the specific inflation models listed above.
Consider a mass-like potential
$V=m^2\phi^2$. If inflation starts near the Planck scale,
the fluctuations in $V$ are $\sim m_{\ss P}^4$ and
our condition becomes $m\ll m_{\ss P}$; 
similarly, for $V=\lambda \phi^4$, the condition
is weak coupling: $\lambda \ll 1$.
Any field with a rather flat potential will thus tend to inflate,
just because typical fluctuations leave it a long
way from home in the form of the potential minimum.
In a sense, inflation is realized by means of `inertial confinement':
there is nothing to prevent the scalar field from reaching the
minimum of the potential -- but it takes a long time to do so,
and the universe has inflated by a large factor in the meantime.

This requirement for weak coupling and/or small mass
scales near the Planck epoch is suspicious, since
quantum corrections will tend to re-introduce the
Planck scale. In this sense, as with the appearance of
the Planck scale as the minimum required field value,
it is not clear that the aim of realizing inflation
in a classical way distinct from quantum gravity has
been fulfilled.

\titlea{3}{Relic fluctuations from inflation}
\titleb{3.1}{Motivation}
We have seen that de Sitter space contains
a true event horizon, of proper size $c/H$.
This suggests that there will be thermal fluctuations
present, as with a black hole, for which the
{\bf Hawking temperature} is $kT_{\ss H}=\hbar c/(4\pi r_s)$.
This analogy is close, but imperfect, and the characteristic
temperature of de Sitter space is a factor of 2 higher:
$$
kT_{\rm d-s}={\hbar H\over 2\pi}.
$$
The details of how these fluctuations translate into density
perturbations after inflation are somewhat technical.
However, we can immediately note that a natural prediction
will be a spectrum of perturbations which are {\it scale invariant}.
This means that the metric fluctuations of spacetime
receive equal levels of distortion from each decade of 
wavelength of perturbation, and may be quantified in terms of
the fluctuations in Netwonian gravitational potential, $\Phi$ ($c=1$):
$$
\boxit{
\delta_{\ss H}^2\equiv \Delta^2_\Phi\equiv {d\,\sigma^2(\Phi) \over d\, \ln k}.
}
$$
The notation $\delta_{\ss H}$ arises because the potential perturbation
is of the same order as the density fluctuation on the scale of the
horizon at any given time.

It is commonly argued that the scale-invariant prediction arises 
because de Sitter space is invariant under time
translation: there is no natural origin of time under
exponential expansion. At a given time, the only length scale
in the model is the horizon size $c/H$, so it is inevitable
that the fluctuations which exist on this scale are the
same at all times. After inflation ceases, the resulting
fluctuations (constant amplitude on the scale of the horizon)
give us the {\bf Zeldovich} or {\bf scale-invariant} spectrum.
The problem with this argument is that it
ignores the issue of how the perturbations
evolve while they are outside the horizon; we have only
really calculated the amplitude for the last generation
of fluctuations -- {\it i.e.} those which are on the
scale of the horizon at the time inflation ends.
Fluctuations generated at earlier times will be inflated
outside the de Sitter horizon, and will re-enter the
FRW horizon at some time after inflation has ceased.

The evolution during this period is a topic where some care
is needed, since the description of these large-scale
perturbations is sensitive to the gauge freedom in
general relativity. A technical discussion is given in
{\it e.g.} Mukhanov, Feldman \& Brandenberger (1992), but there is no
space to do this justice here. Rather, we shall rely on 
simply motivating the result, which is that potential perturbations
re-enter the horizon with the same amplitude they had on
leaving. This may be made reasonable in two ways.
Perturbations outside the horizon are immune to causal effects, so
it is hard to see how any large-scale non-flatness in spacetime could
`know' whether it was supposed to grow or decline.
More formally, we shall show below that small potential
perturbations preserve their value, provided they are on scales
where pressure effects can be neglected, and that this
critical scale corresponds to the horizon.
We therefore argue that the inflationary
process produces a universe which is fractal-like in the
sense that scale-invariant fluctuations correspond to a
metric which has the same `wrinkliness' per log length-scale.
It then suffices to calculate that amplitude on one scale --
{\it i.e.} the smallest one where super-horizon evolution
is not an issue.
It is possible to alter this scale-invariant prediction
only if the expansion is non-exponential; we have seen that such
deviations plausibly do exist towards the end of inflation.

To anticipate the detailed treatment, the inflationary
prediction is of a horizon-scale amplitude 
$$
\boxit{
\delta_{\ss H} = {H^2 \over 2\pi\,\dot\phi}
}
$$
which can be understood as follows. Imagine that the main effect of
fluctuations is to make different parts of the universe 
have fields which are perturbed by an amount $\delta\phi$.
In other words, we are dealing with various copies of
the same $\phi(t)$ rolling behaviour, but viewed at
different times
$$
\delta t = {\delta\phi\over \dot\phi}.
$$
These universes will then finish inflation at different
times, leading to a spread in energy densities.
The horizon-scale density amplitude is given by the
different amounts that the universes have expanded following
the end of inflation:
$$
\delta_{\ss H}\sim H\; \delta t = {H^2\over 2\pi\, \dot\phi},
$$
where the last step uses 
the crucial input of quantum field theory, which
is to say that the rms $\delta\phi=H/2\pi$. This
result will be derived below, but it is immediately
reasonable on dimensional grounds (in natural units,
the field has the dimension of a temperature).

\titleb{3.2}{The fluctuation spectrum}
We now need to go over this vital result in rather
more detail. First, consider the equation of motion
obeyed by perturbations in the inflaton field. The
basic equation of motion is 
$$
\ddot\phi +3H\dot\phi - \nabla^2\phi + V'(\phi)=0,
$$
and we seek the corresponding equation for the
perturbation $\delta\phi$ obtained by starting
inflation with slightly different values of $\phi$
in different places. Suppose this perturbation takes
the form of a comoving plane-wave perturbation of amplitude
$A$: $\delta\phi=A\exp[i{\bf k\cdot x} - ik t/a]$; the
perturbed field $\delta\phi$ obeys the same
equation of motion as the main field.
If the slow-roll conditions are also assumed,
so that $V'$ may be treated as a constant, we get
$$
\ddot{[\delta\phi]} + 3H\dot{[\delta\phi]} + (k/a)^2 \delta\phi=0,
$$
which is a standard wave equation for a massless
field evolving in an expanding universe.

Having seen that the inflaton perturbation behaves in
this way, it is not much work to obtain the quantum
fluctuations which result in the field at late times
({\it i.e.} on scales much larger than the de Sitter
horizon). First consider the fluctuations in flat
space: the field would be expanded as
$$
\phi_k=\omega_k a_k + \omega^*_k a^\dagger_k,
$$
and the field variance would be
$$
\bra{0} \; |\phi_k|^2\; \ket{0} = |\omega_k|^2.
$$
To solve the general problem, we only need to
find how the amplitude $\omega_k$ changes as the
universe expands. The idea is to start from the
situation where we are well inside the horizon
($k/a \gg H$), in which case flat-space quantum
theory will apply, and end at the point of interest
outside the horizon (where $k/a \ll H$).

Before finishing the calculation, note the critical
assumption that the initial state is the vacuum:
in the modes that will eventually be relevant for
observational cosmology, we start with not even one
quantum of the inflaton field present. Is this
smuggling fine-tuning of the initial conditions
back in through the back door?
Given our ignorance of the exact conditions in the
primordial chaos from which the inflationary phase is
supposed to emerge, it is something of a matter of taste
whether this is seen as being a problem. Certainly, if
the initial state is close to equilibrium at temperature
$T$, this is easily understood, since all initial scales
are given in terms of $T$. In natural units, the energy
density in $V(\phi)$ and radiation will be $\sim T^4$,
and the proper size of the horizon will be $\sim T^{-2}$.
In the initial state, the inflaton occupation number will
be $1/2$ for very long wavelengths, and will fall for
proper wavelengths $\ls T^{-1}$. Now, remember that $T$
is in units of the Planck temperature, so that $T\sim 10^{-4}$
for GUT-scale inflation. That means that perturbations of
scale smaller than $T$ times the horizon would start with
$n\simeq 0$ for a thermal state. However, this is only the initial
state, and we expect that occupation number will be an
adiabatic invariant which is constant for a given
comoving wavelength. Thus, after $\ln (1/T)$ $e$-foldings
of inflation ({\it i.e.} a few), every mode that
remains inside the horizon will have the required
zero occupation number. Of course, the motivation for
a thermal initial state is weak, but the main point
can be made in  terms of energy density. If inflation
is to happen at all, $V(\phi)$ must dominate, and it
cannot do this if the inflaton fluctuations exist
down to zero wavelength because the effective radiation
density would then diverge. It thus seems reasonable
to treat any initial state that inflates as rapidly
entering a vacuum state.

It is also worth noting in passing that these fluctuations
in the scalar field can in principle affect the progress
of inflation itself. They can be thought of as adding a
random-walk element to the classical rolling of the
scalar field down the trough defined by $V(\phi)$. In cases where
$\phi$ is too close to the origin for inflation to persist
for sufficiently long, it is possible for the quantum
fluctuations to push $\phi$ further out -- creating further
inflation in a self-sustaining process. This is the
concept of {\bf stochastic inflation} (Linde 1986, 1989).

Returning now to the calculation, we want to know how the
mode amplitude changes as the wavelength passes through
the horizon. Initially, we have the (expanding) flat-space result
$$
\omega_k=a^{-3/2}\, [2k/a]^{-1/2}\, e^{-ikt/a}.
$$
The powers of scale factor, $a(t)$, just allow for 
expanding the field in comoving wavenumbers $k$.
The field amplitude contains a normalizing factor of $V^{-1/2}$,
$V$ being a proper volume, hence the $a^{-3/2}$ factor, if we use 
comoving $V=1$.
Another way of looking at this is that the proper number density of
inflatons goes as $a^{-3}$ as the universe expands. With this boundary condition,
it straightforward to check by substitution 
that the following expression satisfies the evolution equation:
$$
\boxit{
\omega_k=a^{-3/2}\, [2k/a]^{-1/2}\, e^{-ik/aH}\; (1+iaH/k)
}
$$
(remember that $H$ is a constant, so that $(d/dt)[aH]=H\dot a=aH^2$ {\it etc.}).
At early times, when the horizon is much larger than the
wavelength, $aH/k \ll 1$, and so $\omega_k$ is the flat-space 
result, except that the time dependence looks a little odd,
being $\exp[-ik/aH]$. However, since 
$(d/dt)[k/aH]=-k/a$, we see that the oscillatory term has a
leading dependence on $t$ of the desired $kt/a$ form. In the limit
of very early times, the period of oscillation is $\ll H^{-1}$, so
$a$ is effectively a constant from the point of view of the
epoch where quantum fluctuations dominate.

At the opposite extreme, $aH/k \gg 1$, the fluctuation
amplitude becomes frozen out at the value
$$
\bra{0} \; |\phi_k|^2\; \ket{0} = {H^2\over 2k^3}.
$$
The initial quantum zero-point fluctuations in the field
have been transcribed to a constant classical fluctuation
which can eventually manifest itself as large-scale structure.
The fluctuations in $\phi$ depend on $k$ in such a way that
the fluctuations per decade are constant:
$$
{d\,(\delta\phi)^2 \over d\, \ln k} = {4\pi k^3\over (2\pi)^3}\;
\bra{0} \; |\phi_k|^2\; \ket{0} = \left({H\over 2\pi}\right)^2.
$$
This completes the argument. The rms value of fluctuations
in $\phi$ can be used as above to deduce the
power spectrum of mass fluctuations well after inflation
is over. In terms of the variance per $\ln k$ in 
potential perturbations, the answer is
$$
\boxit{
\eqalign{
\delta_{\ss H}^2\equiv  \Delta^2_\Phi(k) &= {H^4\over [2\pi\dot\phi]^2} \cr
H^2 &= {8\pi\over 3}\, {V\over m_{\ss P}^2} \cr
3H\dot\phi &= -V', \cr
}
}
$$     
where we have written once again the exact relation
between $H$ and $V$ and the slow-roll condition, since
manipulation of these three equations is often required in derivations.

This result calls for a number of comments.
First, if $H$ and $\dot\phi$ are both constant
then the predicted spectrum is exactly 
scale-invariant, with some characteristic
inhomogeneity on the scale of the horizon.
As we have seen, exact de Sitter space
with constant $H$ will not be strictly correct
for most inflationary potentials; nevertheless,
in most cases the main points of the analysis still go through. 
The fluctuations in $\phi$ start as normal flat-space
fluctuations (and so not specific to de Sitter space),
which change their character as they are advected beyond
the horizon and become frozen-out classical fluctuations.
All that matters is that the Hubble parameter is roughly
constant for the few $e$-foldings that are required for
this transition to happen. If $H$ does change with time,
the number to use is the value at the time that a mode of
given $k$ crosses the horizon. Even if $H$ were to be
made precisely constant, there remains the dependence on
$\dot\phi$, which again will change as different scales
cross the horizon. This means that different inflationary
models display different characteristic deviations from
a nearly scale-invariant spectrum, and this is discussed
in more detail below.
Two other characteristics of the perturbations are
more general: they should be Gaussian and adiabatic in
nature. A Gaussian density field is one for which the
joint probability distribution of the density at any
given number of points is a multivariate Gaussian.
The easiest way for this to arise in practice is for
the density field to be constructed as a superposition of
Fourier modes with independent random phases; the Gaussian
property then follows from the Central Limit Theorem.
It is easy to see in the case of inflation that this
requirement will be satisfied: the quantum commutation
relations only apply to modes of the same $k$, so that
modes of different wavelength behave independently and
have independent zero-point fluctuations. 
Finally, the principal result of the inflationary
fluctuations in their late-time classical guise is
as a perturbation to curvature, and it is not easy
to see how to produce the separation in behaviour between
photon and matter perturbations which is needed for
isocurvature modes. Towards the end of inflation, the
universe contains nothing but scalar field and whatever
mechanisms that generate the matter/antimatter asymmetry
have yet to operate. When they do, the result will be
a universal photon/baryon ratio but with a total density
modulated by the residual inflationary fluctuations --
adiabatic initial conditions, in short. 

Inflation thus makes a relatively firm prediction about the statistical
character of the initial density perturbations, plus a 
somewhat less firm prediction for their power spectrum.
With sufficient ingenuity, the space of predictions can
be widened; isocurvature perturbations
can be produced at the price of introducing additional
inflation fields and carefully adjusting the coupling
between them (Kofman \& Linde 1987), and breaking the
Gaussian character of the fluctuations is also possible
in such multi-field models (Yi \& Vishniac 1993) --
essentially because all modes in field 2 can respond
coherently to a fluctuation in field 1, in much the same way
as non-Gaussian perturbations are generated by cosmic strings.
However, a nearly scale-free Gaussian
adiabatic spectrum is an inevitable result in the simplest
models with a single inflaton; if the theory is to have
any predictive power and not to appear contrived, this
is the clear prediction of inflation. As we shall see
in the observational sections, the true state of affairs
seems to be close to this state.

\titlec{}{Inflaton coupling}
The calculation of density inhomogeneities sets an
important limit on the inflation potential.
From the slow-rolling equation, we know that the number of $e$-foldings
of inflation is 
$$
N=\int H\, dt= \int H\, d\phi/\dot\phi=\int 3H^2\; d\phi/V'.
$$
Suppose $V(\phi)$ takes the form $V=\lambda\phi^4$,
so that $N=H^2/(2\lambda \phi^2)$.
The density perturbations can then be expressed as
$$
\delta_{\ss H}\sim {H^2\over \dot\phi} = {3H^3\over V'}
\sim\lambda^{1/2}N^{3/2}.
$$
Since $N\gs 60$, the observed $\delta_{\ss H}\sim 10^{-5}$
requires
$$
\boxit{
\lambda\ls 10^{-15}.
}
$$
Alternatively, in the case of $V=m^2\phi^2$,
$\delta_{\ss H} = 3H^3/(2m^2\phi)$. Since $H\sim\sqrt{V}/m_{\ss P}$,
this gives $\delta_{\ss H}\sim m\phi^2/m_{\ss P}^3\sim 10^{-5}$.
Since we have already seen that $\phi\gs m_{\ss P}$ is needed
for inflation, this gives
$$
\boxit{
m\ls 10^{-5} m_{\ss P}.
}
$$

These constraints appear to suggest a defect in inflation, in that
we should be able to us the theory to {\it explain\/}
why $\delta_{\ss H}\sim 10^{-5}$, rather than using this
observed fact to constrain the theory. The amplitude of
$\delta_{\ss H}$ is one of the most important numbers
in cosmology, and it is vital to know if there is a simple
explanation for its magnitude. Such an explanation does exist
for theories based on topological defects, where we would have
$$
\delta_{\ss H} \sim [E_{\ss GUT}/E_{\ss P}]^2.
$$
In fact, the situation in inflation is similar, since another
way of expressing the horizon-scale amplitude is
$$
\delta_{\ss H}\sim {V^{1/2} \over m_{\ss P}^2\, \epsilon^{1/2}}.
$$
We have argued that inflation will end with $\epsilon$ of order
unity; if the potential were to have the characteristic value
$V\sim E_{\ss GUT}^4$ then this would give the same
prediction for $\delta_{\ss H}$ as in defect theories.
The appearance of a tunable `knob' in inflation theories really arises
because we need to satisfy $\phi\sim m_{\ss P}$ (for enough inflation),
while dealing with the characteristic value $V\sim E_{\ss GUT}^4$ (to be
fair, this is likely to apply only at the start of inflation,
but the potential does not change by a large factor 60 $e$-folds
from the end of inflation unless the total number of $e$-folds
is $\gg 60$). It is therefore reasonable to say that
a much smaller horizon-scale amplitude would need $V\ll E_{\ss GUT}^4$,
{\it i.e.} a smaller $E_{\ss GUT}$ than the conventional value.

This section has demonstrated the cul-de-sac in which
inflationary models now find themselves: the field
which drives inflation must be very weakly coupled --
and effectively undetectable in the laboratory. Instead of Guth's original
heroic vision of a theory motivated by particle
physics, we have had to introduce a new entity into
particle physics which exists only for cosmological
purposes. In a sense, then, inflation is a failure.
However, the hope of a consistent scheme eventually
emerging (plus the lack of any alternative), means
that inflationary models continue to be explored
with great vigour.

\titleb{3.3}{Gravity waves and tilt}
The density perturbations left behind as a residue
of the quantum fluctuations in the inflaton field during
inflation are an important relic of that epoch, but
are not the only one. In principle, a further important
test of the inflationary model is that it also
predicts a background of gravitational waves, whose
properties couple with those of the density fluctuations.

It is easy to see in principle how such waves arise.
In linear theory, any quantum field 
is expanded in a similar way into a sum of oscillators
with the usual creation and annihilation operators;
the above analysis of quantum fluctuations in a scalar field
is thus readily adapted to show that analogous fluctuations will
be generated in other fields during inflation. In fact, the linearized
contribution of a gravity wave $h_{\mu\nu}$ to the Lagrangian
looks like a scalar field $\phi=(m_{\ss P}/4\sqrt{\pi})\, h_{\mu\nu}$, so
the expected rms gravity-wave amplitude is 
$$
h_{\rm rms}\sim H/m_{\ss P}.
$$
The fluctuations in $\phi$ are transmuted into density fluctuations,
but gravity waves will survive to the present day, albeit redshifted. 

This redshifting produces a break in the spectrum of waves. Prior
to horizon entry, the gravity waves produce a scale-invariant
spectrum of metric distortions, with amplitude $h_{\rm rms}$ per
$\ln k$. These distortions are observable via the large-scale
CMB anisotropies, where the tensor modes produce a spectrum with
the same scale dependence as the Sachs-Wolfe gravitational
redshift from scalar metric perturbations. In the scalar
case, we have $\delta T/T\sim \phi/3c^2$ -- {\it i.e.} of
order the Newtonian metric perturbation; similarly, the tensor effect is
$$
\left.{\delta T\over T}\right|_{\ss GW}\sim h_{\rm rms}\ls \delta_{\ss H}\sim 10^{-5}
$$
(where the second step follows because the tensor modes can make no
more than 100\% of the observed CMB anisotropy).
The energy density of the waves is $\rho_{\ss GW}\sim m_{\ss P}^2 h^2 k^2$,
where $k\sim H(a_{\rm entry})$ is the proper wavenumber of the waves.
At horizon entry,  we therefore expect
$$
\rho_{\ss GW}\sim m_{\ss P}^2 \, h_{\rm rms}^2 \, H^2(a_{\rm entry}).
$$
After horizon entry, the waves redshift away like radiation, as $a^{-4}$, and generate
a present-day energy spectrum per $\ln k$ which is constant for modes which
entered the horizon while the universe was radiation dominated (because
$a\propto t^{1/2}\Rightarrow H^2a^4=\rm const$). What is the density parameter
of these waves? In natural units, $\Omega=(8\pi/3) \rho/(H^2m_{\ss P}^2)$,
so $\Omega_{\ss GW}\sim  h_{\rm rms}^2$ at the time of horizon entry --
at which epoch the universe was radiation dominated with $\Omega_r=1$ to
an excellent approximation. 
Thereafter, the wave density maintains
a constant ratio to the radiation density, since both redshift as $a^{-4}$,
giving the present-day density as
$$
\boxit{
\Omega_{\ss GW}\sim \Omega_r \, [H/m_{\ss P}]^2 \sim 10^{-4} V/m_{\ss P}^4.
}
$$
Therefore, just as with density perturbations in dark matter, the
gravity-wave spectrum displays a break between constant metric fluctuations
on super-horizon scales to constant density fluctuations on small scales.
If gravity waves make an important contribution to CMB anisotropies,
$h_{\rm rms}\sim 10^{-5}$ and
$\Omega_{\ss GW}\sim 10^{-14}$ is expected.

A gravity-wave background of a similar flat
spectrum is also predicted from cosmic strings (see section 10.4 of
Vilenkin \& Shellard 1994). Here, the prediction is
$$
\Omega_{\ss GW}\sim 100\, [G\mu/c^2]\, \Omega_r,
$$
where $\mu$ is the mass per unit length. A viable string
cosmology requires $G\mu/c^2\sim 10^{-5}$, so $\Omega_{\ss GW}\sim 10^{-7}$
is expected -- much higher than the inflationary prediction.

The part of the spectrum with periods of order years
would perturb the emission from pulsars through
fluctuating gravitational redshifts, and the
absence of this modulation sets a bound of $\Omega \ls 10^{-7}$,
so that $V\ls 10^{-2} m_{\ss P}^4$. 
However, this is still a very long way from the interesting inflationary
level of $\Omega_{\ss GW}\ls 10^{-14}$. 
Although the strains implied by this level of relic gravity-wave
background are tiny, it is not completely inconceivable that
space-based versions of the same interferometer technology being used
to search for kHz-period gravity waves on Earth might eventually
reach the required sensitivity. A direct detection of the gravity-wave
background at the expected level would do much the same for the credibility of
inflation as was achieved for the Big Bang itself by Penzias \& Wilson in 1965.

An alternative way of presenting the gravity-wave effect on the
CMB anisotropies is via the ratio between the tensor effect
of gravity waves and the normal scalar Sachs-Wolfe effect,
as first analysed in a prescient paper by Starobinsky (1985).
Express the fractional temperature variance as the contribution of
a given spherical harmonic, $C_\ell$; for a scale-invariant
spectrum, $\ell^2 C_\ell$ is a constant. The tensor and scalar contributions are 
respectively
$$
\ell^2\, C_\ell^{\ss T} \sim h_{\rm rms}^2\sim [H^2/m_{\ss P}^2] \sim V/m_{\ss P}^4.
$$
$$
\ell^2\, C_\ell^{\ss S} \sim \delta_{\ss H}^2 \sim H^2/\dot\phi \sim H^6/(V')^2 \sim 
  {V^3 \over m_{\ss P}^6\, V'{}^2}.
$$
The ratio of tensor to scalar variances of microwave background anisotropies 
is therefore proportional to the inflationary parameter $\epsilon$:
$$
{C_\ell^{\ss T} \over
C_\ell^{\ss S}} \simeq 12.4 \, \epsilon,
$$
inserting the exact coefficient from Starobinsky (1985).
If it could be measured, the gravity-wave contribution to
CMB anisotropies would therefore give a measure of one of the
dimensionless inflation parameters, $\epsilon$. The less
de Sitter-like the inflationary behaviour, the larger
the relative gravitational-wave contribution.

Since deviations from exact exponential expansion
also manifest themselves as density fluctuations
which have spectra that deviate from scale invariance,
this suggest a potential test of inflation. Define the
{\bf tilt} of the fluctuation spectrum as:
$$
{\rm tilt} = (1-n) = -{d \ln \delta^2_{\ss H} \over d\, \ln k}.
$$
We then want to express the tilt in terms of parameters
of the inflationary potential, $\epsilon$ and $\eta$.
These are of order unity when inflation terminates;
$\epsilon$ and $\eta$ must therefore be evaluated when the
observed universe left the horizon, recalling that
we only observe the last 60-odd $e$-foldings of inflation.
The way to introduce scale dependence is to write the
condition for a mode of given comoving wavenumber to
cross the de Sitter horizon
$$
a/k = H^{-1}.
$$
Since $H$ is nearly constant during the inflationary evolution,
we can replace $d/d\, \ln k$ by $d\, \ln a$, and use the slow-roll
condition to obtain
$$
{d\over d\, \ln k}=a\, {d\over da}={\dot\phi\over H}\,{d\over d\phi}
=-{m_{\ss P}^2\over 8\pi}\,{V'\over V}\,{d\over d\phi}.
$$
We can now work out the tilt, since the horizon-scale amplitude is
$$
\delta_{\ss H}^2 = {H^4\over [2\pi\dot\phi]^2} = {128\pi\over 3} \,
{V^3 \over m_{\ss P}^6\, V'{}^2},
$$
and derivatives of $V$ can be expressed in terms of the dimensionless
parameters $\epsilon$ and $\eta$.
The tilt of the density perturbation spectrum is thus
predicted to be
$$
\boxit{
(1-n)=6\epsilon - 2\eta
}
$$

For most models in which the potential is a smooth
polynomial-like function, $|\eta|\simeq |\epsilon|$.
Since $\epsilon$ has the larger coefficient and is
positive by definition, a general but not unavoidable
prediction of inflation is that the spectrum of
scalar perturbations should be slightly tilted
in the sense that $n$ is slightly less than unity.
With a similar level of confidence, one can state
that there is a coupling between this tilt and the
level of the gravity-wave contribution to CMB anisotropies:
$$
{C_\ell^{\ss T} \over
C_\ell^{\ss S}} \simeq 6(1-n)
$$
In principle, this is a distinctive prediction of inflation, but it
is a test which loses power the more closely
the fluctuations approach scale invariance.

It is interesting to put flesh on the bones of this
general expression and evaluate
the tilt for some specific inflationary models.
This is easy in the case of power-law inflation with
$a\propto t^p$ because the inflation parameters are constant:
$\epsilon=\eta/2=1/p$, so that the tilt here is always
$$
(1-n)=2/p
$$
In general, however, the inflation derivatives have to be 
evaluated explicitly on the
largest scales, 60 $e$-foldings prior to the end of inflation, so
that we need to solve
$$
60=\int H\; dt={8\pi\over m_{\ss P}^2} \int_{\phi_{\rm end}}^{\phi}
{V\over V'}\; d\phi.
$$
A better motivated choice than power-law inflation would be a power-law potential
$V(\phi)\propto \phi^\alpha$; many chaotic inflation models
concentrate on $\alpha=2$ (mass-like term) or $\alpha=4$
(highest renormalizable power). Here, $\epsilon=m_{\ss P}^2\alpha^2/(16\pi \phi^2)$,
$\eta=\epsilon \times 2(\alpha-1)/\alpha$, and
$$
60={8\pi\over m_{\ss P}^2} \int_{\phi_{\rm end}}^{\phi}
{\phi\over\alpha}\; d\phi={4\pi\over m_{\ss P}^2\alpha}[\phi^2-\phi_{\rm end}^2].
$$
It is easy to see that  $\phi_{\rm end}\ll\phi$ and that $\epsilon=\alpha/240$,
leading finally to
$$
(1-n)=(2+\alpha)/120.
$$
The predictions of simple chaotic inflation are thus very close to
scale invariance in practice: $n=0.97$ for $\alpha=2$ and $n=0.95$
for $\alpha=4$. However, such a tilt has a significant effect
over the several decades in $k$ from CMB anisotropy
measurements to small-scale galaxy clustering.
These results are in some sense the default inflationary predictions:
exact scale invariance would be surprising, as would large amounts
of tilt. Either observation would indicate that the potential must
have a more complicated structure (or that the inflationary framework
is not correct).

\titlea{4}{Gravitational dynamics of fluctuations}
\titleb{4.1}{Gravitational perturbation theory}
We now summarize briefly some of the basics relating
to the growth of density perturbations in the post-inflationary phase.
The study of perturbations in general relativity can be a rather complicated
and messy subject. Fortunately, most of the essential
physics can be extracted from a Newtonian approach
(not so surprising when one remembers that small
perturbations mean weak gravitational fields).
We start by writing down the fundamental equations
governing fluid motion (non-relativistic for now):
$$
\eqalign{
&{\rm Euler:}\quad {D{\bf v}\over Dt} = -{\del p\over \rho} - \del\Phi\cr
&{\rm Energy:}\quad {D\rho\over Dt} = -\rho {\bf \del\cdot v}\cr
&{\rm Gauss:}\quad \nabla^2\Phi = 4\pi G\rho,\cr
}
$$
where $D/Dt= \partial/\partial t + {\bf v\cdot\del}$ is
the usual comoving derivative.
We now {\bf linearize} these by collecting terms of first
order in perturbations about a homogeneous background:
$\rho=\rho_0 +\delta\rho$ {\it etc.}
The result looks simpler if we define the fractional
density perturbation
$$
\delta\equiv {\delta\rho\over\rho_0}.
$$
Also, when dealing with time derivatives of perturbed quantities,
the full comoving time derivative $D/Dt$ can be replaced by 
$d/dt\equiv \partial/\partial t
+ {\bf v_0\cdot\del}$ -- which is the time derivative for an observer
comoving with the unperturbed expansion of the Universe.
We then can write
$$
\eqalign{
&{d\over dt} {\bf \delta v} = - {\del\, \delta p\over \rho_0}
  -\del\, \delta\Phi - ({\bf \delta v\cdot\del}){\bf v_0}\cr
&{d\over dt} \delta = - {\bf \del\cdot\delta v}\cr
&\nabla^2 \delta\Phi = 4\pi G\rho_0\, \delta.\cr
}
$$
The next step is to translate spatial derivatives into
comoving coordinates:
$$
{\bf x}(t) = a(t) {\bf r}(t)\quad\Rightarrow\quad 
   \del\rightarrow {1\over a}\del,
$$
and to make a similar transformation for peculiar velocity:
$$
{\bf \delta v} = a{\bf u}
$$
(although note that $\bf u$ is still a function of time,
unlike $\bf r$).

When the equations are recast in these variables, there is
only one complicated term to be dealt with:
$({\bf \delta v\cdot\del}){\bf v_0}$ on the rhs of the perturbed Euler equation.
This is best attacked by writing it in components:
$$
[({\bf \delta v\cdot\del}){\bf v_0}]_j=[\delta v]_i\nabla_i[v_0]_j=H\,[\delta v]_j,
$$
where the last step follows because ${\bf v_0}=H\,{\bf x_0}
\Rightarrow \nabla_i[v_0]_j=H\,\delta_{ij}$.
The equations for conservation of momentum and matter then
take the following simple forms in comoving units:
$$
\boxit{
\eqalign{
&\dot{\bf u}+ 2{\dot a\over a}\,{\bf u} = {{\bf g}\over a} -{ {\bf\del}\,\delta p\over \rho_0}\cr
&\dot\delta = -{\bf \del\cdot u}.\cr
}}
$$
The peculiar gravitational acceleration is denoted by $\bf g$. Note that, in the
absence of peculiar accelerations and pressure forces, velocities redshift
away through the `Hubble drag' term $2H{\bf u}$. This behaviour is
reasonable: if we shoot a bullet away from us with a proper
peculiar velocity v, then after time $t$ it is $vt$ away, and its
near neighbours have a recessional velocity $H\, vt$. The proper
velocity thus decays as $\dot v+Hv=0$ or $\dot u+2Hu=0$, because
the bullet is always having to overtake distant galaxies with
progressively higher speeds.

After doing all this, we still have three equations in
four variables ($\delta$, $\bf u$, $\delta\Phi$, $\delta p$).
The system needs an equation of state to be closed, which
may be specified in terms of the sound speed
$$
c_s^2\equiv {\partial p\over \partial \rho}.
$$
If we now think of a plane-wave disturbance $\delta\propto
e^{i{\bf k\cdot r}}$ (so that $\bf k$ is a comoving wave-vector),
then an equation for $\delta$ can be obtained by eliminating $\bf u$
(take the divergence of the perturbed Euler equation  and the
time derivative of the continuity equation and eliminate $\bf\del\cdot\dot u$):
$$
\boxit{
\ddot\delta + 2 {\dot a\over a}\dot\delta =
  \delta \bigl[ 4\pi G\rho_0 - c_s^2 k^2/a^2\bigr].
}
$$

This equation is the one which governs
gravitational amplification of density perturbations.
If we had linearized about a stationary background
and taken $\bf k$ to be a {\it proper\/} wave-vector,
then the equation would have been much easier to derive,
and the result would be just 
$\ddot\delta=\delta(4\pi G\rho_0 - c_s^2 k^2)$, which
has the solutions
$$
\delta(t)= e^{\pm t/\tau};\quad \tau=1/\sqrt{4\pi G\rho_0 - c_s^2 k^2}.
$$
There is a critical proper wavelength,
known as the {\bf Jeans' Length}, at which we switch from
the possibility of exponential growth for long-wavelength
modes to standing sound waves at short wavelengths.
This critical length is
$$
\boxit{
\lambda_{\ss J} = c_s \sqrt{{\pi\over G\rho}},
}
$$
and clearly delineates the scale at which sound waves
can cross an object in about the time needed for
gravitational free-fall collapse.
When considering perturbations in an
expanding background, things are more complex.
Qualitatively, we expect to have no growth when
the `driving term' on the rhs is negative.
However, owing to the expansion, $\lambda_{\ss J}$ will
change with time, and so a given perturbation
may switch between periods of growth and stasis.
These effects help to govern the form of the
perturbation spectrum which is propagated to the
present Universe from early times,
and will be considered in detail shortly.

\titlec{}{Radiation-dominated universes}
At early enough times, the Universe was radiation dominated
($c_s=c/\sqrt{3}$) and the analysis so far does not apply.
It is conventional to resort to general relativity perturbation theory
at this point. However, the fields are still weak, and
so it is possible to generate the results we need by
using special relativity fluid mechanics and linearized Einstein gravity.
For simplicity, assume that pressure gradients are
negligible ({\it i.e.} restrict ourselves to
$\lambda\gg\lambda_{\ss J}$ from the start). The basic
equations are then
$$
\eqalign{
&{\rm Euler:}\quad {D{\bf v}\over Dt} = - \del\Phi\cr
&{\rm Energy:}\quad {D\over Dt}(\rho +p/c^2) =
  {\partial\over\partial t}p/c^2 - (\rho +p/c^2) {\bf \del\cdot v}\cr
&{\rm Gauss:}\quad \nabla^2\Phi = 4\pi G(\rho +3p/c^2).\cr
}
$$
For total radiation domination, $p=\rho c^2/3$ and we can linearize
as before, to obtain
$$
\boxit{
\ddot\delta + 2 {\dot a\over a}\dot\delta =
  {32\pi \over 3} G\rho_0  \delta, 
}
$$
so the net result of all the relativistic corrections is
a driving term which is a factor $8/3$ higher.

\titlec{}{Solutions for $\delta(t)$}
In both matter- and radiation-dominated universes with $\Omega=1$, we 
have $\rho_0\propto 1/t^2$:
$$
\eqalign{
&{\rm Matter\  domination}\ (a\propto t^{2/3}): 4\pi G\rho_0 = {2\over 3t^2}\cr
&{\rm Radiation\ domination}\ (a\propto t^{1/2}): 32\pi G\rho_0/3 = {1\over t^2}\cr
}
$$
Every term in the equation for $\delta$ is thus the product of
derivatives of $\delta$ and powers of $t$, and  a power-law
solution is obviously possible. If we try $\delta\propto t^n$, then
the result is $n=2/3$ or $-1$ for matter domination;
$n=\pm 1$ for radiation domination.
For the growing mode, these can be combined rather conveniently
using the {\bf conformal time} $\eta\equiv\int dt/a$:
$$
\boxit{
\delta \propto \eta^2.
}
$$
Recall that $\eta$ is proportional to the comoving size of the
horizon.

One further way of stating this result is that gravitational
potential perturbations are independent of time (at least while
$\Omega=1$). Poisson's equation tells us that $-k^2\Phi/a^2\propto \rho\,\delta$;
since $\rho\propto a^{-3}$ for matter domination or $a^{-4}$
for radiation, that gives $\Phi\propto \delta/a$ or $\delta/a^2$
respectively -- independent of $a$ in either case. In other words,
the metric fluctuations resulting from potential perturbations are
frozen, at least for perturbations which are outside the horizon.
This conclusion demonstrates the self-consistency of the basic
set of fluid equations which were linearized. The equations for
momentum and energy conservation are always valid, but the correct
relativistic description of linear gravity should be a wave equation, including
time derivatives of $\Phi$ as well as spatial derivatives.
Since we have used only Newtonian gravity, this implies that
any solutions of the perturbation equations in which $\Phi$
varies will not be valid on super-horizon scales. This
criticism does not apply to the growing mode, where $\Phi$ is
constant, but it does apply to decaying modes (Press \& Vishniac 1980).
This difficulty concerning perturbations on scales greater than
the horizon is related to gauge freedom in general relativity and
the fact that the value of density perturbations $\delta$ can
be altered by a coordinate transformation. 
Exchange of light signals can be used to establish a `sensible'
coordinate system, but only on sub-horizon scales. Otherwise, the
results obtained are gauge dependent. We are implicitly using here
what might be called the Newtonian gauge, where the the metric is
expressed as the FRW form with perturbation factors
$(1\pm 2\Phi/c^2)$ in the time and spatial parts respectively.

In models with $\Omega<1$, the growth is slowed. It is possible
to write explicit expressions for $\delta(a)$, but it is
more convenient in practice to use the following 
accurate approximation, due to Carroll {\it et al.}
(1992), which also allows for the effects of vacuum energy:
$$
{\delta(z=0,\Omega)\over\delta(z=0,\Omega=1)} \simeq 
\frac{5}{2}\Omega_m\left[\Omega_m^{4/7}-\Omega_v+
 (1+\half\Omega_m)(1+\frac{1}{70}\Omega_v)\right].
$$
For models without vacuum energy, the growth is
reduced by a factor of approximately $\Omega^{0.65}$;
for flat models with $\Omega_m+\Omega_v=1$, the
growth suppression is less marked -- approximately $\Omega^{0.23}$.

\titlec{}{M\'esz\'aros effect}
What about the case of collisionless matter in a radiation
background? The fluid treatment is not appropriate here,
since the two species of particles can interpenetrate.
A particularly interesting limit is for perturbations
well inside the horizon: the radiation can then be treated as
a smooth, unclustered background which affects only the overall
expansion rate. The perturbation equation is as before
$$
\ddot\delta + 2 {\dot a\over a}\dot\delta = 4\pi G\rho_m  \delta, 
$$
but now $H^2=8\pi G(\rho_m+\rho_r)/3$. If we change variable to
$y\equiv \rho_m/\rho_r=a/a_{\rm eq}$, then the equation becomes
$$
\delta'' + {2+3y\over 2y(1+y)}\, \delta' -{3\over 2y(1+y)}\, \delta
$$
(for $k=0$, as appropriate for early times).
It may be seen by inspection that a growing solution exists with $\delta''=0$:
$$
\boxit{
\delta\propto y + 2/3.
}
$$
It is also possible to derive the decaying mode. This is simple in
the radiation-dominated case ($y\ll 1$): $\delta\propto -\ln y$ is
easily seen to be an approximate solution in this limit.

What this says is that, at early times, the dominant energy of
radiation drives the universe to expand so fast that the matter
has no time to respond, and $\delta$ is frozen at a constant.
At late times, the radiation becomes negligible, and the growth
picks up smoothly to the Einstein-de Sitter $\delta\propto a$ behaviour.
The overall behaviour is therefore similar to the effects of
pressure on a coupled fluid. For scales greater than the horizon,
perturbations in matter and radiation can grow together; this
growth ceases once the perturbations enter the horizon.
However, the explanations are completely different. In the
fluid case, the radiation pressure prevents the perturbations
from collapsing further; in the collisionless case, the photons
have free-streamed away, and the matter perturbation fails to
collapse only because radiation domination ensures that the
universe expands too quickly for the matter to have time to
self-gravitate. Because matter perturbations enter the horizon with
$\dot\delta>0$, $\delta$ is not frozen quite at the horizon-entry
value, and continues to grow until this initial `velocity'
is redshifted away, giving a total boost factor of roughly
$\ln y_{\rm entry}$. This log factor may be seen below in the
fitting formulae for the CDM power spectrum.

\titlec{}{Solutions for {\bf u}}
To discuss velocities, go back to the basic equation for $\bf u$:
$$
{\bf\dot u} + {2\dot a\over a} {\bf u} = {{\bf g}\over a};
$$
${\bf g=-\del}\delta\Phi/a$ is the peculiar gravitational acceleration,
and pressure terms are neglected, so $\lambda\gg\lambda_{\ss J}$.
The peculiar velocity can be decomposed into modes either parallel or
perpendicular to $\bf g$.
The latter are {\bf vorticity modes} which decay.
For the former, we know from the
continuity equation (${\bf\del\cdot u} = -\dot\delta$) that
${\bf\dot u} = (\ddot\delta/\dot\delta){\bf u}$.
Hence, the solution of the above equation for $\bf u$
has $\bf u\propto g$, and may be expressed as
$$
{\bf\delta v} = {2f(\Omega)\over 3 H\Omega}\;{\bf g},
$$
where the function $f(\Omega)\equiv(a/\delta)d\delta/da$.
A very good approximation to this (Peebles 1980) is
$f\simeq\Omega^{0.6}$.
Alternatively, we can work in Fourier terms. This is
easy, as $\bf g$ and $\bf k$ are parallel, so that
${\bf \del\cdot u}=ik{\bf u}$. Thus, directly from
the continuity equation,
$$
\boxit{
{\bf \delta v_k} = -{iHg(\Omega) a\over k} \; \delta_k \; {\bf\hat k}.
}
$$
The $1/k$ factor tells us that cosmological velocities
come predominantly from large-scale perturbations.
Deviations from the Hubble flow are therefore in principle
a better probe of the inhomogeneity of the
universe than large-scale clustering.

\titleb{4.2}{Transfer functions}
There are in essence two ways in which the power spectrum
which exists at early times may differ from that which
emerges at the present, both of which correspond
to a reduction of small-scale fluctuations:

(i) Jeans' mass effects.
Prior to matter-radiation equality, we have already seen
that perturbations inside the horizon are prevented
from growing by radiation pressure.
This leads to an effective `break' of $\Delta n=4$
in the power spectrum at this point:
$$
\eqalign{
\delta_k&\propto \lambda^{-(n+3)/2}\quad \lambda>\lambda_{\ss J}\cr
&\propto \lambda^{-(n-1)/2}\quad \lambda<\lambda_{\ss J}\cr
}
$$
Once $z_{\rm eq}$ is reached, one of two things can happen.
If collisionless dark matter dominates, perturbations on all
scales can grow. If baryonic gas dominates, the Jeans length
remains approximately constant, as follows:
The sound speed, $c_s^2=\partial p/\partial\rho$, may
be found by thinking about the response of matter and radiation
to small adiabatic compressions: $\delta p = (4/9)\rho_r c^2 (\delta V/V)$,
$\delta\rho=[\rho_m+(4/3)\rho_r](\delta V/V)$, implying
$$
c_s^2=c^2\left(3+{9\over 4}{\rho_m\over\rho_r}\right)^{-1} =
  c^2\left(3+{9\over 4}{1+z_{\rm rad}\over 1+z}\right)^{-1}.
$$
Here, $z_{\rm rad}$ is the redshift of equality between matter
and photons ($1+z_{\rm rad}=1.68(1+z_{\rm eq})$ because of the
neutrino contribution).
At $z\ll z_{\rm rad}$, we therefore have $c_s\propto \sqrt{1+z}$.
Since $\rho=(1+z)^3 3\Omega_{\ss B} H_0^2/(8\pi G)$, the {\it comoving\/}
Jeans' length is constant at
$$
\Lambda_{\ss J}={c\over H_0}\left({32 \pi^2\over 27\Omega_{\ss B} (1+z_{\rm rad})}\right)^{1/2}
  = 50\, (\Omega_{\ss B} h^2)^{-1}\;\;{\rm Mpc}.
$$
Thus, in either case,
one of the critical length scales for the power
spectrum will be the horizon distance at $z_{\rm eq}$
($=25000\Omega h^2$ for $T=2.7$ K, counting neutrinos as radiation). 
In the matter-dominated approximation, we get
$$
d_{\ss H}={2c\over H_0}(\Omega z)^{-1/2} = 29\, (\Omega h^2)^{-1}{\rm Mpc}.
$$
The exact answer including radiation is a factor
$\sqrt{2}-1$ times this: $15.7\,(\Omega h^2)^{-1}$ Mpc.

(ii) Damping. In addition to having their growth
retarded, very small perturbation will be erased
entirely, which can happen in one of two ways.
For collisionless dark matter, perturbations are
erased simply by {\bf free streaming}: random
particle velocities cause blobs to disperse.
At early times ($kT>mc^2$), the particles will
travel at $c$, and so any perturbation which has
entered the horizon will be damped.
This process switches off when the particles
become non-relativistic; for massive particles,
this happens long before $z_{\rm eq}$
({\bf Cold Dark Matter}).
For massive neutrinos, on the other hand, it happens
{\it at\/} $z_{\rm eq}$: only perturbations on
very large scales survive in the case of
{\bf hot dark matter}.
In a pure baryon universe, the corresponding
process is called {\bf Silk damping}: the
mean free path of photons due to scattering by the plasma
is non-zero, and so radiation can diffuse out of a
perturbation, convecting the plasma with it.
The typical distance of a random walk in terms
of the diffusion coefficient, $D$, is $x\simeq\sqrt{Dt}$,
which gives a damping length of
$$
\lambda_{\ss S}\simeq\sqrt{\lambda d_{\ss H}}
$$
-- the geometric mean of the horizon size and the
mean free path.
Since $\lambda=1/(n\sigma_{\ss T}) = 44.3 (1+z)^{-3}(\Omega_{\ss B}h^2)^{-1}$
proper Gpc, we obtain a comoving damping length of
$$
\lambda_{\ss S}=16.3\, (1+z)^{-5/4} (\Omega_{\ss B}^2\Omega h^6)^{-1/4}\; {\rm Gpc}.
$$
This becomes close to the Jeans' length by the time of last
scattering, $1+z\simeq 1000$.

Real power spectra thus result from modifications of
any primordial power by a variety of processes:
growth under self-gravitation, effects of pressure
and dissipative processes. 
In general, modes of short wavelength have their
amplitudes reduced relative to those of
long wavelength in this way.
The overall effect is encapsulated in
the {\bf transfer function}, which gives the
ratio of the late-time amplitude of a mode to its
initial value.
The detailed result can be
hard to calculate, mainly because we have a mixture of
matter (both collisionless dark particles and
baryonic plasma) and relativistic particles
(collisionless neutrinos and collisional photons)
which does not behave as a simple fluid.
Particular problems are caused by the change in
the photon component from being a fluid tightly coupled
to the baryons by Thomson scattering, to being
collisionless after recombination. Accurate
results require a solution of the Boltzmann
equation to follow the evolution in detail.
The transfer function is thus a by-product of
elaborate numerical calculations of microwave
background fluctuations.
Nevertheless, once we possess the transfer function, it
is a most valuable tool. The evolution of
linear perturbations back to last scattering
obeys the simple relations summarised above, and
it is easy to see how structure in the Universe
will have changed during the matter-dominated
epoch.

It is thus invaluable in practice to have some
accurate analytic formulae which fit the
numerical results for transfer functions.
We give below results for some common models in the
form of the transfer function needed to produce a
scale-invariant power spectrum at large wavelength, $\Delta^2\propto k^4 T_k^2$.
For adiabatic models, $T_k$ is the true transfer
function; for isocurvature models, 
this is not the case and $T_k$ is proportional to
$1/k^2$ times the true transfer function.
We assume $\Omega_{\ss B}\ll\Omega$, so that all lengths
scale with the horizon size at matter-radiation equality,
leading to the definition $q\equiv k/(\Omega h^2 {\rm Mpc}^{-1})$. 
We consider the cases of (A) Adiabatic CDM; (B) Adiabatic
massive neutrinos (1 massive, 2 massless); (C) Isocurvature
CDM; these expressions come
from Bardeen {\it et al.} (1986; BBKS).
$$
\eqalign{
{\rm (A)}\quad T_k&={\ln(1+2.34q)\over 2.34 q}[1+3.89q+(16.1q)^2 +(5.46q)^3
  +(6.71q)^4]^{-1/4}\cr
{\rm (B)}\quad T_k&=\exp(-3.9q -2.1 q^2)\cr
{\rm (C)}\quad T_k&=(1+ [15.0q + (0.9q)^{3/2} + (5.6q)^2]^{1.24})^{-1/1.24}\cr
}
$$

%

These models contain some hidden variables.
Since the characteristic length-scale  in the transfer function
depends on the horizon size at matter-radiation equality,
the temperature of the CMB enters. In the above formulae,
it is assumed to be exactly 2.7 K; for other values, the
characteristic wavenumbers scale $\propto T^{-2}$.
For these purposes massless neutrinos count as radiation,
and three species of these contribute a total density
which is 0.68 that of the photons.

There is also the question of the baryon contribution: the
above expressions assume pure dark matter, which is
unrealistic. At least for CDM models, a non-zero baryonic
density lowers the apparent dark-matter density parameter.
We can define an apparent shape parameter for the transfer function, $\Gamma^*$,
where 
$$
\boxit{
q\equiv (k/h\; {\rm Mpc}^{-1})/ \Gamma^*,
}
$$
and
$\Gamma^* = \Omega h$
in a model with zero baryon content. 
Peacock \& Dodds (1994) showed that the effect of
increasing $\Omega_{\ss B}$ was to preserve the CDM-style
spectrum shape, but to shift to lower values of $\Gamma^*$.
This shift was generalized to models with $\Omega\ne 1$ by
Sugiyama (1995):
$$
\Gamma^*=\Omega h\, \exp[-\Omega_{\ss B}(1+1/\Omega)].
$$

\titleb{4.3}{N-body models}
This is a good place to discuss how to use the above
equations of motion to carry out a nonlinear
evolution of the density field. This is usually
done by means of the {\bf N-body simulation}, in which
the density field is represented by the sum of a
set of fictitious discrete particles. The equations of motion
for each particle depend on solving for the gravitational
field due to all the other particles, finding the
change in particle positions and velocities over
some small time step, moving and accelerating
the particles, and finally re-calculating the
gravitational field to start a new iteration.
Using comoving units for length and velocity
(${\bf v}=a{\bf u}$), we have from above the
equation of motion
$$
{d\over dt}{\bf u}=-2{\dot a\over a}{\bf u} - {1\over a^2}{\bf\del}\Phi,
$$
where $\Phi$ is the Newtonian potential.
The time derivative is already in the form of
a total (or convective) derivative, as required
for particle motions, rather than the partial $\partial /\partial t$.
If we change time variable from $t$ to $a$, this becomes
$$
{d\over d\, \ln a}[a^2{\bf u}] ={a\over H}\,{\bf g}
={G\over aH}\; \sum_i m_i\; {{\bf x_i-x}\over |{\bf x_i-x}|^3}.
$$
Here, the gravitational acceleration has been written exactly
by summing over all particles, but this becomes prohibitive
for very large numbers of particles. Since the problem is
to solve Poisson's equation, a faster approach is to use
Fourier methods, since this allows the use of the
FFT algorithm. If the density perturbation field
(not assumed small) is expressed as $\delta=\sum \delta_k\exp[-i{\bf k\cdot x}]$,
then Poisson's equation becomes $-k^2\Phi_k=4\pi G a^2\bar\rho\, \delta_k$, and
the required $k$-space components of ${\bf\del}\Phi$ are just
$$
[{\bf\del}\Phi]_k=-i\Phi_k{\bf k}.
$$
If we finally eliminate
density in terms of $\Omega$, the equation of motion for a
given particle is
$$
\eqalign{
{d\over d\, \ln a}[a^2{\bf u}] &=\sum {\bf F}_k\exp[-i{\bf k\cdot x}];\cr
{\bf F}_k &=-i{\bf k}\;{3\Omega H a^2\over 2 k^2}\; \delta_k.\cr
}
$$

\titlec{}{Boxes and grids}
The efficient way of performing the required Fourier
transforms is by averaging the data onto a grid and
using the FFT algorithm, both to perform the transformation
of density, and to perform the (three) inverse transforms to
obtain the real-space force components from their $k$-space
counterparts. 
This leads to the simplest
$N$-body algorithm: the {\bf particle-mesh (PM) code}.
The only complicated part of the algorithm is the
procedure for assigning mass to gridpoints, and
interpolating the force as evaluated on the grid
back onto the particles (for consistency,
the same procedure must be used for both these steps).
The most naive method is simply to bin the data:
{\it i.e.} associate a given particle with whatever
gridpoint happens to be nearest. There are a
variety of more subtle approaches (see
Hockney \& Eastwood 1988; Efstathiou {\it et al.} 1985),
but whichever strategy is used, the resolution of
a PM code is clearly limited to about the size of
the mesh. To do better, one can use
a {\bf particle-particle-particle-mesh (P$\bf ^3$M) code},
also discussed by the above authors. Here, the
direct forces are evaluated between particles in the
neighbouring cells, with the grid estimate being used
only for particles in more distant cells. A similar
effect, although without the use of the FFT, is achieved
by {\bf tree codes} ({\it e.g.} Hernquist Bouchet \& Suto 1991).

In practice, however, the increase in resolution from
these methods is limited to a factor of a few. This is
because each particle in a cosmological $N$-body
simulation in fact stands for a large number of less
massive particles. Close encounters of these spuriously
large particles can lead, through three-body processes,
to the formation of unphysical close massive binaries.
To prevent this, the forces must be {\bf softened}: set
to a constant below some critical separation, rather than
rising as $1/r^2$. If there are already few particles
per PM cell, the softening must be some significant
fraction of the cell size, so there is a limit to the
gain over pure PM.
For example, consider a box of side $50h^{-1}$ Mpc, which
is the smallest that can be used to simulate the observed
universe without serious loss of power from the omitted
long-wavelength modes. A typical size of calculation
might use $128^3$ particles on a Fourier mesh of the
same size, so that the mean density is one particle
per cell, and the cell size is of order that of the
core of a rich cluster. To use such a simulation to study cluster
cores means we are interested in overdensities of
$10^3$ -- $10^4$, or typical interparticle separations
of 0.05 -- 0.1 of a cell. To avoid collisional effects,
the pairwise interaction must be softened on this scale.
A much larger improvement in resolution is only justified
in regions of huge overdensity ($\sim 10^6$ for a 100-fold
increase in resolution over PM). The overall message is
that $N$-body simulations in cosmology are severely limited
by mass resolution, and that this limits the spatial
resolution that can be achieved while still modelling
the evolution of the true collisionless fluid.

\titlec{}{Units}
From a practical point of view, it is convenient to
change to a new set of units which incorporate the
size of the computational box, and allow the simulation to
be rescaled to different physical situations.
Let the side of the box be $L$; it is clearly convenient
measure length in terms of $L$
and velocities in terms of the expansion velocity
across the box:
$$
\eqalign{
{\bf X}&= {\bf x}/L \cr
{\bf U}&= \delta {\bf v}/(HLa)= {\bf u}/HL.\cr
}
$$
Since, for $N$ particles the density is $\rho=Nm/(aL)^3$,
the mass of the particles and the gravitational
constant can be eliminated and the
equation of motion can be cast in an attractively
dimensionless form:
$$
\boxit{
{d\over d\, \ln a}[f(a){\bf U}] 
={3\over 8\pi}\,\Omega(a) f(a)\; {1\over N} \sum_i  {{\bf X_i-X}\over |{\bf X_i-X}|^3}.
}
$$
The function $f(a)$ is proportional to $a^2H(a)$, and has an
arbitrary normalization -- {\it e.g.} unity at the initial epoch.
If the forces are instead evaluated on a grid, 
dimensionless wavenumbers ${\bf K=k}\,L$ are used, and the corresponding
equation becomes
$$
\eqalign{
{d\over d\, \ln a}[f(a){\bf U}] &=\sum {\bf F}_K\exp[-i{\bf K\cdot X}];\cr
{\bf F}_K &=-i{\bf K}\;{3 \over 2}\;{\delta_K\over K^2}\; 
\Omega(a) f(a).\cr
}
$$

Particles are now moved according to $d{\bf x}={\bf u}\; dt$,
which becomes
$$
d{\bf X}={\bf U}\; d\ln a
$$
in our new units.
It only remains to set up the initial conditions;
this is easy to do if the  initial epoch is
at high enough redshift that $\Omega=1$, since then
${\bf U}\propto a$ and the initial displacements
and velocities are related by
$$
\Delta{\bf X}=\bf U.
$$
In the case where the initial conditions are
specified late enough that $\Omega$ is significantly
different from unity, this can be modified by using
the linear relation between density
and velocity perturbations: for a given $\Delta$, the
corresponding velocity scales as $\Omega^{0.6}$ (see above).
It is not in fact critical that the density fluctuations
be very small at this time: this is related to a
remarkable approximation for nonlinear dynamics
due to Zeldovich.

\titleb{4.4}{Hierarchical density fields}
We have seen that the perturbations which
survive recombination are of two distinct classes:
in some (such as cold dark matter), primordial
fluctuations survive on very small scales
({\bf small-scale damping}); in other cases,
such as hot dark matter or adiabatic baryons, the perturbation
field is dominated by fluctuations on scales $\sim$
the horizon at $z_{\rm eq}$ ({\bf large-scale damping}).
The consequences for galaxy formation are radically
different.
In the former case, non-linear collapse of sub-galactic
mass units can be the first event which occurs after
recombination. These will then cluster together in a
{\bf hierarchy}, forming successively more massive
systems as time progresses. Hierarchies are also
known as {\bf bottom-up} pictures for galaxy formation.
Conversely, in large-scale damping, pancakes of
cluster or supercluster size are the first structures
to form. Galaxies must be presumed to form through
dissipative processes occurring in the shocked gas
which forms at pancake collapse ({\bf top down}).

As described, it is clear that top-down pictures
are unappealing in that the physics of galaxy
formation is likely to be very complex and messy.
Reality {\it is\/} often like this: the formation
of stars is a good example of a fundamental process
where is is hard to understand what is going on.
In contrast, the computational simplicity of
hierarchies has led to much more detailed work 
being performed on them.
Theoretical prejudice aside, however, the Universe
{\it looks\/} like a hierarchy, displaying many small groups
of galaxies ({\it e.g.} the Milky Way's own Local Group), which exist
within superclusters which are only mildly nonlinear.

It may seem that such a situation cannot be analysed
within the bounds of linear theory, but a way
forward was identified by
Press \& Schechter (1974; PS).
The critical assumption in the PS analysis is that, 
even if the field is non-linear,
the amplitude of large-wavelength modes in the final field
will be close to that predicted from linear theory.
For this to be true requires the `true' large-scale power
to exceed that generated via non-linear coupling
of small-scale modes, which turns out to require a
spectral index $n<1$ (Williams {\it et al.} 1991a).
We now proceed by recognising that, for a massive
clump to undergo gravitational collapse, the average
overdensity in a volume containing that mass should (as usual) exceed
some threshold, $\delta_c$,  of order unity.
The location and properties of these bound
objects can thus be estimated by an artificial smoothing (or filtering)
of the initial linear density field. If the filter function has
some characteristic length $R_f$, then the typical size of
filtered fluctuations will be $\sim R_f$ and they can be
assigned a mass $M \sim \rho_0 R_f^3$. The exact analytic
form of the filter function is arbitrary and is often
taken to be a Gaussian for analytic convenience. 

The argument now proceeds in integral terms. For a given $R_f$, the
probability that a given point lies in a region with $\delta > \delta_c$
(the critical overdensity for collapse) is
$$
p(\delta > \delta_c\ | R_f) = \half 
\left[1 - \erf \left(
  {\delta_c\over\sqrt{2}\ \sigma(R_f)}\right)\right],
$$
where $\sigma(R_f)$ is the linear rms in the filtered version of $\delta$.
The PS argument now takes this to be proportional to the probability
that a given point has ever been processed through 
a collapsed object of scale $>R_f$. 
This is really assuming that the only objects which
exist at a given epoch are those which have just collapsed:
if a point has $\delta > \delta_c$ for a given $R_f$, then it
will have $\delta = \delta_c$ when filtered on some larger scale
and will be counted as an object of the larger scale. 
The problem with this argument is that half the mass
remains unaccounted for: this was amended by PS simply by 
multiplying the probability by a factor 2.
Note that this procedure need not be confined to Gaussian
fields; all we need is the functional form of
$p(\delta>\delta_c\ |R_f)$. The analogue of the factor of 2 problem 
remains: what to do with the points having $\delta<0$.

This integral probability is related to the mass function $f(M)$
(defined such that $f(M)dM$ is the comoving number density of objects in the 
range $dM$)
via 
$$
Mf(M)/\rho_0 = |dp/dM|,
$$ 
where $\rho_0$ is the total comoving
density. Thus,
$$
\boxit{
{M^2f(M)\over\rho_0} = {2\delta_c \over \sqrt{2\pi}\ \sigma}
  \ \left|{d\,\ln\sigma \over d\,\ln M}\right| \ \exp (-\half\delta_c^2/\sigma^2).
}
$$
We have expressed the result in terms of the
{\bf multiplicity function}: $M^2f(M)/\rho_0$ is the fraction of
the mass which is carried by objects in a unit range of $\ln M$.
For power-law spectra, this function takes a very simple
form:
$$
{M^2f(M)\over\rho_0} = {n+3\over 6}\sqrt{{2\over\pi}}\;\,\nu\, e^{-\nu^2/2},
$$
Where $\nu$ is the threshold in units of the rms density fluctuation.
The multiplicity function thus always has the same shape
(a skew-negative hump around $\nu\simeq 1$); changing the
spectral index only alters the mass scale via $\nu=(M/M_c)^{(n+3)/6}$.

\titlec{}{Random walks and conditional mass functions}
The factor of 2 `fudge' has long been recognized as
the crucial weakness of the PS analysis.
What one has in mind is that the mass from lower-density
regions accretes onto collapsed objects, but it does
not seem correct for this to cause a doubling of the total
number of objects. 
Recent work has shed some light on the origin of this
problem (Peacock \& Heavens 1990; Bond {\it et al.} 1991).
To see where the error crept in, consider the
{\bf random trajectory} taken by the filtered field at some
fixed point as a function of filtering radius. This
starts at $\delta=0$ at $R=\infty$, and develops
fluctuations of increasing amplitude as we move to smaller
$R$. Thus, if $\delta<\delta_c$ at a given point, it is
quite possible that it will exceed the threshold at some other
point -- indeed, if the field variance diverges as 
$R\rightarrow 0$, it is inevitable that the threshold
will be exceeded. So, instead of ignoring all points
below threshold at a given $R$, we should find the
{\bf first upcrossing} of the random trajectory:
the largest value of $R$ for which $\delta=\delta_c$.

This analysis is most easily performed for one particular
choice of filter: sharp truncation in $k$-space.
Decreasing $R$ then corresponds to adding in new
$k$-space shells, all of which are independent for
a Gaussian field. The trajectory is then just a
random walk, and the solution is very easy.
Consider a point on the walk which has reached the
threshold; its subsequent motion will be symmetric,
so that it is equally likely to be found above the
threshold as below at some smaller $R$. The probability of
never having crossed the threshold 
(the {\bf survival probability}) is then obtained
by reflection of the Gaussian above threshold:
$$
{dP_{\ss S}\over d\delta} = {1\over\sqrt{2\pi}\;\sigma}
\left[\exp\left(-{\delta^2\over 2\sigma^2}\right)-
 \exp\left(-{(\delta-2\delta_c)^2\over 2\sigma^2}\right)\right].
$$
Integrating this up to get the probability of
having crossed the threshold at least once gives
$$
1-P_{\ss S}=
1-\erf \left(
  {\delta_c\over\sqrt{2}\ \sigma}\right),
$$
which is just twice the unconstrained probability of
lying above the threshold -- thus supplying the
missing factor of 2.
Unfortunately, the above analysis is only valid for this
special choice of filter: using $k$-space filters
which are differentiable leads to just the
original PS form (without the factor 2) at high mass, with the surplus
probability being shifted to low masses, so that the shape of
the function changes. Which filter is the best
choice, we cannot say in advance; we are left with 
the empirical fact that the PS formula does fit
$N$-body results quite well.

One useful extension of the random-walk model is that
it allows a calculation of the {\bf conditional
multiplicity function}: given a particle in a system
of  mass $M_0$ at some epoch $a_0$, what was the
distribution of masses where that particle resided
at some earlier epoch $a_1$?
This is now just the random walk with two absorbing
barriers, at $\delta_c/a_0$ and $\delta_c/a_1$;
we want the probability of not having crossed
the second, subject to having crossed the first
at $M_0$. The solution for the integral
mass distribution is
$$
P(>\mu)=1-\erf\left[{\nu(t^{-1}-1)\over\sqrt{2(\mu^{-1}-1)}}\right],
$$
where $\nu\equiv\delta_c/a_0\sigma_0$; $t\equiv a_1/a_0$;
$\mu\equiv\sigma^2(M_0)/\sigma^2(M)$.
This function tells us that
the early histories of particles which end up in different mass
systems are markedly different over a large range of expansion factor.
This may provide a way of understanding many of the
systematics of galaxy systems in terms of their
merger histories: see Bower (1991); Lacey \& Cole (1993).

The way to deduce a merger {\it rate\/} from the above
is to consider the limit of the
conditional mass function at two nearly equal epochs: the
time derivative of the conditional function must be related to the  
merger rate. First, we need to invert the above reasoning, which
gives $P(M_1,z\mid M_0, z=0)$. Conditional probability
definitions then imply
$$
P(M_0\mid M_1)= { P(M_1\mid M_0) \; P(M_0) \over P(M_1)}.
$$
We now have the probability that an object of mass $M_1$ at redshift
$z$ becomes incorporated into one of mass $M_0$ at redshift $z=0$.
The merger rate $\cal M$, at which an object of mass $M_0$
accretes objects of mass $\Delta M$, is then
$$
{\cal M}(M_0, \Delta M)=
\left.{d^2 P(M_0\mid M_1)\over dM_0\; dt}\right|_{t_1\rightarrow t_0}.
$$

These analytical descriptions of how mass gathers into
clumps in hierarchical clustering have been shown to describe
the results of numerical simulations extremely well
(Bond {\it et al.} 1991) and provide a powerful tool for
understanding the growth of cosmological structure.
For example, it is now possible to answer the apparent
paradox of the simple Press-Schechter theory, where
all collapsed objects considered are those which formed
only at the instant considered ({\it i.e.} sit exactly
at the threshold $\delta=\delta_c$), even though common
sense says that some objects must have formed at high
redshift and survived unchanged to the present day. The
conditional mass function indeed allows us to define
something close to the typical formation epoch for a clump, which
we might take as the time when the probability is 0.5
that the precursor mass is at least half the final mass.
The conditional mass function says that this time
depends on the final mass of the clump; if it is $\gg M^*$,
the formation epoch will be in the recent past, whereas
low-mass clumps survive for longer.
As a specific example, this calculation implies that massive clusters 
form extremely late in an Einstein-de Sitter model.
About 30\% of Abell clusters will have doubled their mass since 
as recently as $z=0.2$ (as against only 5\% if $\Omega=0.2$).
The observed frequency of substructure in clusters is used on
this basis by Lacey \& Cole to argue that $\Omega\gs 0.5$
(although lower $\Omega$ is allowed in models with vacuum energy,
since what matters for this analysis is the linear growth suppression factor).

\titlec{}{Application to galaxy clusters}
The most important practical application of the PS
formalism is to rich clusters. As already discussed, these are the
most massive non-linear systems in the current Universe, so
a study of their properties should set constraints
on the shape and normalization of the power
spectrum on large scales.
These issues are discussed by {\it e.g.} Henry \& Arnaud (1991),
who sidestep the issue of what mass to assign to a cluster
by using the observed distribution of temperatures
(see below for the relation between mass and virial temperature).
Fitting their data with the PS form for top-hat filtering
and $\delta_c=1.69$ they deduce a {\it linear-theory} rms in
$8h^{-1}$ Mpc spheres of $\sigma_8=0.59\pm 0.02$ for $\Omega=1$.
Similarly, White, Efstathiou \& Frenk (1993)
deduce $\sigma_8 =0.57\pm 0.06$, although they rely on the
central masses of Abell clusters being known when calibrating this number.
These calculations have been checked against numerical simulations, so
the result therefore seems rather robust and well-determined.
For models with low density,
the required normalization rises: for a lower mean density,
more collapsed systems are needed to yield the observed number of
objects of a given mass. The approximate scaling with $\Omega$ is
approximately the same as is required in the cosmic virial theorem
to keep velocity dispersions constant (reasonably enough, given that
these determine the cluster masses):
$$
\boxit{
\sigma_8\simeq (0.5 - 0.6)\; \Omega^{-0.5}.
}
$$
This knowledge of $\sigma_8$ as a measure of the true mass inhomogeneity
is a fundamental cosmological datum. 

\titleb{4.5}{Galaxy formation}
\titlec{}{Cooling and galaxy formation}
This discussion of mass functions really applies
only to `haloes' of
collisionless dark matter.
Things are more complex for
baryonic matter, where we must ask if the 
matter has been able to {\bf dissipate} and turn into stars.
This question was analyzed in a classic paper by Rees \& Ostriker
(1977), and has been reconsidered in the context of CDM by
Blumenthal {\it et al.} (1984).

Mergers will heat gas up to the virial temperature
via shocks; in order for the gas to form stars, it must
be able to undergo {\bf radiative cooling} -- to dispose of this thermal energy.
Clearly, the redshift of collapse clearly needs
to be sufficiently large that there is time 
for an object to cool between its formation
at redshift $z_{\rm cool}$ (when $\delta\rho/\rho \simeq \delta_c$) and the
present epoch.  
We shall show below that $z_{\rm cool}$ is a function of mass;
it is therefore possible to put cooling into the Press-Schechter
machinery simply by using the
{\it mass-dependent} threshold $\nu(M) = \delta_c
[1+z_{\rm cool}(M)]/\sigma_0(M)$ in the mass function.

The cooling function for a plasma in thermal
equilibrium has been calculated by
Raymond, Cox \& Smith (1976).
For an H + He plasma with $Y=0.25$
and some admixture of metals,
their results for the cooling time 
($t_{\rm cool}\equiv 3kT/2\Lambda(T) n$) 
may be approximated as roughly
$$
t_{\rm cool}/{\rm years} = 1.8\times 10^{24} \left({\rho_{B}
  \over M_\odot {\rm Mpc}^{-3}}\right)^{-1}\,
  \left(T_8^{-1/2} + 0.5f_m T_8^{-3/2}\right)^{-1},
$$
where $T_8\equiv T/10^8 K$.
The $T^{-1/2}$ term represents bremsstrahlung
cooling and the $T^{-3/2}$ term approximates
the effects of recombination radiation.
The parameter $f_m$ governs the metal content:
$f_m=1$ for solar abundances; $f_m\simeq 0.03$ for no metals.
In this model where so far dissipation has not been considered,
the baryon density is proportional to the total density, the collapse
of both resulting from purely gravitational processes.
$\rho_{\ss B}$ is then a fraction $\Omega_{\ss B}/\Omega$ of
the virialized total density. This is itself
some multiple $f_c$ of the background density at
virialization (which we refer to as `collapse'):
$$
\rho_c = f_c\, \rho_0\, (1+z_c)^3.
$$
The virialized potential energy for constant density
is $3GM^2/(5r)$, where the radius satisfies
$4\pi\rho_c r^3/3 =M$. This energy must equal $3MkT/(\mu m_p)$,
where $\mu=0.59$ for a plasma with 75\% Hydrogen by mass. Hence,
using $\rho_0 = 2.78\times 10^{11} \Omega h^2\, M_\odot
{\rm Mpc}^{-3}$, we obtain
$$
\boxit{
T_{\rm virial}/K= 10^{5.1} 
(M/10^{12}M_\odot)^{2/3}\, (f_c \Omega h^2)^{1/3}\, (1+z_c).
}
$$
So, for $\Omega=1$, we must solve $f_t\, t_{\rm cool} = 
{2\over 3} H_0^{-1} [1 - (1+z_c)^{-3/2}]$.
If only recombination cooling was important, the solution to this
would be
$$
\eqalign{
& (1+z_c) = (1+M/M_{\rm cool})^{2/3}\cr
& M_{\rm cool}/M_\odot = 10^{13.1}\, f_t^{-1} f_m f_c^{1/2}\Omega_{\ss B} \Omega^{-1/2}\cr
}
$$
For high metallicity, where bremsstrahlung only dominates at
$T\gs 10^8 K$, this equation for $z_c$ will be a reasonable approximation
up to $z_c\simeq 10$, at which point Compton cooling will
start to operate.
Given that we expect at least some enrichment rather
early in the progress of the hierarchy, we shall keep
things simple by using just the above expression for $z_c$.

We see that cooling is rapid for low masses, where the
luminous and dark mass functions are expected to coincide.
Given that cooling of massive
objects is ineffective, probability in the mass
function must therefore accumulate at intermediate masses:
the numbers of faint galaxies relative
to bright are decreased.   If $M_{\rm cool} \ll M_c$, 
then there is a power-law region between these two masses which
differs from the PS slope: 
$M^2f(M) \propto M^{\left({n+3\over 6}\right)+{2\over 3}}$;
{\it i.e.} there is an effective change in $n$ to $n+4$.
This may be relevant for the galaxy luminosity function, where
the faint-end slope is close to constant numbers per magnitude.
For constant mass-to light ratio, this implies $M^2f(M)\propto M$, and
apparently requires $n=3$. Alternatively, a spectral index more in
accord with large-scale structure observations of $n\ls -1$ gives
a Press-Schechter slope much steeper than the observed galaxy
luminosity function. There have been many full and
complicated studies of galaxy formation that show in detail
why this is not really as much of a paradox as
it initially seems, but the above simplified discussion of cooling
illustrates the main way in which the naive analysis goes wrong.

%

\titlea{5}{Testing inflation against galaxy clustering}
\titleb{5.1}{Clustering statistics}
\titlec{}{Fourier analysis of density fluctuations}
It is often convenient to consider building up a general
field by the superposition of many modes.
For a flat comoving geometry, the natural tool for
achieving this is via Fourier analysis.
For other models, plane waves are not a complete
set and one should use instead the eigenfunctions
of the wave equation in a curved space. Normally
this complication is neglected: even in an open
Universe, the difference only matters on scales
of order the present-day horizon.

How do we make a Fourier expansion of an infinite
density field? If the field was periodic
within some box of side $L$, then we would just
have a sum over wave modes:
$$
F({\bf x}) = \sum F_{\bf k} e^{-i{\bf k\cdot x}}.
$$
Now, if we let the box become arbitrarily large, then the
sum will go over to an integral which incorporates the
density of states in $k$-space -- exactly as in
statistical mechanics:
The Fourier relations in $n$ dimensions are thus
$$
\boxit{
\eqalign{
F(x)&=\left({ L \over 2\pi}\right)^n \int F_k(k) \exp -i{\bf
k\cdot x}\; d^n k\cr
F_k(k)&=\left({1\over  L }\right)^n \int F(x) \exp
i{\bf k\cdot x}\; d^n x.\cr }
}
$$
One advantage of this particular Fourier convention
is that the definition of convolution is just a simple
volume average, with no gratuitous factors of $(2\pi)^{-1/2}$:
$$f*g\equiv {1\over L^n}\int f({\bf x-y}) g({\bf y}) d^ny.$$ 
Although one can make all manipulations on density fields
which follow
using either the integral or sum formulations, it is
usually easier to use the sum. This saves having to introduce
$\delta$-functions in $k$-space.
For example, if we have $f=\sum f_k \exp(-ikx)$, the obvious
way to extract $f_k$ is via $f_k=(1/ L )\int f\exp(ikx)\; dx$:
because of the harmonic boundary conditions, all oscillatory
terms in the sum integrate to zero, leaving only $f_k$ to be
integrated from 0 to $ L $. There is less chance of
committing errors of factors of $2\pi$ in this way than
considering $f=( L /2\pi)\int f_k \exp(-ikx)\; dk$ and then
using $\int\exp[i(k-K)x]\; dx=2\pi\delta_{\ss D}(k-K)$.

\titlec{}{Correlation functions and power spectra}
As an immediate example of the Fourier
machinery in action, consider the important quantity
$$
\xi({\bf r}) \equiv \left\langle \delta({\bf x})\delta({\bf x+ r})\right\rangle,
$$
which is the autocorrelation function of the density field --
usually referred to simply as the {\bf correlation function}.
The angle brackets indicate an averaging over the normalization
volume $V$. If we express $\delta$ as a sum, and note that
reality means we can replace one of the two $\delta$'s by
its complex conjugate, then we obtain
$$
\xi = \left\langle \sum_{\bf k}\sum_{\bf k'}\delta_{\bf k}
  \delta^*_{\bf k'} e^{i{\bf (k'-k)\cdot x}} e^{-i{\bf k\cdot r}}\right\rangle.
$$
An alternative way of obtaining this is to use the
relation between modes with opposite wavevectors
which holds for any real field:
$\delta_{\bf k}(-{\bf k})=\delta^*_{\bf k}({\bf k})$.
By the periodic boundary conditions, however, all the cross terms
with $\bf k'\ne k$ average to zero. Expressing the
remaining sum as an integral, we have
$$
\boxit{
\xi({\bf r})=
{V \over (2\pi)^3}\int|\delta_{\bf k}|^2 e^{-i{\bf k\cdot r}} d^3 k.
}
$$
In short, the correlation function is the Fourier transform of
the {\bf power spectrum}. We shall hereafter often use the 
alternative notation $P(k)\equiv |\delta_k|^2$.

Now, in an isotropic universe, the density perturbation
spectrum should contain no preferred direction,
and so we must have an {\bf isotropic power spectrum}:
$|\delta_{\bf k}|^2({\bf k}) = |\delta_k|^2(k)$.
We can therefore perform the angular integral:
introduce spherical polars
with the polar axis along $\bf k$, and use the reality
of $\xi$ so that $e^{-i{\bf k\cdot x}}\rightarrow
\cos(kr\cos\theta)$.
In three dimensions, this yields
$$
\xi(r) = {V\over (2\pi)^3}\int |\delta_k|^2\, {\sin kr\over kr}\, 4\pi k^2\; dk.
$$
The 2D analogue of this formula is
$$
\xi(r) = {A\over (2\pi)^2}\int |\delta_k|^2\, J_0(kr)\, 2\pi k\; dk.
$$

We shall usually express
the power spectrum in dimensionless form, as the variance per $\ln k$
($\Delta^2 =d\sigma^2/d\ln k \propto k^3 P[k]$):
$$
\boxit{
\Delta^2(k)\equiv {V\over (2\pi)^3} \, 4\pi k^3\, P(k)
={2\over \pi}k^3\int_0^\infty\xi(r)\,
{\sin kr\over kr}\, r^2\, dr.
}
$$
This gives a more easily visualisable meaning to the power
spectrum than does the quantity $V P(k)$, which has
dimensions of volume: $\Delta^2(k)=1$ means that there
are order unity density fluctuations from modes
in the logarithmic bin around wavenumber $k$.

\titlec{}{Power-law spectra}
The above shows that the power spectrum is a vital quantity
in cosmology, but how can we predict its functional form?
For decades, this was thought to be impossible, and so
a minimal set of assumptions was investigated.
In the absence of a physical theory, we should not assume
that the spectrum contains any preferred length scale,
otherwise we should then be compelled to explain it.
This means that the spectrum must be a featureless
power law:
$$
\boxit{
|\delta_k|^2 \propto k^n
}
$$
The index $n$ governs the balance between large-
and small-scale power.
The meaning of different values of $n$ can be seen by
imagining the results of filtering the density
field by passing over it a box of some characteristic
size, $x$, and averaging the density over the box. 
This will filter out waves with $k\gs 1/x$,
leaving a variance $\langle\delta^2\rangle \propto \int_0^{1/x}
k^n 4\pi k^2dk\propto x^{-(n+3)}$. Hence, in terms of a mass
$M\propto x^3$, we have
$$
\delta_{\rm rms} \propto M^{-(n+3)/6}.
$$

Similarly, a power-law spectrum implies a power-law
correlation function.
If $\xi(r)=(r/r_0)^{-\gamma}$,
the corresponding 3D power spectrum is
$$
\Delta^2(k)={2\over\pi}\,(kr_0)^{\gamma}\, \Gamma(2-\gamma) \,
   \sin {(2-\gamma)\pi\over 2}
\equiv \beta (kr_0)^\gamma
$$
($=0.903 (kr_0)^{1.8}$ if $\gamma=1.8$).
This expression is only valid for $n<0$ ($\gamma<3$);
for larger values of $n$, $\xi$ must become
negative at large $r$ (because $P(0)$ must vanish,
implying $\int_0^\infty \xi(r)\, r^2\, dr=0$).
A cutoff in the spectrum at large $k$ is needed
to obtain physically sensible results.

What general constraints can we set on the value of $n$?
Asymptotic homogeneity clearly requires $n>-3$.
An upper limit of $n<4$ comes from an argument due to
Zeldovich.
Suppose we begin with a totally
uniform matter distribution and then group it
into discrete chunks as uniformly as possible.
It can be shown that conservation of momentum
in this process means that we cannot create
a power spectrum which goes to zero at small
wavelengths more rapidly than $\delta_k\propto k^2$.
Thus, discreteness of matter produces
the {\bf minimal spectrum}: $n=4$.

More plausible alternatives lie between these extremes.
The value $n=0$ corresponds to {\bf white noise}:
the same power at all wavelengths.
This is also known as the {\bf Poissonian}
power spectrum, because it corresponds to fluctuations
between different cells which scale as $1/\sqrt{M_{\rm cell}}$
(see below). A density field created by throwing
down a large number of point masses at random would
therefore consist of white noise.
Particles placed at random within cells, one per cell,
create an $n=2$ spectrum on large scales.
Practical spectra in cosmology, conversely, often
have negative effective values of $n$ over a large
range of wavenumber. In this sense, large-scale
structure is much more `real' than the simple white-noise
fluctuations familiar in other contexts.

\titlec{}{The Zeldovich spectrum}
Most important of all is the {\bf scale-invariant}
spectrum, which corresponds to the value $n=1$, {\it i.e.}
$\Delta^2\propto k^4$. To see where the name arises,
consider perturbations in gravitational potential:
$$
\nabla^2\delta\Phi= 4\pi G\rho_0\delta
\Rightarrow \delta\Phi_k = -4\pi G\rho_0\delta_k/k^2.
$$
The two powers of $k$ pulled down by $\nabla^2$ mean
that, if $\Delta^2\propto k^4$ for matter, then
$\delta^2_\Phi$ is a constant.
Since potential perturbations govern the flatness
of spacetime, this says that the scale-invariant
spectrum corresponds to a metric which is
fractal-like: it has the same degree of
`wrinkliness' on each resolution scale.
The total curvature fluctuations diverge, but only
logarithmically at either extreme of wavelength.

Another way of looking at this spectrum is in terms
of perturbation growth balancing the
scale-dependence of $\delta$: $\delta\propto x^{-(n+3)/2}$.
We know that $\delta$ viewed on a given
comoving scale will increase with 
the size of the horizon:
$\delta\propto r_{\ss H}^2$. 
At an arbitrary time, though, the only natural length
provided by the Universe (in the absence of non-gravitational
effects) is the horizon itself:
$$
\delta(r_{\ss H})\propto r_{\ss H}^{-(n-1)/2}.
$$
Thus, if $n=1$, the growth of $r_{\ss H}$ and $\delta$ with time
cancel out so that the universe always looks the same
when viewed on the scale of the horizon; such a universe
is a fractal in the sense of always appearing the same
under the magnification of cosmological expansion.
This spectrum is often known as the {\bf Zeldovich}
spectrum (sometimes hyphenated with Harrison and
Peebles, who also invented it independently).

\titlec{}{Filtering and moments}
A common concept in the manipulation of cosmological
density fields is that of {\bf filtering}: convolution of
the density field with some {\bf window function}: $\delta\rightarrow\delta * f$.
Many observable results can be expressed in this form.
Some common 3D filter functions are
$$
\eqalign{
&{\rm Gaussian}: f={V\over (2\pi)^{3/2} R_{\ss G}^3}e^ {-r^2/2R_{\ss G}^2}\Rightarrow
  f_k=e^{-k^2R_{\ss G}^2/2}\cr
&{\rm Top-hat}: f={3V\over 4\pi R_{\ss T}^3}\quad (r<R_{\ss T})\Rightarrow
  f_k={3\over y^3}[\sin y - y\cos y]\quad (y\equiv k R_{\ss T})\cr
}
$$
Note the factor of $V$ in the definition of $f$; this is needed
to cancel the $1/V$ in the definition of convolution.
For some power spectra, the difference in these
filter functions at large $k$ is unimportant, and we
can relate them by equating the expansions near $k=0$,
where $1-|f_k|^2\propto k^2$. This equality requires
$$
R_{\ss T}=\sqrt{5}\;R_{\ss G}.
$$

We are often interested not in the convolved field itself,
but in its variance, for use as a statistic ({\it e.g.} to
measure the rms fluctuations in the number of
objects in a cell). By the convolution
theorem, this means we are interested in a {\bf moment} of the
power spectrum times the squared filter transform.
We shall generally use the following notation:
$$
\boxit{
\sigma_n^2\equiv {V\over (2\pi)^3}\int P(k)\, |f_k|^2\;k^{2n}\; d^3 k;
}
$$
the filtered variance is thus $\sigma_0^2$ (which we shall often
denote by just $\sigma^2$).
Moments may also be expressed in terms of the
correlation function over the sample volume:
$$
\sigma^2=\int\!\!\int\xi(|{\bf x-x'}|)\; f({\bf x})\,f({\bf x'})
\; d^3x\; d^3x'.
$$
To prove this, it is easiest to start from the
definition of $\sigma^2$ as an integral over the
power spectrum times $|f_k|^2$, write out the
Fourier representations of $P$ and $f_k$,
and use $\int \exp i{\bf k\cdot(x-x'+r)}\; d^3k=(2\pi)^3\delta_{\ss D}^
{(3)}({\bf x-x'+r})$.
Finally, it is also sometimes convenient to express things in terms of
derivatives of the correlation function at zero
lag. Odd derivatives vanish, but even derivatives give
$$
\xi^{(2n)}(0) = (-1)^n\; {\sigma_n^2\over 2n+1}.
$$

\titlec{}{Normalization}
For scale-invariant spectra, a
natural amplitude measure is the (constant)
gravitational potential variance per unit $\ln k$:
$$
\epsilon^2\equiv{V\over (2\pi)^3}\,4\pi k^3
|\Phi_k|^2/c^4 ={9\over 4}
\left({c k\over H_0}\right)^{-4}\, \Delta^2(k).
$$

Two more commonly encountered measures relate to the
clustering field around 10 Mpc. One is $\sigma_8$:
the rms density variation when smoothed in spheres
of radius $8h^{-1}$ Mpc; this is observed to be
very close to unity.
The other is an integral over the correlation function:
$$
J_3\equiv\int_0^r \xi(y)\; y^2 dy = \int \Delta^2(k) W(k)\; {dk\over k},
$$
where $W(k)=(\sin kr - kr\cos kr)/k^3$.
The canonical value of this is $J_3(10h^{-1}{\rm Mpc})=277h^{-3}$ Mpc
(from the CfA survey: Davis \& Peebles 1983). 
It is sometimes more usual to use instead the dimensionless
volume-averaged correlation function $\bar \xi$:
$$
\bar\xi(r)={3\over 4\pi r^3}\int_0^r\xi(x)\;4\pi x^2\;dx =
 {3\over r^3} J_3(r).
$$
The canonical value then becomes $\bar\xi(10h^{-1}{\rm Mpc})=0.83$;
this measure is clearly very close in content to $\sigma_8=1$.

A point to beware of is that the normalization of a theory
is often quoted in terms of a value of these parameters
extrapolated according to {\it linear\/} time evolution. Since
the observed values are clearly non-linear, there is
no reason why these two definitions should match exactly.
Even more confusingly, it is quite common in the
literature to find the linear value of $\sigma_8$
called $1/b$, where $b$ is a {\bf bias parameter}.
The implication is that $b\ne 1$ means that light
does not follow mass; this may well be true in reality,
but with this definition, nonlinearities will
produce $b\ne 1$ even in models where mass traces light
exactly. Use of this convention is not recommended.

\titleb{5.2}{Non-linear evolution}
How is the power spectrum altered
by nonlinear evolution? As with most nonlinear
questions, this cannot be answered analytically in
general, but some aspects of the scaling of the problem
are well understood.

A useful trick is to think of
the density field under full nonlinear evolution
as consisting of a set of collapsed,
virialized clusters. What is the density profile of
one of these objects? At least at separations smaller
than the clump separation, the density profile of the
clusters will be of the same form as the correlation
function, since this just measures the number density
of neighbours to a given galaxy. Thus, a power-law
correlation $\xi(r)\propto r^{-\gamma}$ may be thought
of as arising from the $\rho\propto r^{-\gamma}$ haloes
of clumps in the density field.

In this picture, it is easy to see how $\xi$ will evolve
with redshift, since clusters are virialized
objects which do not change as the universe expands. We 
have {\bf stable clustering}: $\xi$ is fixed in {\it proper\/}
terms apart from a $(1+z)^{-3}$ scaling owing to the
changing mean density of unclustered galaxies which
dilute the clustering at high redshift.
Thus, with $\xi\propto r^{-\gamma}$, we obtain the
comoving evolution 
$$
\xi(r,z)\propto (1+z)^{\gamma-3}\quad {\rm(non-linear)}.
$$
Since the observed $\gamma\simeq 1.8$, this implies slower
evolution than is expected in the linear regime:
$$
\xi(r,z)\propto (1+z)^{-2}\quad {\rm(linear)}.
$$
This argument does not so far give a relation between
the non-linear slope $\gamma$ and the index $n$ of the
linear spectrum.
However, these two rates of evolution must match to give the
same prediction for the evolution of $r_0$ --
the length-scale of nonlinearity -- since this is where
$\xi$ will break from the linear $\xi\propto r^{-(n+3)}$ to
the nonlinear $\xi \propto r^{-\gamma}$. The linear and
non-linear predictions for the evolution of $r_0$ are
respectively
$r_0\propto (1+z)^{-2/(n+3)}$ and
$r_0\propto (1+z)^{-(3-\gamma)/\gamma}$, so that
$\gamma=(3n+9)/(n+5)$, or in terms of an effective index
$\gamma=3+n_{\rm eff}$:
$$
\boxit{
n_{\rm eff}=-{6\over 5+n}.
}
$$
The power spectrum resulting from power-law
initial conditions will evolve self-similarly
with this index. Note the narrow range
predicted:
$-2<n_{\rm eff}<-1$ for $-2<n<+1$,
with an $n=-2$ spectrum having the
same shape in both linear and nonlinear regimes.

Whether this evolution has been seen or not is presently
controversial.
Efstathiou {\it et al.} (1991) have observed a very low
amplitude for the angular clustering of galaxies at
$B\simeq 26$, inferring that (if $\Omega=1$) the
clustering must evolve very rapidly -- at about the
linear-theory rate. However, their fields are very small
and it is possible their small result is not representative.
An indication that clustering may not decline this
rapidly is given by the observed clustering of quasars
at $z\simeq 1$; Shanks \& Boyle (1994) find the
relatively high value $r_0=7h^{-1}$ Mpc.

For many years it was thought that only the limiting
cases of extreme linearity or nonlinearity could be
dealt with analytically, but in a marvelous
piece of alchemy, Hamilton {\it et al.} (1991; HKLM)
gave a universal analytical formula for accomplishing
the linear $\leftrightarrow$ nonlinear mapping.
The conceptual basis of their method can be
understood with reference to the spherical
collapse model. For $\Omega =1$ (the only case
they considered), a spherical clump
virializes at a density contrast of order
100 when the linear contrast is of order unity.
The trick now is to think about the density
contrast in two distinct ways. To make a
connection with the statistics of the density field,
the correlation function $\xi(r)$ may be
taken as giving a typical clump profile.
What matters for collapse is that the integrated
overdensity reaches a critical value, so one should
work with the volume-averaged correlation function
$\bar\xi(r)$. A density contrast of
$1+\delta$ can also be thought of as arising
through collapse by a factor $(1+\delta)^{1/3}$ in radius,
which suggests that a given non-linear correlation
$\bar\xi_{\ss NL}(r_{\ss NL})$ should be thought
of as resulting from linear correlations on a
linear scale 
$$
r_{\ss L} =[1+\bar\xi_{\ss NL}(r_{\ss NL})]^{1/3} r_{\ss NL}.
$$
This is one part of the HKLM procedure.
The second part, having translated scales as above,
is to conjecture that the nonlinear correlations are
a universal function of the linear ones:
$$
\bar\xi_{\ss NL} (r_{\ss NL}) = f_{\ss NL}[\bar\xi_{\ss L}(r_{\ss L})].
$$
The asymptotics of the function can be deduced
readily. For small arguments $x\ll1$, $f_{\ss NL}(x) \simeq x$;
the spherical collapse argument suggests
$f_{\ss NL}(1)\simeq 10^2$.
Following collapse, $\bar\xi_{\ss NL}$ depends on scale
factor as $a^3$ (stable clustering), whereas
$\bar\xi_{\ss L}\propto a^2$; the large-$x$ limit is therefore
$f_{\ss NL}(x)\propto x^{3/2}$. HKLM deduced from numerical
experiments a numerical fit that interpolated
between these two regimes, in a manner that
empirically showed negligible dependence on
power spectrum.

To use this method with power spectra, we can
use the relations
between $\bar\xi(r)$ and $\xi(r)$
$$
\eqalign{
\bar\xi(r) = &  {3\over r^3}\int_0^r \xi(x)\; x^2\;dx \cr
\xi(r)    = &  {d\, [r^3\, \bar\xi(r)]\over d[r^3]}, \cr
}
$$
followed by the Fourier relations between $\xi(r)$
and $\Delta^2(k)$, to obtain
$$
\eqalign{
\bar\xi(r) = & \int_0^\infty\Delta^2(k)\;{dk\over k}\;{3\over (kr)^3}
[\sin kr- kr \cos kr] \cr
\Delta^2(k) = & {2k^3\over 3\pi}\int_0^\infty\bar\xi(r)\, r^2\,dr
\;{1\over (kr)} [\sin kr- kr \cos kr], \cr
}
$$
where the last relation holds provided that $\bar\xi(r)\rightarrow 0$
faster than $r^{-2}$ at large $r$ (i.e. a spectrum
which asymptotically has $n>-1$, a valid assumption
for spectra of practical interest).

However, these equations are often difficult to
use stably for numerical evaluation;
it is better to work directly in terms of
power spectra.
The key idea here is that $\bar\xi(r)$ can often
be thought of as measuring the power at some
effective wavenumber: it is obtained as
an integral of the product of $\Delta^2(k)$,
which is often a rapidly rising function, and
a window function which cuts off rapidly at $k\gs 1/r$:
$$
\eqalign{
\bar\xi(r) &= \Delta^2(k_{\rm eff}) \cr
k_{\rm eff} &\simeq  2/r, \cr
}
$$
where $n$ is the effective power-law index of the power
spectrum. This approximation for
the effective wavenumber is within 20 per cent of the
exact answer over the range $-2<n<0$. 
In most circumstances, it is therefore an excellent
approximation to use the HKLM formulae directly to
scale wavenumbers and powers:
$$
\boxit{
\eqalign{
&  \Delta^2_{\ss NL}(k_{\ss NL}) = f_{\ss NL}
[\Delta^2_{\ss L}(k_{\ss L})] \cr
&  k_{\ss L} = [1+\Delta^2_{\ss NL}(k_{\ss NL})]^{-1/3} k_{\ss NL}. \cr
}
}
$$
Even better, it is not
necessary that the number relating $1/r$ and
$k_{\rm eff}$ be a constant over the whole
spectrum. All that matters is that the number
can be treated as constant over the limited range 
$r_{\ss NL}$ to $r_{\ss L}$. This means that
the deviations of the above formulae from the
exact transformation of the HKLM procedure
are only noticeable in cases where the
power spectrum deviates markedly from a
smooth monotonic function, or where either
the linear or nonlinear
spectra are very flat ($n\ls -2$). 

What about models with $\Omega\ne 1$?
The argument that leads
to the $f_{\ss NL}(x)\propto x^{3/2}$ asymptote in the
nonlinear transformation is just that linear
and nonlinear correlations behave as $a^2$ and $a^3$
respectively following collapse. If collapse
occurs at high redshift, then $\Omega =1$ may be
assumed at that time, and the nonlinear correlations still
obey the $a^3$ scaling to low redshift. All that
has changed is that the linear growth is
suppressed by some $\Omega$-dependent factor $g(\Omega)$.
It then follows that the large-$x$ asymptote of 
the nonlinear function is
$$
f_{\ss NL}(x) \propto [g(\Omega)]^{-3}\,x^{3/2}.
$$
According to Carroll, Press \& Turner (1992), the
required growth-suppression factor
may be approximated almost exactly by
$$
g(\Omega) =\frac{5}{2}\Omega_m\left[\Omega_m^{4/7}-\Omega_v+
(1+\Omega_m/2)(1+\Omega_v/70)\right]^{-1},
$$
where we have distinguished matter ($m$) and vacuum
($v$) contributions to the density parameter explicitly. 

Peacock \& Dodds (1996) suggested the following generalization of
the HKLM method, using the following fitting formula for the
nonlinear function (strictly, the one which applies to the
power spectrum, rather than to $\bar\xi$):
$$
f_{\ss NL}(x) =x \; \left[{ 1+B\beta x +[A x]^{\alpha\beta} \over
1 + ([A x]^\alpha g^3(\Omega)/[V x^{1/2}])^\beta}\right]^{1/\beta}.
$$
$B$ describes a second-order deviation from linear
growth; $A$ and $\alpha$ parameterise the power-law which
dominates the function in the quasilinear regime; $V$ is
the virialization parameter which gives the amplitude of the
$f_{\ss NL}(x) \propto x^{3/2}$ asymptote; $\beta$ softens
the transition between these regimes.

\fig{The generalization of the HKLM function
relating nonlinear power to linear power, for the
cases $n=0$, $-1$ and $-2$, and with $\Omega=1$ and 0.2 (dotted lines).
The nonlinear power increases for lower $\Omega$ and
for more negative $n$, but in a nearly universal way for $n\ge -1$.
The fitting formula is shown for models with zero vacuum energy
only, but what matters in general is
just the $\Omega$-dependent linear growth
suppression factor.}
{7.0}

HKLM's suggestion was that $f_{\ss NL}$ might be 
independent of the form of the linear spectrum,
but Jain, Mo \& White (1995) showed that this is not true, especially when the linear spectrum
is rather flat ($n\ls-1.5$). Peacock \& Dodds (1996)
find that an excellent fit (illustrated in Figure \lastfig) is given by the
following spectrum dependence of the expansion coefficients:
$$
A=0.542\,(1+n/3)^{-0.685}
$$
$$
B=0.097\,(1+n/3)^{-0.224}
$$
$$
\alpha=3.235\,(1+n/3)^{-0.236}
$$
$$
\beta=0.659\,(1+n/3)^{-0.356}
$$
$$
V=11.54\,(1+n/3)^{-0.371}.
$$
The more general case of curved spectra such as CDM can be
dealt with very well by using the tangent spectral index at
each linear wavenumber:
$$
n_{\rm eff}\equiv {d\ln P \over d \ln k}
$$

Note that the cosmological model does not enter anywhere
in these parameters. It is present in the fitting formula
only through the growth factor $g$, which governs the
amplitude of the virialized portion of the spectrum.
This says that all the quasilinear features of the
power spectrum are independent of the cosmological
model, and only know about the overall level of power.
This is not surprising to the extent that quasilinear
evolution is well described by the
Zeldovich approximation, in which the final positions of
particles are obtained by extrapolating their initial
displacements by some universal time-dependent factor.

\titleb{5.3}{Redshift-space effects}
Although a huge amount of astronomical effort has been invested
in galaxy redshift surveys, with the aim of mapping the
three-dimensional galaxy distribution, the results are
not a true 3D picture.  The radial axis of redshift
is modified by the Doppler effects of peculiar velocity:
$1+z\rightarrow (1+z)(1+v/c)$. Since the peculiar velocities
arise from the clustering itself, the apparent clustering
pattern in {\bf redshift space}, where redshift is
assumed to arise from ideal Hubble expansion only,
differs systematically from that in {\bf real space}.

The distortions caused by working in redshift
space are relatively simple to analyse if we assume
we are dealing with a distant region of space
which subtends a small angle, so that radial
distortions can be considered as happening along
one Cartesian axis. In this case, the apparent amplitude of
any linear density disturbance is readily obtained
from the usual linear relation between
displacement and velocity (Kaiser 1987)
$$
\boxit{
\delta_{\rm obs}= \delta\,\left(1+{\Omega^{0.6}\mu^2\over b}\right),
}
$$
where $\mu$ is the cosine of the angle between the wavevector
and the line of sight ($\mu={\bf\hat r\cdot\hat k}$).
The parameter $b$ allows for bias: the set of objects
under study may be more clustered than the mass
($\delta=b\delta_{\rm mass}$). 
The function $f(\Omega)\simeq \Omega^{0.6}$ is the
well-known velocity-suppression factor due
to Peebles, which is in practice a function
of $\Omega_m$ only, with negligible dependence
on the vacuum density (Lahav {\it et al.} 1991).
Redshift-space effects depend on the combination
$$
\beta\equiv \Omega^{0.6}/b.
$$
The anisotropy arises because mass flows 
from low-density regions onto high density sheets,
and the apparent density contrast of the
pattern is thus enhanced in
redshift space if the sheets lie near the plane
of the sky. If we average this anisotropic effect
by integrating over a uniform distribution of $\mu$,
the net boost to the power spectrum is
$$
|\delta_k|^2 \rightarrow b^2\,|\delta_k|^2\;
\left(1+{2\over3}\beta + {1\over 5}\beta^2\right).
$$

On small scales, this is not valid. The main effect
here is to reduce power through the radial smearing
due to virialized motions  and the associated
`finger-of-God' effect. This is hard to treat
exactly because of the small-scale velocity correlations.
A simplified model was introduced by Peacock (1992)
in which the small-scale velocity field is taken
to be an incoherent Gaussian scatter with 1D rms
dispersion $\sigma$. This turns out to be quite a
reasonable approximation, because the observed pairwise
velocity dispersion is a very slow function of separation,
and is all the better if
the redshift data are afflicted by significant
measurement errors (which should be included in $\sigma$).
This model is just a radial convolution,
and so the $k$-space effect is
$$
\delta_k\rightarrow \delta_k\, \exp[-k^2\mu^2\sigma^2/2].
$$
This effect in isolation gives an average isotropic factor of
$$
|\delta_k|^2 \rightarrow |\delta_k|^2\; {\sqrt{\pi}\over2}
\, {\erf(k\sigma)\over k\sigma}
$$
and produces only mild damping (one power of $k$ at
large $k$).
This last feature is true whatever damping function is
assumed, since at large $k$ most surviving signal comes
from $\mu\simeq 0$. An alternative model is to use the
observation that the pairwise distribution of velocity differences
is well-fit by an exponential (probably a superposition of
Gaussians of different widths from different clumps).
For this, the $k$-space damping function is a Lorentzian:
$$
\delta_k\rightarrow \delta_k\, \left[1+k^2\mu^2\sigma^2/2\right]^{-1}.
$$
In either case, note that the damping depends on $\mu$ as does
the Kaiser factor: both
are anisotropic in $k$ space and they interfere
when averaging to get the mean power.
What happens in this case is that the linear Kaiser boost to the power
is lost at large
$k$, where the result is the same as for $\beta=0$
(because the main contribution at large $k$ comes from small $\mu$).

In practice, the relevant value of $\sigma$ to choose is
approximately $1/\sqrt{2}$ times the pairwise dispersion $\sigma_p$
seen in galaxy redshift surveys
(to this should be added in quadrature any errors in measured
velocities).  
At $1h^{-1}$ Mpc separation, the pairwise dispersion is approximately
$$
\sigma_p\simeq 300 - 400\;{\rm kms^{-1}}
$$
(Davis \& Peebles 1983;
Mo, Jing \& B\"orner 1993; Fisher {\it et al.} 1994).
We therefore expect wavenumbers $k\gs 0.3\,h\;{\rm Mpc}^{-1}$
to be seriously affected by redshift-space smearing.

\titlec{}{Measuring the cosmological constant}
There is an alternative potential source of clustering
anisotropy in redshift space, which is geometrical in nature.
We have to turn positions on the sky and redshifts into Cartesian
coordinates using the following quantities
$$
A(z) \equiv 
R_{0}{dr\over dz} =  {c\over H_{0}} {1\over \sqrt{\Omega_v  + \Omega_m (1+z)^{3}}}
$$
and $B(z)\equiv R_0 S_k(r)$. We normally assume the
the Einstein-de Sitter model
$$
R_{0} {dr\over dz}(z) =  {c\over H_{0}} {1\over (1+z)^{3/2}} \equiv A_{0}(z)
$$
$$
R_{0}r(z) = 2 {c\over H_{0}}\left(1- {1\over \sqrt{1+z}}\right) \equiv B_{0}(z),
$$
but if this is not correct, then contours of $\xi(r)$ will be squashed by a factor $F$:
$$
F(z) = 
  {A/A_{0} \over
   B/B_{0}}.
$$
If $\Lambda=0$, this distortion is small, but not for significant vacuum energy:
for $\Omega_m=0.2$, $F\simeq 1.3$ for $z\gs 1$. Detection of this
distortion in {\it e.g.} quasar clustering would be an attractive
means of detecting $\Lambda$. However, this effect interferes with the
dynamical distortions: for a power-law spectrum, the distorted spectrum is
$$
\eqalign{
P^{S}(k,\mu)
\propto k^{n} & \left[1+\mu^{2}\left( {1\over F^{2}}-1\right)\right]^{ {n-4\over 2}} \cr
  & \times \left[1+\mu^{2}\left( {\beta+1\over F^{2}}-1\right)\right]^{2} 
D[k\mu\sigma_{\rm p}],
}
$$
where $D[k\mu\sigma_{\rm p}]$ is the redshift smearing function
(Ballinger, Peacock \& Heavens 1996). To first order in $\mu^{2}$,
$\beta_{\rm eff} \simeq -0.5n(F-1)$.
However, the $\mu^4$ dependencies are different, so this method may
still be attractive with large databases.

\titlec{}{Real-space clustering}
There are a number of methods available which avoid altogether
the need to deal with the complications of redshift space.
These deal with either pure two-dimensional clustering,
as in angular correlations, or the use of projected
correlations in redshift surveys.

An important relation is that between the angular and
spatial power spectra. In outline, this is derived
as follows. The perturbation seen on the sky is
$$
\delta({\bf \hat q})=\int_0^\infty \delta({\bf y})\;
y^2\phi(y)\; dy,
$$
where $\phi(y)$ is the {\bf selection function}, normalized 
such that $\int y^2\phi\, dy=1$, and $y$ is comoving distance.
The form $\phi\propto y^{-1/2}\exp-(y/y^*)^2$ is often
taken as a reasonable approximation to the Schechter function.
A flat Universe ($\Omega=1$) is assumed.
Now write down the Fourier expansion of $\delta$. The
plane waves may be related to spherical harmonics
via the expansion of a plane wave in Spherical Bessel
functions $j_\ell$
$$
e^{ikr\cos\theta}=\sum_0^\infty (2\ell+1)\; i^\ell\; P\ell(\cos\theta)\;
j_\ell(kr),
$$
plus the spherical harmonic addition theorem
$$
P_\ell(\cos\theta)={4\pi\over2\ell+1}\sum_{m=-\ell}^{m=+\ell}
Y_{\ell m}^*({\bf \hat q})
Y_{\ell m}({\bf \hat q'}),
$$
where ${\bf \hat q\cdot\hat q'}=\cos\theta$.
These relations yield the desired result:
$$
\boxit{
\left\langle|a_\ell^m|^2\right\rangle
=4\pi\int\Delta^2(k)\;{dk\over k}\;\left[
\int y^2\phi(y)\,j_\ell(ky)\;dy\right]^2.
}
$$

What is the analogue of this formula for small angles?
Rather than manipulating large-$\ell$ Bessel
functions, it is easier to start again from the correlation
function. By writing as above the overdensity observed at a
particular direction on the sky as a radial integral
over the spatial overdensity, with a weighting of
$y^2\phi(y)$, we see that the angular correlation function is
$$
\langle\delta({\bf\hat q_1})\delta({\bf\hat q_2})\rangle=
\int\!\!\!\int\langle\delta({\bf y_1})\delta({\bf y_2})\rangle
\; y_1^2y_2^2\phi(y_1)\phi(y_2)\;dy_1\;dy_2.
$$
We now change variables to the mean and difference of the
radii, $y\equiv(y_1+y_2)/2$; $x\equiv(y1-y2)/2$. If the depth of the
survey is larger than any correlation length, we only get
signal when $y_1\simeq y_2\simeq y$. If the selection function
is a slowly-varying function, so that the thickness of the
shell being observed is also of order the depth, the integration
range on $x$ may be taken as being infinite.
For small angles, we then obtain {\bf Limber's equation}:
$$
\boxit{
w(\theta)= \int_0^\infty y^4\phi^2\, dy\ \int_{-\infty}^\infty
  \xi(\sqrt{x^2+y^2\theta^2})\; dx.
}
$$
Theory usually supplies a prediction about the linear
density field in the form of the power spectrum, 
and so it is
convenient to recast Limber's equation:
$$
w(\theta)= \int_0^\infty y^4\phi^2\, dy\ \int_0^\infty
  \pi\, \Delta^2(k)\, J_0(ky\theta)\, dk/k^2.
$$
The power-spectrum version of Limber's equation is already in
the form required for relation to the angular
power spectrum ($w=\int \Delta^2_\theta J_0(K\theta) dK/K$), and
so we obtain the direct small-angle relation between
spatial and angular power spectra:
$$
\boxit{
\Delta^2_\theta={\pi\over K}\int\Delta^2(K/y)\; y^5\phi^2(y)\; dy.
}
$$
This is just a convolution in log space, and is considerably
simpler to evaluate and interpret than the $w-\xi$ version
of Limber's equation.

Finally, note that it is easy to make allowance for spatial
curvature in the above discussion. All that is needed is to
replace the $\Omega=1$ volume element $y^2\, dy$ by its
generalised counterpart, $y^2 dy/(1-ky^2)^{1/2}$.

When working with redshift surveys and treating redshift as a radial
coordinate, the presence of peculiar velocities or redshift errors
causes the correlation function to be convolved in the radial direction:
$$
\eqalign{
\xi(r_p,\pi) &= \int_{-\infty}^\infty\xi_{\rm true}
(r_p,r)\; f(\pi-r)\; dr\cr
  &= {r_0^\gamma\over\sqrt{2\pi}\,\sigma_v}
\int_{-\infty}^\infty [r_p^2+(\pi-x)^2]^{-\gamma/2}\;
e^{-x^2/2\sigma_v^2}\; dx,\cr
}
$$
where the latter expression applies for power-law clustering
and a Gaussian dispersion. Treating the convolving function
$f(\Delta r)$ as a zero-mean scatter ignores large-scale
streaming, which becomes more important on larger scales.
However, a Fourier analysis is clearer in this regime.
Looking at the elongation of $\xi(r_p,\pi)$ in the redshift direction allows
the pairwise velocity dispersion to be estimated; it comes out
at a {\it relative\/} dispersion of between 300 \& 400 kms$^{-1}$ for pairs
of $\sim 1$ Mpc separation (Davis \& Peebles 1983; Fisher {\it et al.} 1994).

The effects of peculiar velocities may be evaded by using the
correlation function evaluated explicitly as a 2D function 
of transverse ($r_p$) and radial ($\pi$) separation.
The projection along the redshift axis is then independent
of the velocities
$$
w(r_p)=\int_{-\infty}^\infty\xi(r_p,\pi)\;d\pi =
2\int_{r_p}^\infty\xi(r)\, {r\, dr\over (r^2-r_p^2)^{1/2}},
$$
and has the Abel integral inverse
$$
\xi(r)=-{1\over\pi}\int_r^\infty w'(y)\, {dy\over (y^2-r^2)^{1/2}}.
$$

Improved signal-to-noise in projected correlations can be obtained
in the case of a {\bf sparse-sampled redshift survey} (Kaiser 1986a), where there is
a large catalogue of angular positions from which redshifts are
measured for some fraction (either the brighter members, or a random
subset).
Saunders {\it et al.} (1992) used the angular cross-correlation between
the 1-in-6 QDOT IRAS-galaxy redshift survey and its  parent catalogue
to obtain the statistic
$$
\Xi(r)=2\int_0^\infty \xi[(r^2+x^2)^{1/2}]\; dx
= 2\int_r^\infty \xi(y)\; {y\; dy\over \sqrt{y^2-r^2}}.
$$
At first sight, it is not very attractive to
use this to infer the power spectrum, because the window
function involved is extremely broad:
$$
{1\over r} \Xi(r)= \int \Delta^2(k)\; {dk\over k}\;
\left[ {\pi\over kr}\, J_0(kr) \right].
$$
However, useful results were obtained by Saunders {\it et al.} (1992) from a multi-stage
process where $\Xi(r)$ is first deprojected to obtain $\xi(r)$,
which can then be integrated to yield $\bar\xi(r)$. This has
a compact window function, and can be directly related to the power spectrum.

\titleb{5.4}{Bias}
One of the major advances of cosmology in the 1980s
was the realization that the distribution of galaxies
need not trace the underlying density field.
The main motivation for such a view may be traced
to 1933 and Zwicky's measurement of the dark matter in
the Coma cluster. A series of ever more detailed
studies of cluster masses have confirmed his original
numbers: if the Coma mass-to-light ratio is
Universal, then the density parameter of the Universe
is $\Omega = 0.1$ -- 0.2. Those who argued that
the value $\Omega=1$ was more natural (a greatly
increased camp after the advent of inflation)
were therefore forced to postulate that
the efficiency of galaxy formation was enhanced
in dense environments: {\bf biased galaxy formation}.
This probably remains the strongest argument for the
reality of bias.

A weaker argument surfaced
at around the same time as inflation through the discovery of
large voids in the galaxy distribution.
There was a reluctance to believe that such vast
regions could be truly devoid of matter --
although this was at a time before the discovery
of large-scale velocity fields.
This tendency was given further stimulus
through the work of Davis, Efstathiou, Frenk \& White (1985),
who were the first to calculate $N$-body models
with the CDM spectrum. Since the CDM spectrum
curves slowly between effective indices
of $n=-3$  and $n=1$, the correlation function
clearly steepens with time. There is therefore
a unique epoch when $\xi$ will have the observed
slope of $-1.8$. Davis {\it et al.} identified this
epoch as the present and then noted that, for $\Omega=1$, it
implied a rather low {\it amplitude\/} of fluctuations:
$r_0=1.3h^{-2}$ Mpc. An independent argument for this low
amplitude came from the size of the peculiar velocities
in CDM models: if given an amplitude corresponding to the
$\sigma_8\simeq 1$ seen in the galaxy distribution, the
pairwise dispersion was $\sigma_p\simeq 1000$ -- $1500\kms$,
some 3 -- 4 times the observed value.
What seemed to be required was
a galaxy correlation function which was an amplified
version of that for mass. This was exactly the
phenomenon analysed for Abell clusters by Kaiser (1984), and
thus was born the idea of {\bf high-peak bias}: bright
galaxies form only at the sites of high peaks in the
initial density field. This was developed in some
analytical detail by Bardeen {\it et al.} (1986), and was implemented
in the simulations of Davis {\it et al.}, leading
to the conclusion that the $\Omega=1$ $h=1/2$ CDM model now gave
a good match to observation.

Since the mid-1980s, fashion has moved in the direction of
low-$\Omega$ universes, which removes many of the original 
arguments for bias. However, the lesson of the attempts
to save the $\Omega=1$ universe is that cosmologists have
learned to be wary of assuming that light traces mass.
The assumption is that the galaxy density field is guilty
of bias, until it is shown to be innocent. 

As was shown by Kaiser (1984), the high-peak model produces a linear
amplification of large-wavelength modes.
This is likely to be a general feature of other
models for bias, so it is useful to introduce the
{\bf linear bias parameter}:
$$
\boxit{
\left.{\delta\rho\over\rho}\right|_{\rm galaxies}
 = b\left.{\delta\rho\over\rho}\right|_{\rm mass}.
}
$$
This seems a reasonable assumption when $\delta\rho/\rho\ll 1$.
Galaxy clustering on large scales therefore allows us to determine 
mass fluctuations only if we know the value of $b$.
For example, the normalization in scale-invariant models
may be specified by $\epsilon$ (the rms potential fluctuation
per $\ln k$), but we can only measure the combination 
$\epsilon b$.
However, the coinage of the bias parameter has been debased
by its use in the mildly non-linear regime, where misleading
definitions such as $b=1/\sigma_8$ are to be found.

Even the linear relation cannot be taken for granted, however;
if galaxy formation is not an understood process, then in principle studies of
galaxy clustering may tell us nothing useful about the
statistics of the underlying potential fluctuations against
which we would like to test inflationary theories.
It is possible to construct models ({\it e.g.} Bower {\it et al.} 1993)
in which the large-scale modulation of the galaxy density
is entirely non-gravitational in nature.

\titlec{}{Mechanisms for bias}
Why should the galaxy distribution be biased at all?
In the context of the high-peak model, attempts were
made to argue that the first generation of objects
could propagate disruptive signals, causing
neighbours in low-density regions to be `still-born'.
However, it turned out to be hard to make such
mechanisms operate: the energetics and required scale of the
phenomenon are very large (Rees 1985; Dekel \& Rees 1987).
A more promising idea, known as {\bf natural bias},
was introduced by White {\it et al.} (1987).
This relied on the idea that, in an overdense region,
an object of a given mass will collapse sooner
and thus have a higher density and circular velocity.
Application of a circular-velocity threshold then yields
a bias towards high-density regions.
White {\it et al.} argued that such an effect was to
be expected owing to the {\bf Tully-Fisher effect}:
a tight correlation between luminosity and circular
velocity for spiral galaxies.

However, the problem with this model is that it still
apparently predicts no bias if a strict selection by
mass is performed. What is needed is some way in
which star formation is biased (perhaps by epoch
dependent efficiency) in order to produce more stars
in the galaxies which collapse earlier.
A general discussion of this problem was given by Cole \& Kaiser (1989):
suppose an object collapsing at redshift $z$ generates a
stellar luminosity
$$
L\propto M^\alpha (1+z)^\beta.
$$
Cole \& Kaiser show that a perfect Tully-Fisher relation
then requires $\beta=3\alpha/2$.
This is easily proved from the usual expression for the
virial velocity dispersion resulting from
gravitational collapse: $V\propto M^{1/3}(1+z_c)^{1/2}$.
The above condition removes any redshift dependence
and leaves $V\propto L^{1/3\alpha}$.
The conventional Tully-Fisher slope of $1/4$ then implies
$\alpha=4/3$, $\beta=2$. The natural bias mechanism
implicitly depends on a strong epoch dependence of
star-forming efficiency.
In fact, Cole \& Kaiser argue that
a somewhat stronger epoch dependence ($\beta\gs 3$) is required to
achieve sufficient bias to understand cluster mass-to-light
ratios. However, Peacock (1990) showed that such a high
value would predict a large scatter in the Faber-Jackson
relation between luminosity and velocity dispersion for ellipticals.
The data are much more closely consistent with $\beta\simeq 1$.

In short, there appears to be little evidence for
traditional bias schemes where one tinkers with the
efficiency of star formation. There remains the
alternative that galaxies were born unbiased but
subsequently migrated into clusters more quickly
than dark matter. These dynamical schemes are
currently attracting the most attention
(West \& Richstone 1988; Carlberg {\it et al.} 1990;
Carlberg 1991; Couchman \& Carlberg 1992).
The idea of bias being a phenomenon largely
confined to clusters fits well with the emerging
picture of large-scale structure: away from 
clusters, it seems that all types of galaxies
follow the same overall `skeleton' of large-scale structure, independent
of Hubble type (Thuan {\it et al.} 1987;  
Babul \& Postman 1990; Mo, McGaugh \& Bothun 1994). 
Even luminosity segregation is a very weak and controversial
effect (Valls-Gabaud {\it et al.} 1989; Loveday {\it et al.} 1995).
This is further evidence against earlier pictures in which
the voids were filled with mass, but failed to produce bright galaxies.
Increasingly, it seems that
the voids really have been largely emptied by gravity, as implied
by the large-amplitude peculiar velocities on these scales.
It would not be surprising if the formation of all classes of
galaxy was suppressed in these regions of very low density,
but it is implausible that the voids contain more than a
small fraction ($\sim 10$\%) of the total mass in the universe,
so the implied degree of bias would not be large.

\titlec{}{Outstanding issues}
To sum up the present position, the pictures of the galaxy distribution
obtained with  different sets of tracer objects have certain features in
common (mainly the behaviour in low-density
regions: the existence of filaments and voids),
but diverge greatly in regions of high density.
This {\bf morphological segregation} is a long-established
phenomenon ({\it e.g.} Dressler 1980); it is known that
E/S0 galaxies increase from perhaps 20\% of the
galaxy population in the field to almost 100\% in the
cores of rich clusters where the overdensities are $\sim 10^3$.
This is closely related to the relative proportions
of optically-selected and IRAS galaxies as a function
of density. At high overdensities, the fraction of optical galaxies
which are IRAS galaxies declines by a factor $\simeq 3$ -- 5
from the mean
(Strauss {\it et al.} 1992), reflecting the fact that IRAS
galaxies are mainly spirals.

We need to be able to decide on physical
grounds which tracer is more likely to follow
the mass in regions of high density, and
in some cases this is quite easy.
For example, it is well understood how and why
the spatial correlations of rich clusters will
be an amplified version of the underlying
density correlations (at least for Gaussian
statistics). Given this, it is a safe bet
that the clustering properties of elliptical
galaxies also over-estimate the density correlations;
the known phenomenon of morphological segregation
means that ellipticals are closely associated
with clusters. We know in both the cases of
clusters  and ellipticals that we are excluding
the low-density universe simply through our
observational selection, so an unrepresentative
answer would be expected. 

Much harder is the
critical decision between optically-selected
galaxies or IRAS galaxies. The former give dynamical
determinations consistently in the region of
$\Omega=0.2$ if optical light traces mass, whereas
the latter favour $\Omega=1$, because
IRAS galaxies peak up
less strongly than optically-selected galaxies in
high-density regions, as discussed above.
So, which (if either) of these fields follows
the mass? The problem for the committed believer in
$\Omega=1$ is that it is much more plausible that
it is the IRAS result which is corrupted. Morphological
segregation is thought to arise because spiral
discs have trouble surviving in high-density
environments, and it is plausible that the
X-ray emitting gas in clusters is all that remains
of material which might have made spiral discs in
a less troubled habitat. We conclude that either
$\Omega\simeq0.2$, or that some other mechanism has operated
to boost the optical light in clusters, which is then
fortuitously cancelled by the suppression of IRAS
emission from these regions. 
This interpretation is made still more contrived by
an accounting of the total baryonic material in clusters.
Within the central $\simeq 1$ Mpc, the masses in stars,
X-ray emitting gas and total dark matter can be determined
with reasonable accuracy (perhaps 20\% rms), and allow a
minimum baryon fraction to be determined:
$$
\boxit{
{ M_{\rm Baryons} \over M_{\rm Total} } \gs
0.009 +0.050\,h^{-3/2}
}
$$
(White {\it et al.} 1993). This equation is often referred to
as the {\bf baryon catastrophe}, for the following reasons.
Assume for now that the baryon fraction in  clusters is representative
of the whole universe, and adopt
the primordial nucleosynthesis prediction of $\Omega_{\ss B}h^2=0.0125$.
This gives an equation for $\Omega$:
$$
\Omega \ls 0.25 \, [h^{1/2}+0.18 h^2]^{-1},
$$
which is a limit varying between 0.21 and 0.33 for $h$
between 0.5 and 1. This is a catastrophe for the
Einstein-de Sitter universe, in that clusters have to
be biased not only in the light they emit, but also in
the sense of containing a larger baryon fraction than the average.
However, producing a large-scale separation of dark matter
and baryons on scales which are little past the
turn-round phase is very difficult. If the density
parameter is really unity, it appears that the nucleosynthesis
density must be too low by at least a factor 3.

Since there continue to be compelling reasons to
expect $\Omega=1$, a high priority
in current cosmological research will continue to be to
produce a convincing mechanism for bias,
and to detect its traces in galaxy properties.
However, this should not blind us to the fact that
the simplest interpretation of the existing evidence is
that the Universe has $\Omega<1$..

\titleb{5.5}{Power-spectrum data}
The history of attempts to quantify galaxy clustering 
goes back to Hubble's proof that the distribution of galaxies
on the sky was non-uniform.
The major post-war landmarks were the angular analysis of
the Lick catalogue, described in Peebles (1980), and the
analysis of the CfA redshift survey (Davis \& Peebles 1983).
It has taken some time to obtain data on samples which
greatly exceed these in depth, but several pieces of
work appeared around the start of the 1990s which
clarified many of the discrepancies between different surveys, and which paint a
relatively consistent picture of large-scale structure
(see Peacock \& Dodds 1994).

Clustering results are often published in the
form of the variance ($\sigma^2$) of $\delta$ as a function of scale --
using either cubical cells of side $\ell$
(Efstathiou {\it et al.} 1990b) 
and Gaussian spheres of radius $R_{\ss G}$
(Saunders {\it et al.} 1991).
For a power-law spectrum ($\Delta^2\propto k^{n+3}$), we have
for the Gaussian sphere
$$
\sigma^2=\Delta^2\left(k=\left[ \half \left({\textstyle 
{\scriptstyle n+1\over\scriptstyle 2}}
\right) ! \right]^{1/(n+3)}
 / R_{\ss G} \right).
$$
For $n\ls 0$, this formula also gives a good approximation to
the case of cubical cells, with $R_{\ss G}\rightarrow \ell/\sqrt{12}$.
The result is rather insensitive to assumptions
about the power spectrum, and just says that the variance
in a cell is mainly probing waves with $\lambda\simeq 2\ell$.
Since we know the shape of $\Delta^2$ reasonably well,
we can get very accurate effective wavenumbers and
plot the $\sigma^2$ values on the $\Delta^2$ -- $k$
plane directly using these $k_{\rm eff}$ values.
Such a compilation of results is shown in Figure \nextfig.

\fig{(a) The raw power spectrum data in the form 
$\Delta^2\equiv d\sigma^2/d\ln k$; all
data with the exception of the APM power
spectrum are in redshift space. The two
lines shown for reference are the transforms of the
canonical real-space correlation functions
for optical and IRAS galaxies ($r_0=5$ and $3.78
\;h^{-1}\,\rm Mpc$ and slopes of 1.8 and 1.57
respectively).
(b) The power-spectrum data, individually
linearized assuming $\Omega=b_{\ss I}=1$. There is
an excellent degree of agreement, particularly
in the detection of a break around $k=0.03h$.
}{7.0}

There is a wide range of power measured, ranging
over perhaps a factor 20 between the real-space
APM galaxies and the rich Abell clusters.
Are these measurements all consistent with 
one Gaussian power spectrum for mass fluctuations?
The corrections for redshift-space distortions and
nonlinearities can be applied to these data to
reconstruct the linear mass fluctuations, subject to
an unknown degree of bias.
The reconstruction analysis has available eight
datasets containing 91 distinct $k-\Delta^2$ pairs.
The modelling has available five free parameters
in the form of $\Omega$ and the four bias parameters
for Abell clusters, radio galaxies, optical
galaxies and IRAS galaxies ($b_{\ss A}$, $b_{\ss R}$, 
$b_{\ss O}$, $b_{\ss I}$); however,
only two of these really matter:
$\Omega$ and a measure of the overall level of
fluctuations. For now, we take the IRAS bias parameter to
play this latter role. Once these two are specified,
the other bias parameters are well determined --
principally from the linear data at small $k$,
and have the ratios
$$
b_{\ss A}:b_{\ss R}:b_{\ss O}:b_{\ss I} =
4.5:1.9:1.3:1, 
$$
to within 6 per cent rms.

The various reconstructions of the linear power spectrum 
for the case $\Omega=b_{\ss I}=1$ 
are shown superimposed in Figure \lastfig, and display
an impressive degree of agreement.
This argues very strongly that what we measure
with galaxy clustering has a direct relation to
mass fluctuations, rather than the large-scale
clustering pattern being an optical illusion
caused by non-uniform galaxy-formation
efficiency (Bower {\it et al.} 1993). If this were
the case, the shape of spectrum inferred from
clusters should have a very different shape at
large scales, contrary to observation.

\fig{The linearized power-spectrum data of
Figure \llastfig, averaged over
bins of width 0.1 in $\log_{10}k$,
compared to various CDM models.
These assume scale-invariant initial conditions,
with the same large-wavelength normalization.
Different values of the fitting
parameter $\Omega h=0.5$, 0.45, \dots 0.25, 0.2
are shown. The best-fit model has
$\Omega h=0.25$ and a normalization 
of $\sigma_8({\rm IRAS)}=0.75$.}
{7.0}

\titlec{}{Large-scale power-spectrum data and models}
It is interesting to ask if the power spectrum
contains any features, or whether it is consistent
with a single smooth curve. A convenient description is in terms
of the CDM power spectrum, which is $\Delta^2(k)\propto k^{n+3}T_k^2$.
We shall use the BBKS approximation for the transfer
function:
$$
\eqalign{
T_k= &  {\ln(1+2.34q)\over 2.34q} \;\times \cr
     &  [1+3.89q+(16.1q)^2 +(5.46q)^3 +(6.71q)^4]^{-1/4},\cr
}
$$
where $q\equiv k/[\Omega h^2\; {\rm Mpc}^{-1}]$. Since observable
wavenumbers are in units of $h\,{\rm Mpc}^{-1}$, the shape parameter
is the apparent value of $\Omega h$. 
This scaling applies for models with zero baryon content,
but there is an empirical scaling (Sugiyama 1995) that can account for this:
$$
T_k(k)=T_{\ss BBKS}(k/[\Omega h^2\,\exp -(\Omega_{\ss B}+\Omega_{\ss B}/\Omega)]).
$$
The symbol $\Gamma^*$ is used 
to refer to $\Omega h$ in the BBKS spectrum as an empirical
fitting parameter, on the understanding that it would
mean the above combination if CDM models were taken literally.
Fitting this spectrum to the large-scale linearised data 
of Figure \lastfig\ requires the parameters
$$
\Gamma^* \simeq   0.25 + 0.3({1/n}-1),
$$
$$
\sigma_8({\rm IRAS}) \simeq 0.75,
$$
in agreement with many previous arguments suggesting
that an apparently low-density model is needed.
For any reasonable values of $h$ and baryon density,
a high-density CDM model is not viable. Even a high degree
of `tilt' in the primordial spectrum (Cen {\it et al.} 1992)
does not help change this conclusion unless $n$ is set so low that
major difficulties result when attempting to account for
microwave-background anisotropies.

The fit of this model is illustrated in Figure \lastfig,
which makes it clear that the problem with
CDM is the {\it shape\/} of the power spectrum, rather
than the absolute amount of power at large scales. The
linear transfer function does not bend sharply enough
at the break wavenumber if a high-density $\Gamma^*=0.5$ model is adopted. 
An important general lesson can also be drawn from the
lack of large-amplitude features in the power spectrum.
This is a strong indication that collisionless
matter is deeply implicated in forming large-scale
structure. Purely baryonic models contain
large bumps in the power spectrum around
the Jeans' length prior to recombination
($k\sim 0.03\Omega h^2\;\rm Mpc^{-1}$), whether
the initial conditions are isocurvature or
adiabatic ({\it e.g.} Section 25 of Peebles 1993).
It is hard to see how such features can
be reconciled with the data, beyond a `visibility' of
perhaps 20 -- 30\%.

\titlec{}{Small-scale clustering data}
It should clearly be possible to reach stronger
conclusions by using the data at larger $k$.
However, here the assumption of linear bias is clearly very weak,
and what is needed is a model for the scale dependence of the
bias. Mann, Peacock \& Heavens (1996) argue that an empirical
approach can be taken here, even lacking the physics of bias.
We know that, if $\Omega=1$, the density of light in high-density
regions must receive an additional enhancement. There is also
the possibility that galaxy formation may be suppressed in voids.
General arguments (Coles 1993) indicate that such {\bf local bias}
should produce a systematic steepening of the correlation function,
so that the effective bias is larger on small scales.
Experiments with such local density-field modifications on numerical datasets suggests
that the effect on the power spectrum is of a steepening which
can be roughly approximated by
$$
1+\Delta^2(k) \rightarrow [1+b_1\Delta^2]^{b_2},
$$
where $[b_1 b_2]^{1/2}$ would be the bias parameter
in the linear regime. 
Such an expression can fit bias schemes from high-peak bias as
in Davis {\it et al.} (1985) or the `physical bias' seen in
full hydrodynamical simulations ({\it e.g.} Cen \& Ostriker 1992).
Generally the steepening is not so extreme, and it is hard
to find models where $b_2$ exceeds $b_1$.

Irrespective of {\it a priori} considerations, such an expression
accounts well for the difference in real-space clustering of
APM and IRAS galaxies, and can map one onto the other with almost
uncanny precision (see Figure \nextfig), with a relative bias
of $b_1=1.2$, $b_2=1.1$ or $b_{\ss APM}/b_{\ss IRAS}\simeq 1.15$.
Of particular interest is the inflection in the spectrum around
$k\simeq 0.1 \hompc$, which seems likely to be real, since it
is seen in two rather different datasets, which probe different
regions of space.

\fig{(a) The real-space power spectra of APM and IRAS
galaxies, as deduced by Baugh \& Efstathiou (1993; 1994) and
Saunders {\it et al.} 1992. The APM data have been boosted by
a factor 1.2 because clustering evolution
was not allowed for in the Baugh \& Efstathiou data.
(b) The same with a two-parameter scale-dependent boost to the IRAS data.
The agreement is outstanding, except for the three largest-scale IRAS points,
and these can be seen to be too high from a comparison with the redshift-space
results.}
{7.0}

Attempting to fit the small-scale clustering data now
complicates the picture from large scales, since the nonlinear
extrapolation of the $\Gamma^*=0.25$ model is not consistent
with the small-scale clustering. Figure \nextfig\ compares three
different models with the data: Einstein-de Sitter and open and
flat $\Omega=0.3$ models. There is a tension between
the data and all of these models, in that it seems impossible
to fit both large scales and small scales simultaneously.
Without bias, the correct amplitude of small-scale clustering
requires $\sigma_8\simeq 0.7$; this greatly under-predicts
the clustering at $k\simeq 0.1 \hompc$ unless $\Gamma^*\ls 0.1$
is adopted, in which case the power at  $k\simeq 0.02 \hompc$
is greatly exceeded. The $\Omega=1$ model needs a lower
normalization, and so does not exceed the
small-scale data, but it suffers from related problems of shape.
Any hypothetical bias which would scale the mass spectrum to
that of light would need to be non-monotonic, with a smaller
effect at $k\simeq 2 \hompc$ than on larger scales.

\fig{The clustering data for optical galaxies, compared to three
models with $\Gamma^*=0.25$: (a) $\Omega=1$, $\sigma_8=0.5$;
(b) $\Omega_m=0.3$, $\Omega_v=0$ $\sigma_8=1$.
(c) $\Omega_m=0.3$, $\Omega_v=0.7$ $\sigma_8=1$.
Linear spectra are shown dotted; evolved nonlinear
spectra are solid lines. All of these models are chosen
with a normalization which is approximately correct for
the rich-cluster abundance and large-scale peculiar
velocities. In all cases, the shape of the spectrum is wrong.
The high-density model would require a bias which is not a monotonic
function of scale, whereas the low-density models exceed
the observed small-scale clustering.
}{7.0}

What then is required of a linear power spectrum that would
fit the data? None of the CDM-like alternatives considered above
explain the inflection at $k\simeq 0.1 \hompc$, and it is
unlikely to be produced by any bias scheme, since these
always tend to give a smooth scale dependence. The general
conclusion is therefore that there must be a relatively
sharp break in the linear spectrum around this point.

A further general point which emerges from this plot is that it is
something of a puzzle that the clustering data continue as an
unbroken $n\simeq -1$ power law up to $\Delta^2\sim 10^3$.
As the regime of virialized clustering is reached, a break to a
flatter slope would be expected; only the open models fail
to show this feature, which is a robust prediction for any
smooth linear spectrum. What is needed is a small linear
growth-suppression factor $g(\Omega)$, with $\Omega_m\ls 0.5$
for open models or $\ls 0.1$ for flat models
This is far from being a conclusive argument
for open models, but it does require a coincidence from the
bias mechanism: that the galaxy correlations should be steepened just
where the mass correlations are saturating.

We can implement these ideas with a simple empirical model,
which works extremely well.
Consider a spectrum in the form of a break between two power laws:
$$
\Delta^2(k)={(k/k_0)^\alpha\over 1+(k/k_1)^{\alpha-\beta}}.
$$
As shown in Figure \nextfig,  this matches the data very nicely,
if we choose the parameters
$$
\eqalign{
k_0 &  = 0.3\; h\, {\rm Mpc}^{-1} \cr
k_1 &  = 0.05\; h\, {\rm Mpc}^{-1} \cr
\alpha &  = 0.8 \cr
\beta &  = 4.0. \cr
}
$$
A value of $\beta=4$ corresponds to a scale-invariant
spectrum at large wavelengths, whereas the effective
small-scale index is $n=-2.2$. 
The linear spectrum is not required to be non-zero for $k\gs 1 \hompc$,
and so a variety of other possibilities would be made to work,
including those with short-wavelength cutoffs.

\fig{An empirical double power-law model for the power spectrum
provides an extemely good fit to the optical-galaxy power
spectrum but requires an open universe and a sharp
break in the spectrum to a rather flat ($n<-2$) high-$k$ behaviour.
}
{7.0}

\titlec{}{CMB anisotropies}
A consistent model must match the normalization of the mass 
fluctuations on large scales inferred from fluctuations in
the Cosmic Microwave Background. In making this comparison,
it is important to be clear that the CMB fluctuations
depend only on the very large-scale $P\propto k^n$ portion
of the spectrum. Predictions of smaller-scale
fluctuations such as the amplitude $\sigma_8$ then
require additional information in the form of
the shape parameter $\Gamma^*$. 
Rather than quoting the $\sigma_8$ implied by the CMB, it
is therefore clearer to give the large-scale normalization
separately, with $\sigma_8$ then depending on the
choice of $\Gamma^*$.

Bunn, Scott \& White (1995) and
White \& Bunn (1995) discuss the large-scale normalization from
the 2-year COBE data in the context of CDM-like models.
The final 4-year COBE data favour slightly lower results, and
we scale to these in what follows.
For scale-invariant spectra and $\Omega=1$, the best normalization is
$$
\Delta^2(k)=(k/0.0737\, h\, {\rm Mpc}^{-1})^4,
$$
equivalent to $Q_{\rm rms}=18.0\, \mu{\rm K}$, or $\epsilon=3.07\times 10^{-5}$
in the notation of Peacock (1991), with an rms error in density fluctuation
of 8\%. 

For low-density models, a naive analysis as in PD suggests that the
power spectrum should depend on $\Omega$ and the growth factor $g$ as
$P\propto g^2/\Omega^2$. Because of time dependence of gravitational
potential (integrated Sachs-Wolfe effect) and spatial curvature, this
expression is not exact, although it captures the main effect. From the
data of White \& Bunn (1995), a better approximation is
$$
\Delta^2(k)\propto {g^2\over \Omega^2}\; g^{0.7}.
$$
This applies for low-$\Omega$ models both with and without
vacuum energy, with a maximum error of 2\% in density fluctuation
provided $\Omega>0.2$ (and gives the same $\sigma_8$ values as
G\'orski {\it et al.} (1995), when the appropriate $\Gamma^*$ corrections
are made, to within 3\%).
Since the rough power-law dependence of
$g$ is $g\simeq \Omega^{0.65}$ and $\Omega^{0.23}$ for open and flat
models respectively, we see that the implied density fluctuation
amplitude scales approximately as $\Omega^{-0.12}$ and $\Omega^{-0.69}$
for these two cases. The dependence is very weak for open models,
but vacuum energy implies very much larger fluctuations.
These results are illustrated for CDM spectra in Figure \nextfig, which
shows $\sigma_8$ as a function of $\Gamma^*$ for three models.
For $\Gamma^*=0.25$, open low-density models are close to the
required $\sigma_8=1$, whereas flat models have an amplitude
perhaps a factor 2 too high. Einstein-de Sitter models have
$\sigma_8=0.65$, which is only slightly high.

\fig{The clustering normalization $\sigma_8$ as a function of $\Gamma^*$,
predicted from COBE assuming scale-invariant primordial fluctuations.
Note that flat low-density models require a much larger normalization
than do open models.}
{7.0}

What if we tilt the spectrum?
For tilt, one evaluates $\sigma_8$ as in the above
no-tilt case, and then scales as
$$
\sigma_8 \propto \exp[2.3(n-1)];
$$
since $\sigma_8$ measures the power spectrum at 
$k\simeq 0.2 \hompc$, this corresponds to
saying that COBE determines the spatial power spectrum at an
effective wavenumber of about $0.002\,h\,{\rm Mpc}^{-1}$.
If gravity waves are included with the usual inflationary
coupling between wave amplitude and tilt, the effect
increases to
$$
\sigma_8 \propto \exp[4.3(n-1)].
$$
The flat models can then have $\sigma_8$ reduced by the required factor of 2,
but a substantial
degree of tilt is needed ($n\simeq 0.70$ or 0.84, the latter figure including
gravity waves). These are significantly larger degrees of tilt than would be
expected from at least the simplest inflationary models.

In any case, the CDM model is, as we have seen, in some
difficulty as a general description of the spectrum. A more
robust datum is probably the power on the largest
reliable scales:
$$
\Delta^2_{\rm opt}(k=0.02 \hompc) \simeq 0.005,
$$
which is to be compared to a COBE scale-invariant prediction
of 0.0054. Scaling as $\Omega^{-0.24}$ or $\Omega^{-1.38}$
boosts this by a factor of 1.5 (open $\Omega=0.2$) or
9.2 (flat $\Omega=0.2$). The former factor is within the
plausible effect of a transfer function, but the latter is not.
The required tilt to remove the additional factor 6 in power
is gross: $n=0.2$, pivoting the spectrum about $k=0.002\,h\,{\rm Mpc}^{-1}$.
Allowing for gravity waves improves this to $n=0.7$, but the conclusion
remains that low-density flat models require an extremely large
degree of tilt in order to be viable.

\titlec{}{What does it all mean?}
What then is the interpretation of the spectrum?
A CDM spectrum with $\Gamma^*\simeq 0.25$
is not consistent with $\Omega=1$ and any plausible
estimate for $h$. However, even the low-$\Gamma^*$
spectra are probably of the wrong shape, so it is not clear
if one can argue from the best-fitting $\Gamma^*$
for $\Omega<1$.

Interesting alternatives with high density are
either mixed dark matter (Holtzman 1989; Klypin {\it et al.} 1993),
or non-Gaussian pictures such as cosmic strings + HDM,
where the lack of a detailed prediction for
the power spectrum helps ensure that the model is
not yet excluded (Albrecht \& Stebbins 1992).
Isocurvature CDM is also attractive in that it
gives a rather sharper bend at the break scale,
as the data seem to require. However, this model conflicts
strongly with limits on CMB anisotropies, and cannot be correct.
(Efstathiou \& Bond 1986).
Mixed dark matter seems rather ad hoc, but may
be less so if it is possible to produce both
hot and cold components from a single particle,
with a Bose condensate playing the role of
the cold component (Madsen 1992; Kaiser, Malaney
\& Starkman 1993). The main problems with an MDM model are ones
generic to any model with a very flat high-$k$ spectrum in
a high-$\Omega$ universe: difficulty in forming high-redshift
objects and difficulty in achieving a steep correlation
function on small scales.

Alternatively, if the good fit of a low-density CDM transfer function is taken
literally, then perhaps this is a hint that the
epoch of matter-radiation equality needs to be delayed.
An approximate doubling of the number of relativistic degrees of
freedom would suffice -- but  this would
do undesirable violence to primordial
nucleosynthesis: any such boost would have
to be provided by a particle which decays after nucleosynthesis.
The apparent value of $\Omega h$ depends on the mass
and lifetime of the particle roughly as
$$
\left.\Omega h\right|_{\rm apparent} =
\Omega h\; [1+(m^2_{\rm keV} \tau_{\rm years})^{2/3}]^{-1/2}
$$
(Bardeen, Bond \& Efstathiou 1987; Bond \& Efstathiou 1991),
so a range of masses is possible.
Apart from making the observed large-scale structure, 
such a model yields a small-scale enhancement 
of power which could lead to early galaxy formation
(McNally \& Peacock 1995).

Of course, the simplest alternative is to admit that the
above attempts to save the Einstein-de Sitter model are too
contrived, and that $\Omega_m$ is $<1$. This would make a
low-$\Gamma^*$ model easier to understand, but it
introduces few new possible ways to alter the shape
of the linear power spectrum. There are of course the
oscillatory features expected in the power spectra of
baryon-dominated universes ({\it e.g.} Section 25 of Peebles 1993),
but these occur at too small $k$. A possibility would be
warm dark matter with a cutoff at $k\simeq 0.2 \hompc$,
but it may well be that something entirely new is needed.

Apart from lessening the
large-scale structure problems, low densities make life easier in two
other ways: the universe is made older for a given $h$ and
strong bias need not be accounted for.
The main difficulty which will need to be overcome in order
for the reality of a low-density universe to be accepted
is to understand the values for $\Omega$ which have emerged
from attempts to us large-scale peculiar velocities to
`weigh' the universe ({\it e.g.} Dekel 1994. The amplitude of such 
motions should not be a problem, as they should scale with
$\delta v\propto \Omega^{0.6} \sigma_8$, and the normalization
inferred from nonlinear systems is $\sigma_8\propto \Omega^{-0.5}$.
However, there is a problem in the degree of bias inferred from
comparing such velocities with the density field. In principle,
this allows one to determine $\beta\equiv \Omega^{0.6}/b$, and
values of $\beta\simeq 1$ have been inferred for IRAS galaxies.
If $\Omega\simeq 0.2$, this would require $b\simeq 0.5$.
However, Loveday {\it et al.} (1996) have found the much lower
value $\beta\simeq 0.5$ from redshift-space distortions in the
Stromlo-APM redshift survey. If these lower values are confirmed,
the strength of the case for low $\Omega$ and little bias would
become overwhelming.

This leaves unresolved the distinction between an open model and one in
which a significant vacuum energy keeps the inflationary
$k=0$ prediction (Efstathiou {\it et al.} 1990b).
The general case $\Omega_v\ne0$ and $k\ne 0$ is also a
logical possibility, but not an appealing one.
As constraints on $h$, galactic ages, and details of
the CMB measurements improve, there is
the happy possibility that a clear discrimination between these
alternatives may soon be reached.

\titleb{5.6}{Gaussianity}

\titlec{}{Skewness}
How are we to distinguish observationally whether the
density field of the Universe is Gaussian?
We need to look at more subtle statistics than
just the power spectrum. In principle, one might
use the higher-order $n$-point correlation functions,
since these are directly related to the power spectrum for a
Gaussian field.
On small scales, nonlinear evolution must produce non-Gaussian
behaviour. For any $\sigma$, a Gaussian density distribution will produce
a tail to unphysical $\delta<-1$ values. The lognormal model
$\delta\rightarrow \exp[\delta-\sigma^2/2]$ is the simplest
analytical modification which cures this problem (Coles \& Jones 1991). 
This distribution is skew, with a tail towards large values of $\delta$.
The question is whether the density field contains a greater degree
of non-Gaussianity than that induced by gravitational evolution.

Non-Gaussian behaviour may be measured through
the skewness parameter (not the skewness itself):
$$
S\equiv {\langle \delta^3 \rangle \over
[\langle \delta^2 \rangle ]^2},
$$
which can be calculated through second-order gravitational
perturbation theory, and should be a constant of order unity.
Gazta\~naga (1992) showed that the skewness parameter for the APM
galaxy survey was approximately constant with scale, at the
value expected for nonlinear gravitational evolution of a Gaussian
field. Does this mean (a) that conditions are Gaussian;
(b) that $b=1$? This is possible, but the effects of bias
need to be understood first. As a simple example, consider
a power-law modification of a lognormal field: $\rho' \propto \rho^b$.
Since this is generated by a Gaussian field, it is easy to find the
skewness parameter, which is
$$
S=2 + e^{b^2\sigma^2}.
$$
In the linear regime, the skewness is independent of
$b$ and so this sort of model would not violate the
observation that the moments $\langle \delta^3\rangle$
and $\langle \delta^2\rangle$ are in the correct ratio
for straightforward gravitational evolution without bias.

\titlec{}{Topology}
An interesting alternative probe of Gaussianity was suggested by
Gott, Melott \& Dickinson (1986): the
topology of the density field.
To visualise the main principles, it will
help to think initially about a 2D field.
Two extreme non-Gaussian fields would
consist either of discrete `hotspots'
surrounded by uniform density, or the opposite:
discrete `coldspots'; the picture
in either case is a set of polka dots.
Both of these cases are clearly non-Gaussian
just by symmetry: the contours of average
density will be simply-connected circles
containing regions which are all either
above or below the mean density,
but in a Gaussian field (or any symmetric case),
the numbers of hotspots and coldspots must
balance.

One might think that things would be
much the same in 3D: the obvious alternatives
are `meatball' or `Swiss cheese' models.
However, there is a third topological
possibility: that of the {\bf sponge}.
In our two previous examples, high- and
low-density regions were distinguished by their
{\bf connectivity} (whether it is possible to
move continuously between all points in
a given set). In contrast, 
a sponge has both classes of region being connected:
it is possible to swim to any point through the
holes, or to burrow to any point within the
body of the sponge; filling a sponge with cement
and etching away the sponge produces a cement sponge.
Again, just by the symmetry between overdensity
and underdensity, a Gaussian field in 3D must have
a sponge-like topology.

The above discussion has focused on the
properties of contour surfaces. These properties
can be studied quantitatively via the
{\bf genus}: the number of `holes' in a
surface (zero for a sphere, one for a
doughnut {\it etc.}). This is related to the
Gaussian curvature of the surface,
$K=1/(r_1r_2)$ (where $r_1$ and $r_2$ are the
two principal radii of curvature), via the 
Gauss-Bonnet theorem (see {\it e.g.} Dodson \& Poston 1977)
$$
C\equiv\int K\; dA=4\pi (1-G),
$$
where $G$ is the genus.
Topological results are sometimes
instead quoted in terms of 
the {\bf Euler-Poincar\'e characteristic}, which
is $-2$ times the genus.

For Gaussian fields, the expectation value of
the genus per unit volume (denoted by $g$) is
(see Hamilton, Gott \& Weinberg 1986)
$$
\boxit{
g={1\over 4\pi^2} \left[{-\xi''(0)\over\xi(0)}\right]^{3/2}
  (1-\nu^2)e^{-\nu^2/2}.
}
$$
For the median density contour ($\nu=0$), the
curvature is negative, implying that the surface
has genus greater than unity.
For $|\nu|>1$, however, the curvature is
positive -- as expected if there are no holes.
The contours become simply connected balls
around either isolated peaks or voids.

It is interesting to note that the genus
carries some information about the shape of the
power spectrum, not in the behaviour with $\nu$,
but in the overall scaling. For Gaussian filtering,
$$
g={1\over4\pi^2 R_{\ss G}^3}\left({3+n\over 3}\right)^{3/2} (1-\nu^2)
e^{-\nu^2/2},
$$
and so the effective spectral index can be determined in this way.

A similar procedure can be carried out in 2D
(see Melott {\it et al.} 1989; Coles \& Plionis 1991).
The Gauss-Bonnet theorem is now
$$
C\equiv\int K\; dA=2\pi (1-G),
$$
where the meaning of $1-G$ is the number of isolated
contours minus the number of 
contour loops within other loops (sometimes
the 2D genus is defined with the opposite sign; our
convention follows that in 3D and the signs below are consistent).
The result for the 2D genus per unit area is
$$
\boxit{
g=-{1\over(2\pi)^{3/2}} \left[{-\xi''(0)\over\xi(0)}\right]\; \nu\;
   e^{-\nu^2/2}.
}
$$
The 3D case is analogous but, as always, more messy; see BBKS.

\fig{Results from the Genus analysis applied to 3D redshift data,
taken from Gott  {\it et al.} (1989).
The small `meatball shift' seen here is argued by the authors
to be consistent with non-linear evolution from Gaussian
initial conditions.
It is interesting that the behaviour becomes more nearly
Gaussian as we move to deeper samples which allow larger
filtering lengths. Such plots constitute the strongest
evidence we have that cosmic structure did indeed form via
gravitational instability from Gaussian primordial fluctuations.}
{7.0}

Applications of this method to real data (Figure \lastfig) naturally reveal
departures from Gaussian behaviour -- one wishes to test
whether the initial conditions were Gaussian, realising
that nonlinear evolution will cause the field to become
non-Gaussian. This means that either $N$-body simulations
have to be used to predict the degree of non-Gaussian
behaviour (usually in the `meatball' direction), or one
is confined to smoothing the data heavily to probe only
large linear/angular scales. These should still be Gaussian,
but of course by smoothing over many small regions
there is the danger that the central limit theorem
will produce a Gaussian-like result in all cases.

{

\pretolerance 10000

\titlec{}{Fourier tests}
A third class of test was suggested by Feldman, Kaiser \& Peacock (1994),
and rests on measuring phase correlations  between different
mode amplitudes in the Fourier analysis of redshift surveys.
First note that the two-point function in $k$ space for a
homogeneous random statistical process is always a
delta-function:
$$
\langle \delta_k({\bf k})\delta_k^*({\bf k'})\rangle
={[2\pi]^3\over V} P(k) \; \delta_{\ss D}({\bf k-k'}),
$$
and that this applies for both Gaussian and non-Gaussian processes.
To prove this, write down the definition of $\delta_k$ twice and multiply
for different wavenumbers, using the reality of $\delta$:
$$
\delta_k \delta^*_{k'}= {1\over V^2} \int
\delta({\bf r})\delta({\bf r+x})\; e^{i{\bf k'\cdot x}}\;
d^3 x \; \int e^{i{\bf r\cdot (k-k')}}\; d^3r.
$$
Performing the ensemble average for a stationary statistical
process gives 
$$
\langle\delta({\bf r})\delta({\bf r+x})\rangle =\xi(x),
$$
independent of $r$. The integral over $r$ can now
be performed, showing that
$\langle \delta_k({\bf k})\delta_k^*({\bf k'})\rangle$ vanishes
unless $\bf k=k'$ in the discrete case, or that in the
continuum limit there is a delta-function in $k$ space.

}

This result applies in the limit of an infinite survey.
When there is a limited survey volume, delimited by the
mean density $\bar n({\bf r})$, we know that the Fourier
coefficients are convolved by the transform of $\bar n({\bf r})$.
There will therefore be a coherence length in $k$ space of
order the reciprocal of the survey depth, over which length
Fourier modes will have a significant two-point correlation.
In the generalization where the survey galaxies may be weighted,
Feldman, Kaiser \& Peacock show that the exact expression for the
two-point function is
$$
\langle \delta_k({\bf k})\delta_k^*({\bf k+\delta k})\rangle =
P(k) Q({\bf \delta k}) + S({\bf \delta k}),
$$
where
$$
\eqalign{
Q({\bf k}) &\equiv { \int w^2\bar n^2 \, \exp[i{\bf k\cdot r}]\; d^3r \over
\int w^2\bar n^2 \; d^3r } \cr
S({\bf k}) &\equiv { \int w^2\bar n \, \exp[i{\bf k\cdot r}]\; d^3r \over
\int w^2\bar n^2 \; d^3r } \cr
}
$$
Furthermore, in the case of Gaussian fields only,
this two-point function of amplitudes is simply
related to the two-point function for the power:
$$
\langle \delta P({\bf k})\delta P({\bf k+\delta k})\rangle =
|\langle \delta_k({\bf k})\delta_k^*({\bf k+\delta k})\rangle|^2.
$$

The significance of these results is that they allow direct constraints
to be placed on a large class of non-Gaussian models in which the
character of the linear density fluctuations is Gaussian, but
with a spatial modulation or {\bf intermittency}, so that there 
are `quiet' and `noisy' parts of space:
$$
\delta({\bf r}) \rightarrow \delta({\bf r}) \,[1 + M({\bf r})].
$$
Such a model was proposed on phenomenological grounds by Peebles (1983),
but might also be realised in inflationary models with multiple
scalar fields. The modulating field $M({\bf r})$ acts in the same
way as a mask imposed by observational selection, multiplying the
effective $\bar n$. It therefore convolves the transform of $\bar n$ and
broadens it. The signature of this form of non-Gaussianity is thus
an extended tail of correlated power in the transform of the survey,
and the agreement of the observed and expected power correlations
can be used to set limits on the non-Gaussianity. Figure 
\nextfig\ shows the application of this analysis to the combined
QDOT and 1.2-Jy surveys, and the excellent agreement provides a
strong piece of evidence for the Gaussian
nature of primordial fluctuations.

\fig{The normalized 2-point correlation of the power measured in the
combined QDOT and 1.2-Jy IRAS redshift surveys,
plotted against $k$ separation $\delta k / \hompc$
(Stirling \& Peacock 1996). Wavenumbers $k<0.1 \hompc$
are considered. The observed correlations follow
the expected form very closely, and limit any
modulation of the statistical properties of the
density field on 100 Mpc scales.
}{8.0}
 
\titlea{6}{Conclusions}
\titlec{}{The testability of inflation}
This brief summary of inflationary models has presented
the `party line' of many workers in the field for the
simplest ways in which inflation could happen. It is
a tremendous achievement to have a picture of this
level of detail -- but is it at all close to the truth?

What are the predictions of inflation?
The simplified package is (1) $k=0$; (2) scale-invariant,
Gaussian fluctuations, and these are essentially
the only tests discussed in the 1990 NATO symposium on
observational test of inflation (Shanks {\it {\it et al.} } 1991).
As the observations have hardened, however, there has been a tendency for the
predictions to weaken. The flatness prediction was long
taken to favour an $\Omega=1$ Einstein-de Sitter model, but
timescale problems have moved attention to models  with $\Omega_{\rm matter}
+\Omega_{\rm vacuum}=1$.
More recently, inflationary models have even been proposed
which might yield open universes. This is
achieved not by fine-tuning the amount of inflation, which
would certainly be contrived, but by appealing to the details
of the mechanism whereby inflation ends. It has been
proposed that quantum tunnelling might
create a `bubble' of open universe in a plausible way
(Bucher, Goldhaber \& Turok 1995; G\'orski {\it et al.} 1995). It seems that inflation
can never be disproved by the values of the global cosmological parameters.

A more characteristic inflationary prediction is the
gravity-wave background. Although it is not unavoidable,
the coupling between tilt and the gravity-wave contribution
to CMB anisotropies is a signature expected in many of
the simplest realizations of inflation. An observation of
such a coupling would certainly constitute very powerful
evidence that something like inflation did occur. Sadly,
this test loses its power as the degree of tilt becomes
smaller, which does seem to be the case observationally
(Peacock \& Dodds 1994).
The most convincing verification of inflation would of course
be the direct detection of the predicted flat spectrum of
gravity waves in the local universe. However, barring some
clever new technique, this will remain technologically
unfeasible for the immediate future.

It therefore seems likely that the debate over the truth of
inflation will continue without a clean resolution.
There are certainly points of internal consistency in
the theory which will attract further work: the
form of the potential, and whether it can be maintained
in the face of quantum corrections.
Also, there is one more fundamental difficulty, which
has been evaded until now.

\titlec{}{The $\Lambda$ problem}
All our discussion of inflation
has implicitly assumes that the zero of energy is set
at $\Lambda=0$ now, but there is no known principle of physics
which requires this to be so.
By experiment, $\Omega_{\rm vac}\ls 1$, which 
corresponds to a density in the
region of $10^{100}$ times smaller than the GUT value.
This fine tuning represents one of the major unsolved
problems in physics (Weinberg 1989).

Other ways of looking at the origin of the vacuum
energy include thinking of it as arising from
a Bose-Einstein condensate, or via contributions
from virtual particles.
In the latter case, the energy density would be the
rest mass times the number density of particles.
A guess at this is to set the separation of particles
at the Compton wavelength, $\hbar/mc$, yielding a
density
$$
\rho_{\rm vac}\simeq {m^4 c^3\over \hbar^3}.
$$
This exceeds the cosmological limit unless $m\ls10^{-8}m_e$.
One proposal for evading such a nonsensical
result was made by Zeldovich: perhaps we cannot
observe the rest mass of virtual particles
directly, and we should only count the
contribution of their gravitational interaction.
This gives instead
$$
\rho_{\rm vac}\simeq \left({Gm^2\over c\hbar}\right){m^4 c^3\over \hbar^3},
$$
which is acceptable if $m\ls 100m_e$ -- still far short of any plausible
GUT-scale or Planck-scale cutoff.

We are left with the strong impression that some
vital physical principle is missing.
It is perhaps just as well that the average
taxpayer who funds research in physics does not
realise how much difficulty we have in understanding
even nothing at all!

Fortunately, the existence of a non-zero vacuum density is
not entirely a philosophical conundrum, but is subject to
empirical verification in cosmology. 
If $\Lambda$ exists at above the level of a few tenths of the critical
density, it can be detected by a combination of geometrical
tests, its effect on cosmological ages, and detailed signatures
in CMB anisotropies.
The challenge of confirming or ruling out a cosmologically
significant vacuum energy is therefore developing into one of the
dominant themes of cosmology in the 1990s, and is an area
where we can reasonably hope to reach a decision within the
next few years.

 

\bigskip\noindent
 
\begrefchapter{References}

\ref Albrecht A., Stebbins A., 1992, Phys. Rev. Lett., 69, 2615
\ref Babul A., Postman M., 1990, ApJ, 359, 280
\ref Ballinger W.E., Peacock J.A., Heavens A.F., 1996, MNRAS, submitted.
\ref Bardeen J.M., Bond J.R., Efstathiou G., 1987, ApJ, 321, 28
\ref Bardeen J.M., Bond J.R., Kaiser N., Szalay A.S., 1986, ApJ, 304, 15 (BBKS)
\ref Baugh C.M., Efstathiou G., 1993, MNRAS, 265, 145
\ref Baugh C.M., Efstathiou G., 1994, MNRAS, 267, 323
\ref Blumenthal, G.R., Faber, S.M., Primack, J.R. \& Rees, M.J., 1984. {\it Nature}, {\bf 311}, 517.
\ref Bond J.R., Efstathiou G., 1991, Phys. Lett. B, 265, 245
\ref Bond, J.R., Cole, S., Efstathiou, G. \& Kaiser, N., 1991. {\it Astrophys. J.}, 379, 440
\ref Bower R.G., Coles P., Frenk C.S., White S.D.M., 1993, ApJ, 405, 403
\ref Bower, R.G., 1991. {\it Mon. Not. R. astr. Soc.}, {\bf  248}, 332.   
\ref Brandenberger, R.H., 1990. in {\it Physics of the Early Universe},     proc 36$^{th}$ Scottish Universities Summer School in Physics,     eds Peacock, J.A., Heavens, A.F. \& Davies, A.T. (Adam Hilger), p281.
\ref Bucher M., Goldhaber A.S., Turok N., 1995, {\it Phys. Rev. D}, {\bf 52}, 3314
\ref Bunn, E.F., Scott D., White M., 1995, ApJ, 441, 9
\ref Carlberg, R.G., 1991. {\it Astrophys. J.}, {\bf 367}, 385.
\ref Carlberg, R.G., Couchman, H.M.P. \& Thomas, P.A., 1990. {\it Astrophys. J.}, {\bf 352}, L29.
\ref Carroll S.M., Press W.H., Turner E.L., 1992, ARAA, 30, 499
\ref Cen R., Gnedin N.Y., Kofman L.A.,  Ostriker J.P., 1992, ApJ, 399, L11
\ref Cen R., Ostriker J.P., 1992, ApJ, 399, L113
\ref Cole, S. \& Kaiser, N., 1989. {\it Mon. Not. R. astr. Soc.}, {\bf 237}, 1127.
\ref Coles P., 1993, MNRAS, 262, 1065
\ref Coles, P. \& Jones, B.J.T., 1991.   {\it Mon. Not. R. astr. Soc.}, {\bf 248}, 1.
\ref Coles, P. \& Plionis, M, 1991.   {\it Mon. Not. R. astr. Soc.}, {\bf 250}, 75.
\ref Couchman, H.M.P. \& Carlberg, R.G., 1992. {\it Astrophys. J.}, 389, 453
\ref Davis M., Efstathiou G., Frenk C.S., White S.D.M., 1985, ApJ, 292, 371
\ref Davis, M. \& Peebles, P.J.E., 1983. {\it Astrophys. J.}, {\bf 267}, 465.
\ref Dekel, A. \& Rees, M.J., 1987. {\it Nature}, {\bf 326}, 455.
\ref Dekel, A., 1994, {ARAA}, {32}, 371
\ref Dodson, C.T.J. \& Poston, T., 1977. {\it Tensor Geometry\/} (London; Pitman).
\ref Dressler, A., 1980, ApJ, 236, 351.
\ref Efstathiou G., Davis M., White S.D.M., Frenk C.S., 1995, ApJ Suppl., 57, 241
\ref Efstathiou G., Sutherland W.J., Maddox S.J., 1990a, Nature, 348, 705
\ref Efstathiou, G. \& Bond, J.R., 1986.    {\it Mon. Not. R. astr. Soc.}, {\bf 218},103.
\ref Efstathiou, G., Bernstein, G., Katz, N., Tyson, T. \& Guhathakurta, P., 1991. {\it Astrophys. J.}, 380, 47
\ref Efstathiou, G., Kaiser, N., Saunders, W., Lawrence, A.,   Rowan-Robinson, M., Ellis, R.S. \& Frenk, C.S., 1990b.   {\it Mon. Not. R. astr. Soc.}, {\bf 247}, 10P.
\ref Feldman H.A., Kaiser N.,  Peacock J.A., 1994, ApJ, 426, 23
\ref Fisher K.B., Davis M., Strauss M.A., Yahil A.,  Huchra J.P., 1994, MNRAS, 267, 927
\ref G\'orski K.M., Ratra B., Sugiyama N., Banday A.J., 1995, {\it Astrophys. J.}, 444, L65
\ref Gazta\~naga E., 1992, ApJ, 398, L17
\ref Gott, J.R. III, Melott, A.L. \& Dickinson, M., 1986.  {\it Astrophys. J.}, {\bf 306}, 341.
\ref Gott, J.R. III, {\it et al.} 1989. {\it Astrophys. J.}, {\bf 340}, 625.
\ref Guth, A.H., 1981. {\it Phys. Rev. D}, {\bf 23}, 347
\ref Hamilton A.J.S., Kumar P., Lu E.,  Matthews A., 1991, ApJ, 374, L1 (HKLM)
\ref Hamilton, A.J.S., Gott, J.R. III \& Weinberg, D.H., 1986.   {\it Astrophys. J.}, {\bf 309}, 1.
\ref Henry, J.P. \& Arnaud, K.A., 1991. {\it Astrophys. J.}, {\bf 372}, 410.
\ref Hernquist L., Bouchet F.R., Suto Y., 1991, ApJ Suppl., 75, 231.
\ref Hockney R. W., Eastwood J. W., 1988, ``Computer Simulations Using Particles'', IOP Publishing, Bristol.
\ref Holtzman J.A., 1989, ApJs, 71, 1
\ref Jain B., Mo H.J., White S.D.M., 1995, MNRAS, 276, L25 
\ref Kaiser N., 1984, ApJ,  284, L9
\ref Kaiser N., 1987, MNRAS, 227, 1
\ref Kaiser, N., 1986a. {\it Mon. Not. R. astr. Soc.}, {\bf 219}, 785.
\ref Kaiser N., Malaney R.A., Starkman G.D., 1993, Phys. Rev. Lett., 71, 1128
\ref Klypin A., Holtzman J., Primack J., Reg\H os E., 1993, ApJ, 416, 1
\ref Kofman, L. \& Linde, A., 1987. {\it Nucl. Phys.}, {\bf B282}, 555.
\ref Kolb E.W., Turner M.S., 1990, {\it The Early Universe\/} (Addison-Wesley)
\ref Lacey C., Cole S., 1993, MNRAS, 262, 632
\ref Lahav O., Lilje P.B., Primack J.R., Rees M.J., 1991, MNRAS, 251, 128
\ref Liddle A.R., Lyth D., 1993, Phys. Rep., 231, 1
\ref Linde, A., 1986. {\it Phys. Lett.}, {\bf 175B}, 295.
\ref Linde, A., 1989. {\it Inflation and quantum cosmology}, Academic Press, Boston.
\ref Loveday J., Maddox S.J., Efstathiou G., Peterson B.A., 1995, ApJ, 442, 457
\ref Loveday J., Efstathiou G., Maddox S.J., Peterson B.A., 1996, ApJ, in press
\ref Madsen J., 1992, phys. Rev. Lett., 69, 571
\ref Mann R.G., Peacock J.A., Heavens A.F., 1996, MNRAS, in preparation.
\ref McNally S.J., Peacock J.A., 1995, MNRAS,  277, 143
\ref Melott, A.L., Cohen, A.P., Hamilton, A.J.S.,  Gott, J.R. III \& Weinberg, D.H., 1989. {\it Astrophys. J.}, {\bf 345}, 618.
\ref Mo H.J., Jing Y.P., B\"orner G., 1993, MNRAS, 264, 825
\ref Mo H.J., McGaugh S.S., Bothun G.D., 1994, MNRAS, 267, 129
\ref Mukhanov V.F., Feldman H.A., Brandenberger R.H., 1992, Phys. Rep., 215, 203
\ref Peacock J.A., 1991, MNRAS,  253, 1P
\ref Peacock J.A., 1992, in Martinez V., Portilla M., S\'aez D., eds, New insights into the Universe, Proc. Valencia summer school (Springer, Berlin), p1
\ref Peacock J.A., Dodds S.J., 1994. MNRAS, 267, 1020
\ref Peacock J.A., Dodds S.J., 1996. MNRAS, submitted
\ref Peacock, J.A. \& Heavens, A.F., 1990, {\it Mon. Not. R. astr. Soc.}, {\bf 243}, 133
\ref Peacock, J.A., 1990. {\it Mon. Not. R. astr. Soc.}, {\bf 243}, 517.
\ref Peebles P.J.E., 1980, The Large-Scale Structure of the Universe.  Princeton Univ. Press, Princeton, NJ
\ref Peebles P.J.E., 1993, Principles of physical cosmology.  Princeton Univ. Press, Princeton, NJ
\ref Peebles P.J.E., 1983, ApJ, 274, 1
\ref Press W.H., Schechter P., 1974, ApJ, 187, 425
\ref Press W.H., Vishniac E.T., 1980, ApJ, 239, 1
\ref Raymond, J.C., Cox, D.P. \& Smith, B.W., 1976. {\it Astrophys. J.}, {\bf 204}, 290.
\ref Rees, M.J. \& Ostriker, J.P., 1977. {\it Mon. Not. R. astr. Soc.}, {\bf 179}, 541.
\ref Rees, M.J., 1985. {\it Mon. Not. R. astr. Soc.}, {\bf 213}, 75P.
\ref Saunders W., Rowan-Robinson M., Lawrence A., 1992, MNRAS, 258, 134
\ref Saunders, W., Frenk, C., Rowan-Robinson, M., Efstathiou, G.,     Lawrence, A., Kaiser, N., Ellis, R., Crawford, J., Xia, X.-Y. \&  Parry, I., 1991. {\it Nature}, {\bf 349}, 32.
\ref Shanks T., Boyle B.J., 1994, MNRAS, 271, 753
\ref Shanks T. {\it et al.} (eds), 1991, {\it Observational tests of cosmological inflation}, NATO ASI C264, Kluwer
\ref Starobinsky A.A., 1985, Sov. Astr. Lett., 11, 133
\ref Stirling A.J., Peacock J.A., 1996, MNRAS, in preparation.
\ref Strauss M.A., Davis M., Yahil Y., Huchra J.P., 1992, ApJ, 385, 421
\ref Sugiyama N., 1995, ApJ Suppl, 100, 281
\ref Thuan, T.X., Gott, J.R. III \& Schneider, S.E., 1987. {\it Astrophys. J.}, {\bf 315}, L93.
\ref Valls-Gabaud, D., Alimi, J.-M.,  \& Blanchard, A., 1989. {\it Nature}, {\bf 341}, 215.
\ref Vilenkin, A. \& Shellard, E.P.S., 1994. {\it Cosmic strings and other topological defects}, CUP.
\ref Weinberg, S. 1989.  {\it Rev. Mod. Phys.}, {\bf  61}, 1.
\ref West, M.J. \& Richstone, D.O., 1988. {\it Astrophys. J.}, {\bf 335}, 532.
\ref White M., Bunn E.F., 1995, ApJ, 450, 477
\ref White S.D.M., Efstathiou G., Frenk C.S., 1993, MNRAS, 262, 1023
\ref White, S.D.M., Davis, M., Efstathiou, G. \& Frenk, C.S., 1987.  {\it Nature}, {\bf 330}, 451.
\ref White, S.D.M., Navarro, J.F., Evrard, A.E. \& Frenk, C.S., 1993.  {\it Nature}, {\bf 366}, 429.
\ref Williams, B.G., Heavens, A.F.,  Peacock, J.A. \& Shandarin, S.F., 1991a. {\it Mon. Not. R. astr. Soc.}, {\bf 250}, 458.
\ref Yi, I. \& Vishniac, E.T., 1993. {\it Astrophys. J. Suppl.}, {\bf 86}, 333.
\endref
\byebye